\documentclass[rmp,aps,amsfonts,twocolumn,superscriptaddress,preprintnumbers]
{revtex4}
\usepackage{graphics,epsfig}
\begin{document}
\newcommand{\etal}{{et~al.}}
\def\cite#1{\citep{#1}}
\def\citeasnoun#1{\citet{#1}}
\newcommand\be{\begin{equation}}
\newcommand\ee{\end{equation}}
\def\bea{\begin{eqnarray}}
\def\eea{\end{eqnarray}}
\def\crt{\nonumber \\}
\def\eqn#1#2{\be \label{eq:#1} #2  \ee}
\def\CPbar{\hbox{{\rmCP}\hskip-1.80em{/}}}
\def\ctn#1{{{\cite{#1}}}}
\def\rfn#1{Eq. {({\ref{eq:#1}})}}
\def\rfs#1{Sec.~{\ref{sec:#1}}}
\def\lref#1#2{{\bibitem{#1}#2}}
%
\def\np#1#2#3{Nucl. Phys. {\bf B#1} (#2) #3}
\def\pl#1#2#3{Phys. Lett. {\bf #1B} (#2) #3}
\def\prl#1#2#3{Phys. Rev. Lett. {\bf #1} (#2) #3}
\def\pr#1#2#3{Phys. Rev. {\bf #1} (#2) #3}
\def\aph#1#2#3{Ann. Phys. {\bf #1} (#2) #3}
\def\prep#1#2#3{Phys. Rep. {\bf #1} (#2) #3}
\def\rmp#1#2#3{Rev. Mod. Phys. {\bf #1}}
\def\cmp#1#2#3{Comm. Math. Phys. {\bf #1} (#2) #3}
\def\mpl#1#2#3{Mod. Phys. Lett. {\bf #1} (#2) #3}
\def\ptp#1#2#3{Prog. Theor. Phys. {\bf #1} (#2) #3}
\def\jhep#1#2#3{JHEP {\bf#1}(#2) #3}
\def\jmp#1#2#3{J. Math Phys. {\bf #1} (#2) #3}
\def\cqg#1#2#3{Class.~Quantum Grav. {\bf #1} (#2) #3}
\def\ijmp#1#2#3{Int.~J.~Mod.~Phys. {\bf #1} (#2) #3}
\def\atmp#1#2#3{Adv.~Theor.~Math.~Phys.{\bf #1} (#2) #3}
\def\ap#1#2#3{Ann.~Phys. {\bf #1} (#2) #3}
%
%
\def\Nc{Noncommutative}
\def\nc{noncommutative}
\def\con{conventional} 
%
%
\def\IB{\relax\hbox{$\inbar\kern-.3em{\rm B}$}}
\def\IC{\relax\hbox{$\inbar\kern-.3em{\rm C}$}}
\def\ID{\relax\hbox{$\inbar\kern-.3em{\rm D}$}}
\def\IE{\relax\hbox{$\inbar\kern-.3em{\rm E}$}}
\def\IF{\relax\hbox{$\inbar\kern-.3em{\rm F}$}}
\def\IG{\relax\hbox{$\inbar\kern-.3em{\rm G}$}}
\def\IGa{\relax\hbox{${\rm I}\kern-.18em\Gamma$}}
\def\IH{\relax{\rm I\kern-.18em H}}
\def\IK{\relax{\rm I\kern-.18em K}}
\def\IL{\relax{\rm I\kern-.18em L}}
\def\IP{\relax{\rm I\kern-.18em P}}
\def\IR{\relax{\rm I\kern-.18em R}}
\def\IT{{\bf T}}
\def\IZ{\relax\ifmmode\mathchoice{\hbox{\cmss Z\kern-.4em Z}}{\hbox{\cmss Z\kern-.4em Z}}
{\lower.9pt\hbox{\cmsss Z\kern-.4em Z}} {\lower1.2pt\hbox{\cmsss
Z\kern-.4em Z}} \else{\cmss Z\kern-.4em Z}\fi}
\def\II{\relax{\rm I\kern-.18em I}}
\def\IIa{{\II a}}
\def\IIb{{\II b}}
\def\IX{{\bf X}}
\def\ttb{Type $\II$B string theory}
\def\ndt{{\noindent}}
\def\bx{{\bf G}}
\def\sssec#1{\ndt$\underline{#1}$}
\def\CA{{\cal A}}
\def\CB{{\cal B}}
\def\CC{{\cal C}}
\def\CD{{\cal D}}
\def\CE{{\cal E}}
\def\CF{{\cal F}}
\def\CG{{\cal G}}
\def\CH{{\cal H}}
\def\CI{{\cal I}}
\def\CJ{{\cal J}}
\def\CK{{\cal K}}
\def\CL{{\cal L}}
\def\CM{{\cal M}}
\def\CN{{\cal N}}
\def\CO{{\cal O}}
\def\CP{{\cal P}}
\def\CQ{{\cal Q}}
\def\CR{{\cal R}}
\def\CS{{\cal S}}
\def\CT{{\cal T}}
\def\CU{{\cal U}}
\def\CV{{\cal V}}
\def\CW{{\cal W}}
\def\CX{{\cal X}}
\def\CY{{\cal Y}}
\def\CZ{{\cal Z}}
\def\B#1{{\bf #1}}
\def\BH{\B{H}}
\def\BM{\B{M}}
\def\BN{\B{N}}
\def\Ba{\B{a}}
\def\Bb{\B{b}}
\def\Bk{\B{k}}
\def\Br{\B{r}}
\def\p{\partial}
\def\pb{\bar{\partial}}
\def\TeV{{\rm TeV}}
\def\Mat{{\rm Mat}}
\def\End{{\rm End}}
\def\dir{{\CD}\hskip -6pt \slash \hskip 5pt}
\def\dd{{\rm d}}
\def\Dslash{\rlap{\hskip0.2em/}D}
\def\cb{\bar{c}}
\def\ib{\bar{i}}
\def\jb{\bar{j}}
\def\kb{\bar{k}}
\def\lb{\bar{l}}
\def\mb{\bar{m}}
\def\nb{\bar{n}}
\def\ub{\bar{u}}
\def\wb{\bar{w}}
\def\sb{\bar{s}}
\def\tb{\bar{t}}
\def\vb{\bar{v}}
\def\xb{\bar{x}}
\def\zb{\bar{z}}
\def\Cb{\bar{C}}
\def\Db{\bar{D}}
\def\Tb{\bar{T}}
\def\Zb{\bar{Z}}
\def\half{{1\over 2}}
\def\bra#1{{\langle}#1|}
\def\ket#1{|#1\rangle}
\def\bbra#1{( #1|}
\def\kket#1{|#1 )}
\def\vev#1{\langle{#1}\rangle}
\def\codim{{\mathop{\rm codim}}}
\def\cok{{\rm cok}}
\def\rank{{\rm rank}}
\def\coker{{\mathop {\rm coker}}}
\def\diff{{\rm diff}}
\def\Diff{{\rm Diff}}
\def\Tr{{\rm Tr~}}
\def\tr{{\rm tr~}}
\def\Id{{\rm Id}}
\def\vol{{\rm vol}}
\def\Vol{{\rm Vol}}
\def\c{\cdot}
\def\sdtimes{\mathbin{\hbox{\hskip2pt\vrule height 4.1pt depth -.3pt
width.25pt\hskip-2pt$\times$}}}
\def\ch{{\rm ch}}
\def\Det{{\rm Det}}
\def\DET{{\rm DET}}
\def\Hom{{\rm Hom}}
\def\dim{{\rm dim}}
\def\Re{{\rm Re~}}
\def\Im{{\rm Im~}}
\def\imp{$\Rightarrow$}
\def\danger{{NB:}}
\def\Lie{{\rm Lie}}
\def\mod{{\rm mod}}
\def\lieg{{\underline{\bf g}}}
\def\liet{{\underline{\bf t}}}
\def\liek{{\underline{\bf k}}}
\def\lies{{\underline{\bf s}}}
\def\lieh{{\underline{\bf h}}}
\def\clieg{{\underline{\bf g}}_{\scriptscriptstyle{\IC}}}
\def\cliet{{\underline{\bf t}}_{\scriptstyle{\IC}}}
\def\cliek{{\underline{\bf k}}_{\scriptscriptstyle{\IC}}}
\def\clies{{\underline{\bf s}}_{\scriptstyle{\IC}}}
\def\CCK{K_{\scriptscriptstyle{\IC}}}
\def\inbar{\,\vrule height1.5ex width.4pt depth0pt}
\def\IIa{IIa}
\font\cmss=cmss10 \font\cmsss=cmss10 at 7pt
\def\sdtimes{\mathbin{\hbox{\hskip2pt\vrule height 4.1pt
depth -.3pt width .25pt\hskip-2pt$\times$}}}
\def\ao{{\bf a}}
\def\co{{\bf a}^{\dagger}}

\def\a{{\alpha}}
\def\ap{{l_s^2}}
\def\b{{\beta}}
\def\d{{\delta}}
\def\g{{\gamma}}
\def\e{{\epsilon}}
\def\z{{\zeta}}
\def\ve{{\varepsilon}}
\def\vf{{\varphi}}
\def\kk{{\kappa}}
\def\m{{\mu}}
\def\n{{\nu}}
\def\u{{\Upsilon}}
\def\l{{\lambda}}
\def\s{{\sigma}}
\def\t{{\theta}}
\def\o{{\omega}}
\def\seealso{{See also }}
\def\hepth#1{}
\def\hepph#1{}
\preprint{ITEP-TH-31/01}
\preprint{IHES/P/01/27}
\preprint{RUNHETC-2001-18}
\title{Noncommutative Field Theory}

\author{Michael R. Douglas}
\email{mrd@physics.rutgers.edu}
\affiliation{Department of Physics and Astronomy, Rutgers University,
Piscataway NJ 08855 U.S.A.}
\affiliation{Institut des Hautes Etudes Scientifiques, 35 route des
Chartres, Bures-sur-Yvette France 91440}
\author{Nikita A.~Nekrasov}
\affiliation{Institut des Hautes Etudes Scientifiques, 35 route des
Chartres, Bures-sur-Yvette France 91440}
\affiliation{Institute for Theoretical and Experimental Physics, 117259
Moscow Russia}
\begin{abstract}  
We review the generalization of field theory to space-time with
noncommuting coordinates, starting with the basics and covering
most of the active directions of research.
Such theories are now known to emerge from
limits of M theory and string theory, and to describe
quantum Hall states.
In the last few years they have been studied intensively, and
many qualitatively new phenomena have been discovered, both on the
classical and quantum level.

Submitted to Reviews of Modern Physics. 
\end{abstract}                                                                 

\maketitle
\tableofcontents


\section{INTRODUCTION}
\label{sec:intro}

Noncommutativity is an age-old theme in mathematics and physics.  The
noncommutativity of spatial rotations in three and more dimensions
is deeply ingrained in us.  Noncommutativity is the central
mathematical concept expressing uncertainty in quantum mechanics,
where it applies to any pair of conjugate variables, such as position
and momentum.  In the presence of a magnetic field, even momenta fail
to mutually commute.

One can just as easily imagine that position measurements might fail
to commute, and describe this using noncommutativity of the coordinates.
The simplest
noncommutativity one can postulate is the commutation relations
\begin{equation}
\label{eq:noncom} [ x^i ,x^j ] = i\theta^{ij},
\end{equation}
\def\refnoncom{Eq. (\ref{eq:noncom})}
with a parameter $\theta$ which is an antisymmetric (constant) tensor
of dimension $({\rm length})^2$.

As has been realized independently many times, at least as early as
\cite{Snyder:1947a}, there is a simple modification to quantum field theory
obtained by taking the position coordinates to be noncommuting
variables.  Starting with a conventional field theory Lagrangian and
interpreting the fields as depending on coordinates satisfying
\refnoncom, one can follow the usual development of perturbative
quantum field theory with surprisingly few changes, to define a large
class of ``\nc\ field theories.''

It is this class of theories which our review will focus on.
Until recently, such theories had not been studied very
seriously.  Perhaps the main reason for this is that postulating an
uncertainty relation between position measurements will {\it a
priori} lead to a nonlocal theory, with all of the attendant
difficulties.  A secondary reason is that noncommutativity of the
space-time coordinates generally conflicts with Lorentz
invariance, as is apparent in \refnoncom.  Although it is not implausible
that a theory defined using such coordinates could be effectively
local on length scales longer than that of $\theta$, it is harder to
believe that the breaking of Lorentz invariance would be unobservable
at these scales.

Nevertheless, one might postulate noncommutativity for a number of
reasons.  Perhaps the simplest is that it might improve the
renormalizability properties of a theory at short distances or even
render it finite.  Without giving away too much of our story, we
should say that this is of course not obvious {\it a priori} and a
\nc\ theory might turn out to have the same or even worse
short distance behavior than a conventional theory.

Another motivation is the long-held belief that in quantum
theories including gravity, space-time must change its nature at
distances comparable to the Planck scale.  Quantum
gravity has an uncertainty principle which prevents one from measuring
positions to better accuracies than the Planck length: the momentum and
energy required to make such a measurement will itself modify the
geometry at these scales \cite{DeWitt:1956a}.
One might wonder if these effects could
be modeled by a commutation relation such as \refnoncom.

A related motivation is that there are reasons to believe
that any theory of quantum gravity will not be local in the
conventional sense.  Nonlocality brings with it deep conceptual and
practical issues which have not been well understood, and one
might want to understand them in the simplest examples first, before
proceeding to a more realistic theory of quantum gravity.

This is one of the main motivations for the intense current activity
in this area among string theorists.  String theory is not local in
any sense we now understand, and indeed has more than one parameter
characterizing this nonlocality: in general, it is controlled by the
larger of the Planck length and the ``string length,'' the average
size of a string.  It was discovered in 
\cite{Connes:1998cr,Douglas:1998fm}
that simple limits of M theory and string theory lead
directly to \nc\ gauge theories, which appear far simpler
than the original string theory yet keep some of this nonlocality.

One might also study \nc\ theories as interesting analogs of
theories of more direct interest, such as Yang-Mills theory.  An
important point in this regard is that many theories of interest in
particle physics are so highly constrained that they are difficult to
study.  For example, pure Yang-Mills theory with a definite simple
gauge group has no dimensionless parameters with which to make a
perturbative expansion or otherwise simplify the analysis.  From this
point of view it is quite interesting to find any sensible and
non-trivial variants of these theories.

Now, physicists have constructed many, many variations of Yang-Mills
theory in the search for regulated (UV finite) versions as well as
more tractable analogs of the theory.  A particularly interesting
example in the present context is the twisted Eguchi-Kawai model
\cite{Eguchi:1983ta,Gonzalez-Arroyo:1983hz},
which in some of its forms, especially that of 
\citeasnoun{Gonzalez-Arroyo:1983ac}, is a \nc\ gauge theory.  
This model was developed in the study of the large $N$ limit
of Yang-Mills theory \cite{'tHooft:1974jz} and we will see that
\nc\ gauge theories show many analogies to this limit 
\cite{Filk:1996dm,Minwalla:1999px}, suggesting that they should play an
important role in the circle of ideas relating large $N$ gauge theory
and string theory \cite{Polyakov:1987ez,Aharony:1999ti}.

\Nc\ field theory is also known to appear naturally in condensed
matter theory.  The classic example (though not always discussed using
this language) is the theory of electrons in a magnetic field
projected to the lowest Landau level, which is naturally thought of as
a \nc\ field theory.  Thus these ideas are relevant to the theory of
the quantum Hall effect \cite{Girvin:1987a}, and indeed, \nc\ geometry
has been found very useful in this context \cite{Bellissard:1994a}.
Most of this work has treated noninteracting electrons, and it seems
likely that introducing field theoretic ideas could lead to further
progress.

It is interesting to note that despite the many physical motivations
and partial discoveries we just recalled, \nc\ field theory
and gauge theory was first clearly formulated by mathematicians
\cite{Connes:1987a}.
This is rather unusual for a theory of significant interest to physicists;
usually, as with Yang-Mills theory, the flow goes in the other direction.

An explanation for this course of events might be found in the deep
reluctance of physicists to regard a nonlocal theory as having any
useful space-time interpretation.  Thus, even when these theories
arose naturally in physical considerations, they tended to be regarded
only as approximations to more conventional local theories, and not as
ends in themselves.  Of course such sociological questions rarely have
such pat answers and we will not pursue this one further except to
remark that, in our opinion, the mathematical study of these theories
and their connection to \nc\ geometry has played an
essential role in convincing physicists that these are not arbitrary
variations on conventional field theory but indeed a new universality
class of theory deserving study in its own right.  Of course this
mathematical work has also been an important aid to the more prosaic
task of sorting out the possibilities, and is the source for many
useful techniques and constructions which we will discuss in detail.

Having said this, it seems that the present trend is that the
mathematical aspects appear less and less central to the physical
considerations as time goes on.  While it is too early to judge the
outcome of this trend and it seems certain that the aspects which
traditionally have benefited most from mathematical influence will
continue to do so (especially, the topology of gauge field
configurations, and techniques for finding exact solutions), we have
to some extent deemphasized the connections with \nc\ %
geometry in this review.  This is partly to make the material
accessible to a wider class of physicists, and partly because many
excellent books and reviews, starting with \citeasnoun{Connes:1994a}, and
including \cite{Nekrasov:2000ih} focusing on classical solutions of
\nc\ gauge theory, \cite{Konechny:2000dp} focusing
on duality properties of gauge theory on a torus, 
as well as \cite{Varilly:1997qg,Gracia-Bondia:2001tr},
cover the material
starting from this point of view.  We maintain one section which
attempts to give an overview of aspects for which a more mathematical
point of view is clearly essential.

Although many of the topics we will discuss were motivated by and
discovered in the context of string theory, we have also taken the
rather unconventional approach of separating the discussion of \nc\
field theory from that of its relation to string theory, to the extent
that this was possible.  An argument against this approach is that the
relation clarifies many aspects of the theory, as we hope will become
abundantly clear upon reading section \ref{sec:string}.  However it is
also true that string theory is not a logical prerequisite for
studying the theory and we feel the approach we took better
illustrates its internal self-consistency (and the points where this
is still lacking).  Furthermore, if we hope to use \nc\ field theory
as a source of {\it new} insights into string theory, we need to be
able to understand its physics without relying too heavily on the
analogy.  We also hope this approach will have the virtue of broader
accessibility, and perhaps help in finding interesting applications
outside of string theory.  Reviews with a more string-theoretic
emphasis include \cite{Harvey:2001yn}
which discusses solitonic solutions and their relations to string theory.

Finally, we must apologize to the many whose work we were not able to
treat in the depth it deserved in this review, a sin we have tried to
atone for by including an extensive bibliography.

\section{KINEMATICS}
\label{sec:kin}

\subsection{Formal considerations}
\label{sec:formal}

Let us start by defining \nc\ field theory in a somewhat pedestrian
way, by proposing a configuration space and action functional from
which we could either derive equations of motion, or define a
functional integral.  We will discuss this material from a more
mathematical point of view in \rfs{math}.

\ndt {\sl Conventions.} 
Throughout the review we use the
following notations: Latin indices $i,j,k, \ldots$ denote
space-time indices, Latin indices from the begining of the
alphabet $a,b, \ldots$ denote commutative dimensions, Greek
indices $\m,\n, \ldots$ enumerate particles, vertex operators,
etc, while Greek indices from the begining of the alphabet $\a,
\b, \ldots$ denote noncommutative directions.

In contexts where we simultaneously discuss a noncommuting
variable or field and its commuting analog, we will 
use the ``hat'' notation: $x$ is the commuting analog to $\hat x$.
However, in other contexts, we will not use the hat.

\subsubsection{The algebra}

The primary ingredient in the definition is an associative but not
necessarily commutative algebra, to be denoted $\CA$.
The product of elements $a$ and $b$ of $\CA$ will be denoted $ab$,
$a\cdot b$, or $a \star b$.  This last notation (the ``star product'')
has a special connotation, to be discussed shortly.

An element of this algebra will correspond to a configuration of a
classical complex scalar field on a ``space'' $M$.  
Suppose first that
$\CA$ is commutative.  The primary example of a commutative
associative algebra is the algebra of complex valued functions on a
manifold $M$, with addition and multiplication defined pointwise:
$(f+g)(x) = f(x)+g(x)$ and $(f\cdot g)(x) = f(x) g(x)$.  In this case,
our definitions will reduce to the standard ones for field theory on
$M$.

Although the mathematical literature is usually quite precise
about the class of functions (continuous, smooth, etc.) to be
considered, in this review we follow standard physical practice
and simply consider all functions which arise in reasonable
physical considerations, referring to this algebra as $\CA(M)$ or
(for reasons to be explained shortly) as ``$M_0$''.
If more precision is wanted, for most purposes one
can think of this as $C(M)$, the bounded continuous functions on
the topological manifold $M$.

The most elementary example of a \nc\ algebra is $\Mat_n$, the algebra of
complex $n\times n$ matrices.  Generalizations of this which are
almost as elementary are the algebras $\Mat_n(C(M))$ of $n\times n$
matrices whose matrix elements are elements of $C(M)$, and with
addition and multiplication defined according to the usual rules for
matrices in terms of the addition and multiplication on $C(M)$.  This
algebra contains $C(M)$ as its center (take functions times the
identity matrix in $\Mat_n$).

Clearly elements of $\Mat_n(C(M))$ correspond to configurations of a
matrix field theory.  Just as one can gain some intuition about
operators in quantum mechanics by thinking of them as matrices, this
example already serves to illustrate many of the formal features of
\nc\ field theory.  In the remainder of this subsection we introduce
the other ingredients we need to define \nc\ field theory in this
familiar context.

To define a real-valued scalar field, it is best to start with
$\Mat_n(C(M))$ and then impose a reality condition analogous to
reality of functions in $C(M)$.  The most useful in practice is to
take the hermitian matrices $a = a^\dagger$, whose eigenvalues will be
real (given suitable additional hypotheses).  To do this for general
$\CA$, we would need an operation $a \rightarrow
a^\dagger$ satisfying $(a^\dagger)^\dagger=a$ and (for $c\in\IC$)
$(c a)^\dagger = c^* a^\dagger$,
in other words an antiholomorphic involution.

The algebra $\Mat_n(C(M))$ could also be defined as the tensor
product $\Mat_n(\IC) \otimes C(M)$.  This construction generalizes
to an arbitrary algebra $\CA$ to define $\Mat_n(\IC) \otimes \CA$,
which is just $\Mat_n(\CA)$ or $n\times n$ matrices with elements
in $\CA$. This algebra admits the automorphism group $GL(n,\IC)$,
acting as $a \rightarrow g^{-1} a g$ (of course the center acts
trivially). Its subgroup $U(n)$ preserves hermitian conjugation and
the reality condition $a=a^\dagger$.  One sometimes refers to
these as ``$U(n)$ \nc\ theories,'' a bit confusingly.  We will
refer to them as rank $n$ theories.

In the rest of the review, we will mostly consider \nc\ associative
algebras which are related to the algebras ${\CA}(M)$ by deformation
with respect to a parameter $\t$, as we will define shortly.
Such a deformed algebra will be denoted by
$M_{\t}$, so that $M_{0} = {\CA} (M)$.

\subsubsection{The derivative and integral}

A \nc\ field theory will be defined by an action functional of
fields ${\Phi}, {\phi}, {\varphi}, \ldots$ defined in terms of
the associative algebra $\CA$ (it could be elements of
$\CA$, or vectors in some representation thereof). Besides the
algebra structure, to write an action we will need an integral
$\int\Tr$ and derivatives $\p_i$. These are linear operations
satisfying certain formal properties:

(a) The derivative is a derivation on ${\CA}$, $\p_i (AB) = (\p_i
A) B + A (\p_i B).$ With linearity, this implies that the
derivative of a constant is zero.

(b) The integral of the trace of a total derivative is zero,
$\int \Tr \p_i A = 0.$

(c) The integral of the trace of a commutator is zero,
$\int \Tr [A,B] \equiv \int \Tr (A\cdot B - B\cdot A) = 0$.

A candidate derivative $\p_i$ can be written using an element
$d_i\in\CA$; let $\p_i A = [d_i,A]$.  Derivations which can be written
in this way are referred to as inner derivations, while those which
cannot are outer derivations.

We denote the integral as $\int\Tr$ as it turns out that for general
\nc\ algebras, one cannot separate the notations of trace
and integral.  Indeed, one normally uses
either the single symbol $\Tr$ (as is done in mathematics)
or $\int$ to denote this combination; we do not follow
this convention here only to aid the uninitiated.

We note that just as condition (b) can be violated in conventional
field theory for functions which do not fall off at infinity, leading
to boundary terms, condition (c) can be violated for general
operators, leading to physical consequences in \nc\ theory which we
will discuss.

\subsection{Noncommutative flat space-time}
\label{sec:ncflat}

After $\Mat_n(C(M))$, the next simplest example of a \nc\ space is
the one associated to the algebra $\IR^d_\theta$ of all complex
linear combinations of products of $d$ variables $\hat x^i$ satisfying
\begin{equation}
[x^i ,x^j ] = i\theta^{ij} .
\end{equation}
The $i$ is present as the commutator of hermitian operators is
antihermitian.  As in quantum mechanics, this expression is the
natural operator analog of the Poisson bracket determined by the
tensor $\theta^{ij}$, the ``Poisson tensor'' or noncommutativity
parameter.

By applying a linear transformation to the coordinates, the
Poisson tensor can be brought to canonical form. This form depends
only on its rank, which we denote as $2r$. We keep this general as
one often discusses partially \nc\ spaces, with $2r<d$.

A simple set of derivatives $\p_i$ can be defined by the relations
\begin{eqnarray}
\label{eq:rnderivatives}
\p_i x^j &\equiv \delta_i^j \\ ~
\label{eq:rnderivativestwo}
[\p_i,\p_j] &= 0
\end{eqnarray}
and the Leibnitz rule.
This choice also determines the integral uniquely (up to overall
normalization), by requiring that $\int \p_i f = 0$ for any $f$
such that $\p_i f\ne 0$.

We will occasionally generalize \rfn{rnderivativestwo} to
\be
\label{eq:defPhi}
[\p_i,\p_j] = -i\Phi_{ij} ,
\ee
to incorporate an additional background magnetic field.

Finally, we will require a metric,
which we will take to be a
constant symmetric tensor $g_{ij}$, satisfying $\p_i g_{jk} = 0$.
In many examples we will take this to be $g_{ij}=\delta_{ij}$, but
note that one cannot bring both $g_{ij}$ and $\theta^{ij}$ to
canonical form simultaneously, as the symmetry groups preserved by
the two structures, $O(n)$ and $Sp(2r)$,
are different. At best one can bring the metric and
the Poisson tensor to the following form:
\begin{eqnarray}
\label{eq:canon} & g = \sum_{{\a} = 1}^{r} \, dz_{\a}  d{\zb}_{\a}
\, + \, \sum_b dy_b^2 ; \crt & \qquad\quad \ {\t} = {1\over
2}\sum_{\a} \, {\t}_{\a} \ {\p}_{\zb_{\a}} \wedge {\p}_{{z}_{\a}};
\qquad \t_a > 0.
\end{eqnarray}
Here $z_{\a} = q_{\a} + i p_{\a}$ are convenient complex
coordinates. In terms of $p,q,y$ the metric and the commutation
relations \rfn{canon} read as
\begin{eqnarray}
\label{eq:ncm} & [y_a,y_b]= [y_b,q_{\a}] = [y_b, p_{\a}] = 0,
\quad [q_{\a}, p_{\b}] = i {\t}_{\a} \ \d_{\a\b} \crt & \quad ds^2
= dq_{\a}^2 + dp_{\a}^2 + dy_b^2 .
\end{eqnarray}

\subsubsection{Symmetries of $\IR^d_\theta$}

An infinitesimal translation $x^i\rightarrow x^i + a^i$
on $\IR^d_\theta$ acts on functions as $\delta\phi = a^i\p_i \phi$.
For the noncommuting coordinates $x^i$, these
are formally inner derivations, as
\begin{equation}
\label{eq:rndercomm} \p_{i} f = [ -i (\theta^{-1})_{ij} x^j, f ] .
\end{equation}
One obtains global translations by exponentiating these,
\begin{equation} \label{eq:gltr} f ( x^i + {\ve}^i) = e^{-i
{\t}_{ij} {\ve}^i x^j} f (x) e^{i {\t}_{ij} {\ve}^i x^j} .
\end{equation} 

In commutative field theory, one draws a sharp distinction between
translation symmetries (involving the derivatives) and internal
symmetries, such as $\delta \phi = [A,\phi]$.  We see that in \nc\
field theory, there is no such clear distinction, and this is why
one cannot separately define integral and trace. 

One often uses only $[\p_i,f]$ and if so,
\rfn{rndercomm} can be simplified further to the operator substitution
$\p_i \rightarrow -i (\theta^{-1})_{ij} x^j$.  This leads to derivatives
satisfying \rfn{defPhi} with $\Phi_{ij}=-(\theta^{-1})_{ij}$.  

The $Sp(2r)$ subgroup of the rotational
symmetry $x^i \rightarrow R^i_j \ x^j$ which preserves $\theta$,
$R^{i^{\prime}}_{i} R^{j^{\prime}}_{j}
{\t}_{i^{\prime} j^{\prime}} = {\t}_{ij}$
can be obtained similarly, as
\begin{equation} \label{eq:ncrot} 
f ( R^i_j \ x^j ) =
e^{-i A_{ij} x^i x^j} \ f (x) \ e^{i A_{ij} x^i x^j}
\end{equation} 
where
$R = e^{i L}$, $L^i_j = A_{kj} {\t}^{ik}$ and  $A_{ij} = A_{ji}$.
Of course only the $U(r)$ subgroup of this will preserve the Euclidean
metric.  

After considering these symmetries, we might be tempted to go on
and conjecture that 
\be \label{eq:poiss}
\delta \phi = i[\phi,\epsilon]
\ee
for any $\epsilon$ is a symmetry of $\IR^d_\theta$.  However, although
these transformations preserve the algebra structure and the 
trace,\footnote{Assuming certain conditions
on $\epsilon$ and $\phi$; see \rfs{opalg}.}
they do not preserve the derivatives.
Nevertheless they are important and will be discussed in detail below.

\subsubsection{Plane wave basis and dipole picture}

One can introduce several useful bases for the algebra $\IR^d_\theta$.
as we discuss in subsection E.
For discussions of perturbation theory and scattering, the most useful
basis is the plane wave basis, which consists of
eigenfunctions of the derivatives:
\begin{equation}
\label{planewave}
\p_i e^{ik x} = i k_i e^{ik x}.
\end{equation}
The solution $e^{ik x}$ of this linear differential equation is the
exponential of the operator $ik\cdot x$ in the usual operator sense.

The integral can be defined in this basis as
\begin{equation}
\label{eq:planeint} \int {\Tr} \ e^{ik x} = \delta_{k,0}
\end{equation}
where we interpret the delta function in the usual physical way
(for example, its value at zero represents the volume of physical space).

More interesting is the interpretation of the multiplication law in
this basis.  This is easy to compute in the plane wave basis, by
operator reordering:
\begin{equation}
\label{eq:moyalprod}
e^{ikx} \cdot  e^{ik'x} =
 e^{-{i\over 2}\theta^{ij} k_i k'_j} e^{i(k+k')\cdot x}
\end{equation}

\def\refmoyalproduct{Eq. (\ref{eq:moyalphase})}
The combination $\theta^{ij} k_ik'_j$ appearing in the exponent
comes up very frequently and a standard and convenient notation
for it is 
$$ k \times k' \equiv \theta^{ij} k_i k'_j = k
\times_{\t} k^{\prime}, 
$$
the latter notation being used to stress
the choice of Poisson structure.

We can also consider
\begin{equation}
\label{eq:dipoleshift} e^{ikx} \cdot f(x) \cdot e^{-ikx}
= e^{-\theta^{ij} k_i \p_j} f(x) = f(x^i-\theta^{ij}k_j) 
\end{equation}
\def\refdipoleshift{Eq. (\ref{eq:dipoleshift})}
Multiplication by a plane wave translates a general function by
$x^i\rightarrow x^i-\theta^{ij} {k_j}$.  This exhibits the nonlocality
of the theory in a particularly simple way, and gives rise to the
principle that large momenta will lead to large nonlocality.

A simple picture can be made of this nonlocality
\cite{Sheikh-Jabbari:1999vm,Bigatti:1999iz}
by imagining that a plane wave corresponds not to
a particle (as in commutative quantum field theory)
but instead a ``dipole,'' a rigid oriented
rod whose extent is proportional to its momentum:
\begin{equation}
\label{eq:dipolelength} \Delta x^i = \theta^{ij} p_j.
\end{equation}

If we postulate that dipoles interact by joining at their ends,
and grant the usual
quantum field theory relation $p=\hbar k$ between wave number and momentum,
the rule \rfn{dipoleshift} follows immediately.

\begin{figure}
\epsfysize=2in
\epsfbox{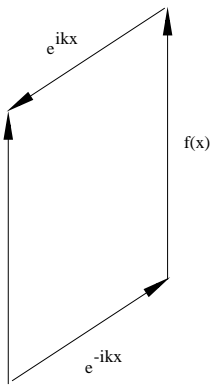}
\caption{The interaction of two dipoles.}
\end{figure}

\subsubsection{Deformation, operators and symbols}

There is a sense in which $\IR^d_\theta$ and the commutative
algebra of functions $C(\IR^d)$ have the same topology and the
same ``size,'' notions we will keep at an intuitive level.  In the
physical applications, it will turn out that $\theta$ is typically
a controllable parameter, which one can imagine increasing from
zero to go from commutative to \nc\ (this does not imply that the
physics is continuous in this parameter, however).  These are all
reasons to study the relation between these two algebras more
systematically.

There are a number of ways to think about this relation.  If $\t$
is a physical parameter, it is natural to think of $\IR^{d}_\t$ as
a deformation of $\IR^d$.  A deformation $M_{\t}$ of $C(M)$ is an
algebra with the same elements and addition law (it is the same
considered as a vector space) but a different multiplication law,
which reduces to that of $C(M)$ as a (multi-)parameter $\t$ goes
to zero. This notion was introduced in \cite{Bayen:1978ha}
as an approach to quantization, and has been much studied since, as we
will discuss in \rfs{math}.
Such a deformed multiplication law is often denoted $f \star g$ or
``star product'' to distinguish it from the original point-wise
multiplication of functions.

This notation has a second virtue,
which is that it allows us to work with $M_{\t}$ in a way which is
somewhat more forgiving of ordering questions.  Namely, we can
choose a linear map $S$ from $M_{\t}$ to $C(M)$, ${\hat f} \mapsto
S[{\hat f}]$, called the ``symbol'' of the operator. We then
represent the original operator multiplication in terms of the
star product of symbols as
\begin{equation}
\label{eq:defstar} {\hat f} {\hat g} = S^{-1}[~ S[\hat f] \star
S[\hat g]~] .
\end{equation}

One should recall that the symbol is not ``natural'' in the
mathematical sense: there could be many valid definitions of $S$,
corresponding to different choices of operator ordering prescription
for $S^{-1}$.

A convenient and standard choice is the Weyl ordered symbol.
The map $S$, defined as a map taking elements of
$\IR^d_\theta$ to $\CA(\IR^d)$ (functions on momentum space), and
its inverse, are
\begin{eqnarray}
\label{eq:weyl}
f(k) \equiv {S[\hat f]}(k) = {1\over (2\pi)^{n/2}}
   \int\Tr e^{-ik\hat x} \hat f(\hat x) \\
\hat f(\hat x) =
 {S^{-1}[f]} = {1\over (2\pi)^{n/2}}\int d^nk\ e^{ik\hat x} f(k)
\end{eqnarray}
Formally these are inverse Fourier transforms, but the first expression
involves the integral \rfn{planeint} on $\CA(\IR_\theta)$, while the
second is an ordinary momentum space integral.

One can get the symbol in position space by performing a second
Fourier transform; e.g.
\begin{equation}
{S[\hat f]}(x) = {1\over (2\pi)^{n}}
 \int d^nk \int\Tr e^{ik(x-\hat x)} \hat f(\hat x) .
\end{equation}
We will freely assume the usual Fourier relation between position and
momentum space for the symbols, while being careful to say (or denote
by standard letters such as $x$ and $k$) which we are using.

The star product for these symbols is
\begin{equation}
\label{eq:moyalphase}
e^{ikx} \star e^{ik'x} =
 e^{-{i\over 2}\theta^{ij} k_i k'_j} e^{i(k+k')\cdot x} .
\end{equation}
Of course all of the discussion in B.2 above still applies, as
this is only a different notation for the same product \rfn{moyalprod}.

Another special case which often comes up is
\begin{equation}
\label{eq:quadraticint} \int\Tr f \star g = \int\Tr f g .
\end{equation}

\subsubsection{The \nc\ torus}

Much of this discussion applies with only minor changes to define
$\IT_\theta^{d}$, the algebra of functions on a \nc\ torus.

To obtain functions on a torus from functions on $\IR^d$, we would
need to impose a periodicity condition, say $f(x^i)=f(x^i+2\pi
n^i)$.  A nice algebraic way to phrase this is to instead define
$\IT^d_{\t}$ as the algebra of all sums of products of arbitrary
integer powers of a set of $d$ variables $U_i$, satisfying
\begin{equation}
\label{eq:torusphase}
U_i U_j = e^{-i\theta^{ij}} U_j U_i .
\end{equation}
The variable $U_i$ takes the
place of $e^{ix^i}$ in our previous notation, and the derivation
of the Weyl algebra from \refnoncom\ is familiar from quantum
mechanics.  Similarly, we take
$$
[\p_i, U_j ] = i \delta_{ij} U_j ,
$$
and
$$
\int\Tr U_1^{n_1} \ldots U_d^{n_d} = \delta_{\vec n,0} .
$$

There is much more to say in this case about the topological aspects,
but we postpone this to \rfs{math}.

\subsection{Field theory actions and symmetries}
\label{sec:ft}

Field theories of matrix scalar fields are very familiar and are
treated in most textbooks on quantum field theory.  The matrix
generalization is essential in discussing Yang-Mills theory.  In a
formal sense we will now make explicit, any field theory
Lagrangian which is written in terms of matrix fields, matrix
addition and multiplication, and the derivative and integral, can
be equally well regarded as a \nc\ field theory
Lagrangian, with the same equations of motion and (classical)
symmetry properties as the matrix field theory.

Let us consider a generic matrix scalar field theory with a
hermitian matrix valued field ${\phi}(x)={\phi}(x)^\dagger$ and
(Euclidean) action
\begin{equation}
\label{eq:genaction} S = \int {\rm d}^d x\ \sqrt{g} \left( {\half}
g^{ij} \Tr \p_i {\phi} \p_j {\phi} + \Tr V({\phi}) \right)
\end{equation}
\def\refgenaction{Eq. (\ref{eq:genaction})}
where $V(z)$ is a polynomial in the variable $z$, $\p_i={\p\over \p x^i}$
are the partial derivatives, and $g^{ij}$ is the metric.
The constraint we require in order to generalize a matrix action
to a \nc\ action is that it be written only using the combination
$\int \Tr$ appearing in \refgenaction;
we do not allow either the integral $\int$ or the trace $\Tr$
to appear separately.  In particular, the rank of the matrix $N$ can
not appear explicitly, only in the form $\Tr 1$ combined with the integral.

Under this assumption, it is an easy exercise to check that if we
replace the algebra $\Mat_N(C(M))$ by a general associative algebra
$\CA$ with integral and derivative satisfying the requirements above,
the standard discussion of equations of motion, classical symmetries
and Noether's theorem, all go through without change.  The point is
that formal manipulations which work for arbitrary matrices of
functions can always be made without commuting the matrices.  Another
way to think about this result is to imagine defining the theory in
terms of an explicit matrix representation of the algebra $\CA$.

Thus the \nc\ theory with action \refgenaction\ has the standard
equation of motion 
$$g^{ij} \p_i \p_j {\phi} = V'({\phi})
$$ 
and
conservation laws $\p_i J^i = 0$ with the conserved current $J^i$
associated to a symmetry $\delta {\phi}(\epsilon,{\phi})$
determined by the usual variational procedures, 
$$
\delta S = \int \Tr J^i \p_i \epsilon .
$$

For example, let us consider the transformations \rfn{poiss}.  In
matrix field theory, these would be the infinitesimal form of a $U(N)$
internal symmetry ${\phi} \rightarrow U^\dagger {\phi} U$.  Although
in more general \nc\ theories $[\p_i,\epsilon] \ne 0$ and these
are not in general symmetries, we can still consider their action, and
by exponentiation define an analogous ``$U(N)$'' action.  We will
refer to this group as $U(\CH)$, the group of unitary operators acting
on a Hilbert space $\CH$ admitting a representation of the algebra
$\CA$.  In more mathematical terms, discussed in \ref{sec:math}, $\CH$
will be a module for $\CA$.

Of course, if we do not try to gauge $U(\CH)$, it could also be broken
by other terms in the action, for example source terms $\int \Tr J
{\phi}$, position-dependent potentials $\int \Tr f V({\phi})$ and so
forth.

Application of the Noether procedure to \rfn{poiss}
leads to a conserved current $T^i$, which for \rfn{genaction} would be
\be \label{eq:poissst}
T^i = i g^{ij} [ {\phi}, {\p}_j {\phi} ] .
\ee
As we discussed in \rfs{ncflat}.1, \rfn{poiss} includes
translations and the rotations which preserve $\theta^{ij}$,
so $T^i$ can be used
to define momentum and angular momentum operators, for example
\be \label{eq:momentum}
P_i = -i(\theta^{-1})_{ij} \int \Tr x^i T^0 ,
\ee
Thus we refer to it
as the ``restricted stress-energy tensor.''

One can also apply the Noether definition to the general variation
$x^i \mapsto x^i + v^i(x)$, to define a more conventional
stress-energy tensor $T_{ij}$, discussed in 
\cite{Gerhold:2000ik,Abou-Zeid:2001up}.
In general, the action of this stress tensor 
changes $\theta$ and the underlying algebra, and its
interpretation has not been fully elucidated at present 
(see Secs. \ref{sec:defq},\ref{sec:ncstring}.2 for related issues).

Finally, there is a stress-energy tensor for \nc\ gauge
theory which naturally appears in the relation to string theory,
which we will discuss in \rfs{gstress} and \rfs{strq}.

We could just as well consider theories containing an arbitrary number
of matrix fields with arbitrary Lorentz transformation properties
(scalar, spinor, vector and so on).  However, at this point we will
only consider a generalization directly analogous to the treatment of
higher spin fields in Euclidean and Minkowski space.  We will discuss
issues related to curved backgrounds later; at present the \nc\
analogs of manifolds with general metrics are not well understood.

Although one can be more general, let us now assume that the
derivatives $\p_i$ are linearly independent and
satisfy the usual flat space relations
$[\p_i,\p_j] = 0$.

Given Poincar\'e symmetry or its subgroup preserving $\theta$,
one can use the \con\ definitions for the action of the
rotation group on tensors and spinors, which we do not repeat here.
In particular, the standard Dirac equation also makes sense over
${\IR}^d_{\t}$ and ${\IT}^d_{\t}$, so spin $1/2$ particles can be
treated without difficulty.

The discussion of supersymmetry is entirely parallel to that for \con\
matrix field theory or Yang-Mills theory, with the same formal
transformation laws.  Constraints between the dimension of space-time
and the number of possible supersymmetries enter at the point we
assume that the derivatives $\p_i$ are linearly independent.  With
care, one can also use the conventional superfield formalism, treating
the anticommuting coordinates as formal variables which
commute with elements of $\CA$ \cite{Ferrara:2000mm}.

Finally, as long as time is taken as commutative, the standard
discussion of Hamiltonian mechanics and canonical quantization goes
through without conceptual difficulty.  On the other hand, \nc\ time
implies nonlocality in time, and the Hamiltonian formalism becomes
rather complicated \cite{Gomis:2000gy}; it is not clear that it has
any operator interpretation.  Although functional integral
quantization is formally sensible, the resulting perturbation theory
is problematic as discussed in \rfs{quant}.  It is
believed that sensible string theories with time-like noncommutativity
exist, discussed in \rfs{exotic}.

\subsection{Gauge theory}
\label{sec:gt}

The only unitary quantum field theories including vector fields are
gauge theories, and the standard definitions also apply in this
context.  However there is a great deal more to say about the
kinematics and observables of gauge theory.

A gauge connection will be a one-form $A_i$, each component of
which takes values in $\CA$ and satisfies $A_i=A_i^\dagger$.
(See \rfs{gaugetop} for a more general definition.)
The associated field strength is
\begin{equation}
\label{eq:fieldstrength}
F_{ij} = \p_i A_j - \p_j A_i + i [A_i, A_j] ,
\end{equation}
which under the gauge transformation
\begin{equation}
\label{eq:gaugetrans}
\delta A_i = \p_i \epsilon + i [A_i, \epsilon]
\end{equation}
transforms as $\delta
F_{ij} = i [F_{ij},\epsilon]$, allowing us to write the gauge
invariant Yang-Mills action
\be \label{eq:yangmills}
S = -{1\over 4g^2} \int \Tr F^2 .
\ee
All this works for the reasons already discussed in \rfs{ft}.

Gauge invariant couplings to charged matter fields can be written in
the standard way using the covariant derivative
\begin{equation}
\label{eq:adjointderiv}
D_i \phi \equiv \p_i \phi + i [A_i,\phi].
\end{equation}

Finite gauge transformations act as
$$
(\p_i + i{A}_i,~F,~\phi)
 \rightarrow {U}^\dagger\ (\p_i + i{A}_i,~F,~\phi)\ {U}
$$
and these definitions gauge the entire $U(\CH)$ symmetry. One can
also use $\Mat_N(\CA)$ to get the \nc\ analog of $U(N)$ gauge theory,
though (at this point) not the other Lie groups.

As an example, we quote the maximally supersymmetric
Yang-Mills (MSYM) Lagrangian in ten dimensions,
from which $\CN=4$
SYM in $d=4$ and many of the simpler theories can be deduced by
dimensional reduction and truncation:
\begin{equation}
\label{eq:msym}
S = \int d^{10}x\
\Tr\left(~ F_{ij}^2 + i\bar\chi_I \Dslash \chi^I ~\right),
\end{equation}
where $\chi$ is a $16$-component adjoint Majorana-Weyl fermion.
This action satisfies all of our requirements and thus leads to a
wide variety of supersymmetric \nc\ theories. Indeed,
noncommutativity is the only known generalization (apart from
adding irrelevant operators and taking limits of this)
which preserves maximal supersymmetry.

Because one cannot separately define integral and trace, the local
gauge invariant observables of conventional gauge theory do not carry
over straightforwardly: only $\int \Tr O$ is gauge invariant.  We now
discuss this point.

\subsubsection{The emergence of space-time}

The first point to realize is that the gauge group in \nc\
theory contains space-time translations.  This is already clear from
the expression \rfn{rndercomm}, which allows us to express a
translation $\delta A_i = v^j \p_j A_i$ in terms of a
gauge transformation \rfn{gaugetrans}
with $\epsilon = v^j (\theta^{-1})_{jk} x^k$.
Actually, this produces
$$
\delta A_i = v^j \p_j A_i + v^j (\theta^{-1})_{ji} ,
$$
but an overall constant shift of the vector potential drops out of
the field strengths and has no physical effect in infinite flat space.

Taking more general functions for $\epsilon$ will produce more
general space-time transformations.  
As position-dependent translations, one might compare these with
coordinate definitions or diffeomorphisms.  To do this, we consider
the products as star products and expand \rfn{moyalphase} in $\theta$,
to obtain
\bea\label{eq:canontrans}
\delta \phi = i[\phi,\epsilon] &= \theta^{ij} \p_i\phi \p_j\epsilon 
+ \CO(\p^2\phi\p^2\epsilon) .\\
\rightarrow \{\phi,\epsilon\} ,
\eea
($\{,\}$ is the Poisson bracket),
so at leading order the gauge group is the group of
canonical transformations preserving $\theta$
(we discuss this further in E.5).
Of course, the higher derivative terms modify this result.
In fact the full gauge group $U(\CH)$ is simpler, as we will see in
E.2.

Another aspect of this unification of space-time and gauge
symmetry is that if the derivative is an inner derivation, we can
absorb it into the vector potential itself. In other words, we can
replace the covariant derivatives $D_i = \p_i + iA_i$ with
``connection operators'' in $\IR^d_\theta$,
\begin{equation}
\label{eq:Cdefn} C_i \equiv (-i\theta^{-1})_{ij} x^j + iA_i
\end{equation}
such that
\begin{equation}
\label{eq:DCsubstitution} D_i \ f \rightarrow [C_i,f] .
\end{equation}
\def\refDCsub{Eq. (\ref{eq:DCsubstitution})}
We also introduce the ``covariant coordinates,''
\bea \label{eq:Ydef} 
Y^i = x^i + {\t}^{ij} A_j (x) .
\eea
If ${\t}$ is invertible, then $Y^i = i{\t}^{ij} C_j$ and this is
just another notation, but the definition makes sense more generally.

In terms of the connection operators, the Yang-Mills field
strength is
\begin{equation}
\label{eq:matrixfield}
F_{ij} = i[D_i,D_j] \rightarrow
i[C_i,C_j] - (\theta^{-1})_{ij} ,
\end{equation}
and the Yang-Mills action becomes a simple ``matrix model'' action,
\begin{equation}
\label{eq:matrixmodel} S = \Tr \sum_{i,j} (i[C_i,C_j] -
(\theta^{-1})_{ij})^2 .
\end{equation}

Now, although we motivated this from \rfn{yangmills}, we could look at this
the other way around, starting with the action \rfn{matrixmodel} as a
function of matrices $C_i$ and postulating \rfn{Cdefn}, to derive
\nc\ gauge theory (and Yang-Mills theory in the limit $\theta\rightarrow 0$)
from a matrix model.
This observation is at the heart of most of the common ways that
\nc\ gauge theory arises in particle physics, as the
action \rfn{matrixmodel} and its supersymmetrization is
simple enough to arise in a wide variety of contexts.  For
example, it can be obtained as a limit of the twisted Eguchi-Kawai model
\cite{Eguchi:1983ta,Gonzalez-Arroyo:1983hz},
which was argued to reproduce the physics of large $N$ Yang-Mills
theory.  The maximally supersymmetric version obtained in the
same way from \rfn{msym}, often referred to as the ``IKKT model''
\cite{Ishibashi:1997xs}, plays an important role in M theory, to be
discussed in \rfs{string}.

Having so effectively hidden it, we might well wonder how
$d$-dimensional space-time is going to emerge again from
\rfn{matrixmodel}.  Despite appearances, we do not want to claim
that \nc\ gauge theory is the same in all dimensions $d$.  Now in the
classical theory, to the extent we work with explicit expressions for
$A_i$ in \rfn{Cdefn}, this is generally not a problem.
However, in the quantum theory, we need to integrate over field
configurations $C_i$.  We will need to argue that this functional
integral can be restricted to configurations which are ``similar to''
\rfn{Cdefn} in some sense.

This point is related to what at first appears only to be a technical
subtlety involving the $\theta^{-1}$ terms in
\rfn{matrixmodel}.  They are there to cancel an extra term
$[(\theta^{-1}x)_i,(\theta^{-1}x)_j]$, which would have led to an
infinite constant shift of the action.  The subtlety is that one could
have made a mistake at this point by assuming that $\Tr [C_i,C_j]=0$,
as for finite dimensional operators.  

Of course the possibility that $\Tr [C_i,C_j]\ne 0$ probably comes as
no surprise, but the point we want to make is that $\Tr [C_i,C_j]$
should be considered a {\it topological} aspect of the configuration.
We will argue in \rfs{opalg} that it is invariant under
any variation of the fields which (in a sense) preserve the
asymptotics at infinity.  This invariant detects the presence of the
derivative operators in $C_i$ and this is the underlying reason 
one expects to consistently identify sectors with a higher dimensional
interpretation in what is naively a zero dimensional theory.

\subsubsection{Observables}

All this is intriguing, but it comes with conceptual problems. The
most important of these is that it is difficult to define local
observables.  This is because, as noted from the start, there is
no way to separate the trace over $\CH$ (required for gauge
invariance) from the integral over \nc\ space.  We can easily
enough write gauge invariant observables, such as 
$$ \int \Tr F(x)^n ,
$$ 
but they are not local.

A step forward is to define the Wilson loop operator.  Given a
path $L$, we write the holonomy operator using exactly the same
formal expression as in conventional gauge theory,
$$
W_L = P \exp i\int_{L} d\sigma A(x(\sigma)) ,
$$
but where the products in
the expansion of the path ordered exponential are star products.
This undergoes the gauge transformation
$$
W_L \rightarrow U^\dagger(x_1) W_{L(x_1,x_2)} U(x_2)
$$
where $x_1$ and $x_2$ are the start
and end points of the path $L_{1,2}$.

We can form a Wilson loop by taking for $L$ a closed loop with
$x_1=x_2$, but again we face the problem that we can only
cyclically permute operators, and thus cancel $U^{-1}(x_1)$ with
$U(x_1)$, if we take the trace over $\CH$, which includes the
integral over \nc\ space.

We can at least formulate multilocal observables with this construction,
such as
$$
\int \Tr O_1(x_1) W[L_{1,2}] O_2(x_2) W[L_{2,3}] \ldots O(x_n) W[L_{n,1}]
$$
with arbitrary gauge covariant operators $O_i$ at arbitrary points $x_i$,
joined by Wilson loops.  This allows us to control the distance between
operators within a single trace, but not to control the distance between
operators in different traces.

Actually one can do better than this, using what are called
{\it open} Wilson loops \cite{Ishibashi:1999hs}.
The simplest example is
$$
W_L(k) \equiv \int \Tr\ W[L_{1,2}] e^{ik\hat x_2} .
$$
If the distance between the end points of $L$ and the momentum
$k$ satisfy the relation \rfn{dipolelength},
$\theta^{ij} k_i = (x_1-x_2)$, this
operator will be gauge invariant, as one can see by using
\rfn{dipoleshift}.

This provides an operator which carries a definite momentum and
which can be used to define a version of local correlation
functions.  There is a pleasing correspondence between its construction
and the dipole picture of subsection B.2; not only can we think of
a plane wave as having a dipole extent, we should think of the two
ends of the dipole as carrying opposite electric charges which for
gauge invariance must be attached to a Wilson line.

The straight line has a preferred role in this construction, and
the open Wilson loop associated to the straight line 
with length determined by \rfn{dipolelength} can be written 
$$ 
W(k) \equiv \int \Tr\ e^{k \times C} = \int \Tr e^{ i k \cdot Y} 
$$
where $Y^i$ are the covariant coordinates of \rfn{Ydef}. This
construction can be used to covariantize local operators as
follows: given an operator $\CO (x)$, transforming in the adjoint,
and momentum $k$, we define 
\bea \label{eq:openwil} 
& W[{\CO}](k)
= \int \Tr e^{ k \times C} \CO(x) = \int \Tr e^{i k_m Y^m (x)} {\CO}(x) \crt
& {\CO}[y] =   \int {\rm d}^d k \int \Tr e^{i k \cdot \left( Y - y
\right) } {\CO} (x) 
\eea

\subsubsection{Stress-energy tensor}
\label{sec:gstress}

As we discussed above, the simplest analog of the stress-energy
tensor in \nc\ field theory is \rfn{poissst}, which generates
the \nc\ analog of canonical transformations on space-time.
However, in \nc\ gauge theory, this operator is the generator of gauge
transformations, so it must be set to zero on physical states.  This
leads to a subtlety analogous to one known in general relativity: one
cannot define a gauge invariant local conserved momentum density.
This is compatible with the difficulties we just encountered in
defining local gauge invariant observables.

One can nevertheless regard \rfn{momentum} as a nontrivial global
conserved momentum.  This is because it corresponds to a formal gauge
transformation with a parameter $\epsilon \sim x^i$ which does not
fall off at infinity (on the torus, it is not even single valued),
and as such can be consistently excluded from the gauge group.
This type of consideration will be made more precise in \rfs{opalg}.

One can make a different definition of stress-energy tensor,
motivated by the relation \rfn{Cdefn} between the connection and the
\nc\ space-time coordinates, as the Noether current associated to
the variation
$$
C_i \rightarrow C_i + a_i(k) e^{i k\cdot Y}
$$
which for the action \rfn{matrixmodel} can easily be seen to produce
\be
\label{eq:gstress}
T_{ij}(k) = \sum_l \int_0^1 ds\ \int\Tr\ e^{i s k\cdot Y} [C_i,C_l]
	e^{i(1-s) k\cdot Y} [C_j,C_l] ,
\ee
which is conserved in the sense that 
$k_m \theta^{mn} T_{nl}(k)=0$ for a solution of the equations of motion.
This appears to be the natural definition in string theory,
as we discuss in \rfs{string}.E.

\subsubsection{Fundamental matter}

Another type of gauge invariant observable can be obtained by
introducing new fields (bosons or fermions) which transform in the
fundamental of the \nc\ gauge group.
In other words, we consider a field $\psi$ which is
operator valued just as before, but instead of transforming under
$U({\CH})$ as $\psi \rightarrow U \psi U^\dagger$, we impose the
transformation law 
$$ 
\psi \rightarrow U \psi . 
$$ 
More generally,
we need to define multiplication $a\cdot \psi$ by any element of
$\CA$, but this can be inferred using linearity.

Bilinears such as $\psi^\dagger \psi$, $\psi^\dagger D_i \psi$ and
so on will be gauge invariant and can be used in the action and
to define new observables, either by enforcing an
equation of motion on $\psi$ or doing a functional integral over $\psi$
\cite{Ambjorn:2000nb,Gross:2000ss,Rajaraman:2000dw}.

Although in a strict sense this is also a global observable,
an important point (which will be central to \rfs{math}.D)
is that one can also postulate an independent rule for multiplication
by $\CA$ (and the unitaries in $\CA$) on the right,
$$\psi \rightarrow \psi a.$$
Indeed, if we take $\psi\in\CA$, we will clearly get a nontrivial
second action of this type, since left and right multiplication are
different.
In this case, we can think of $\psi^\dagger\psi$
as a function on a second, dual \nc\ space.
For each $f\in\CA$ one obtains a gauge invariant
observable $\Tr \psi^\dagger \psi f$, which is local on the dual space
in the same sense that an \nc\ field is a local observable in an
ungauged theory.

Taking $\psi\in\CA$ is a choice.
One could also have taken $\psi\in\CH$,
which does not lead to such a second multiplication law.  The general
theory of this choice is discussed in \rfs{math}.D.

The two definitions lead to different physics.
Let us compare the spectral density.  If we
take the Dirac operator $\gamma^i D_i$ acting on $\psi\in\CH_r$, this has
$d\rho(E)\sim dE E^{r-1}$, as for a field in $r$ dimensions.

If we take the same Dirac operator with $\psi\in\CA$,
we would get infinite spectral density.  A more useful definition is
$$
\Dslash \psi = \gamma^i (D_i \psi - \psi \p_i)
$$
which fixes this by postulating an ungauged right action of translations,
leading to a spectral density appropriate to $2r$ dimensions.

\subsubsection{The Seiberg-Witten map}

Having discovered an apparent generalization of gauge theory, we
should ask ourselves to what extent this theory is truly novel and to
what extent we can understand it as a conventional gauge theory.  This
question will become particularly crucial once we find \nc\ gauge
theory arising from open string theory, as general arguments imply
that open string theory can always be thought of as giving rise to a
conventional gauge theory.  Is there an inherent contradiction in
these claims?

\citeasnoun{Seiberg:1999vs} proposed that not only is there no
contradiction, but that one should be able to write an explicit map
from the \nc\ vector potential to a conventional Yang-Mills vector
potential, explicitly exhibiting the equivalence between the two
classes of theories.

One might object that the gauge groups of \nc\ gauge theory and
\con\ gauge theory are different, as is particularly clear
in the rank $1$ case.  However, this is not an obstacle to the
proposal, as only the physical configuration space -- namely the
set of orbits under gauge transformation -- must be equivalent in
the two descriptions.  It does imply that the map between the two
gauge transformation laws must depend on the vector potential, not
just the parameter.

Thus, the proposal is that there exists a relation between a conventional
vector potential $A_i$ with the standard Yang-Mills gauge transformation
law with parameter $\epsilon$, \rfn{gaugetrans}, and a
\nc\ vector potential $\hat A_i(A_i)$ and gauge transformation
parameter $\hat\epsilon(A,\epsilon)$ with noncommmutative gauge
invariance
$\hat\delta \hat A_i = \p_i \hat\epsilon + i \hat A_i * \hat \epsilon
- i \hat \epsilon * \hat A_i$,
such that
\begin{equation}
\label{eq:swmapdef}
\hat A(A) + \hat\delta_{\hat\epsilon} \hat A(A) =
\hat A(A + \delta_\epsilon A) .
\end{equation}
This equation can be solved to first order in $\theta$
without difficulty.  Writing $\theta=\delta\theta$, we have
\begin{eqnarray}
\label{eq:swmap}
\hat A_i(A) - A_i &= - {1\over 4}\delta\theta^{kl} \{A_k,\p_l A_i+F_{li}\}_{+}
        + \CO(\delta\theta^2) \\
\label{eq:swmaptwo}
\hat \epsilon(A,\epsilon) - \epsilon &=
    {1\over 4}\delta\theta^{kl} \{\p_k \epsilon,A_l\}_{+}
        + \CO(\delta\theta^2)
\end{eqnarray}
where $\{A,B\}_{+}\equiv AB+BA$.
The corresponding first order relation between the field strengths is
$$
\hat F_{ij} - F_{ij} =
 {1\over 4}\delta\theta^{kl} \left(
  2 \{F_{ik},F_{jl}\}_{+} - \{A_k,D_l F_{ij}+\p_l F_{ij} \}_{+} \right).
$$

This result even admits a reinterpretation which defines the map
to all finite orders in $\theta$.  Consider the problem of mapping
a \nc\ gauge field $\hat A^{(\theta)}$ defined with respect to the
star product for $\theta$, to a \nc\ gauge field $\hat
A^{(\theta+\delta\theta)}$ defined for a nearby choice of
$\theta$. To first order in $\delta\theta$, it turns out that the
solution to the corresponding relation \rfn{swmapdef} is
again Eqs. (\ref{eq:swmap}), (\ref{eq:swmaptwo}), now with the
right hand side evaluated using the star product for $\theta$.
Thus these equations can be interpreted as differential equations
(Seiberg-Witten equations) determining the map to all orders.

\rfn{swmap} can be solved explicitly for the case of
a rank one gauge field with constant $F$.  In this case, it reduces to
$$
\delta \hat F = - \hat F \delta\theta \hat F
$$
where Lorentz indices are contracted as in matrix multiplication.
It has the solution (with boundary condition $F$ at $\theta=0$)
\begin{equation}
\label{eq:constmap}
\hat F = \left( 1 + F\theta \right)^{-1} F .
\end{equation}
This result can be used to relate the \con\ and \nc\ gauge
theory actions at leading order in a derivative expansion, as we will
discuss in detail in \rfs{ncstring}.

All this might suggest that the \nc\ framework is merely a simpler
way to describe theories which could have been formulated as conventional
gauge theories, by just applying the transformation
$\hat F\rightarrow F$ to the action.  This however ignores the possibility
that the map might take nonsingular field configurations in the one
description, to singular field configurations in the other.
Indeed, \rfn{constmap} gives an explicit example.
When $F = - \theta^{-1}$, the \nc\ description
appears to break down, as $\hat F$ would have a pole.  Conversely,
$F$ is singular when $\hat F = \theta^{-1}$.

As we continue, we will find many examples in which \nc\ gauge theory has
different singular solutions and short distance properties from \con\
gauge theory, and despite this formal relation between the theories it will
become clear that their physics is in general rather different.

A solution to the Seiberg-Witten equation was recently found 
\cite{Liu:2000mj,Liu:2001pk,Okawa:2001mv,Mukhi:2001vx}.  
Namely, the following inhomogeneous
even degree form on ${\IR}^d$, defined as the integral over the
superspace ${\IR}^{d \vert d}$, is closed: 
\bea
\label{eq:swmapthree} \int {\rm d}^{d} k {\rm d}^{d} {\vartheta}
\, {\Tr}_{\CH} \left( {\exp} {{\CF}\over{2\pi i}} \right) \rho(k) \crt
{\CF} = k_i (Y^i - y^i ) - {\vartheta}_i dy^i  + {\vartheta}_i
{\vartheta}_j [ Y^i , Y^j] \crt 
\eea 
where $y$'s are the coordinates on ${\IR}^d$, and
$\rho(k)$ can be any smooth function such that $\rho(0)=1$
(this slightly generalizes the references, which take $\rho=1$).
This expression has an expansion in
differential forms on $\IR^d$, whose two-form part is the 
\con\ $F+\theta^{-1}$.  
Deeper aspects of this rather suggestive superspace expression 
will be discussed in \citeasnoun{Nekrasov:2001}.

\subsection{Bases and physical pictures}
\label{sec:bases}

The algebra $\IR^{d}_\theta$ of ``functions on \nc\ $\IR^d$,''
considered as a linear space, admits several useful bases.  Since
it is just a product of Heisenberg algebras and commuting
algebras, all of this formalism can be traced back to the early
days of quantum mechanics, as can much of its physical
interpretation.  In quantum mechanics, it appears when one
considers density matrices (the Wigner functional) and free fermi
fluids (in one dimension, this leads to bosonization and
$W_\infty$ algebra).

However we stress that the noncommutativity under discussion here is
{\it not} inherently quantum mechanical.  Rather, it is a formal device
used to represent a particular class of interactions between fields,
which can exist in either classical or quantum field theory.
In particular, an essential difference with the standard quantum
mechanical applications is that these involve linear equations,
while we are going to encounter general nonlinear equations.\footnote{
The special case of the equation $\phi^2=\phi$ defining a projection
can arise as a normalization condition in quantum mechanics.
Also, somewhat similar nonlinear equations appear
in the approximation methods of quantum statistical mechanics.}

\subsubsection{Gaussians and position-space uncertainty}

While the plane wave basis is particularly good for perturbation
theory, nonperturbative studies tend to be simpler in position space.
However, in \nc\ theory the standard position space basis
tends not to be the most convenient, because of the nonlocal nature of
the interactions.  One is usually better off using a basis which
simplifies the product.

One expects the noncommutativity \rfn{noncom} to lead to a
position-space uncertainty principle, which will exclude the
possibility of localized field configurations.  Although there is
truth to this, the point is a bit subtle, as it is certainly possible
to use delta functions ${\d}^{(d)} (x-x_0)$ as a basis (for the
symbols) which from the point of view of the kinetic term is local.

Of course, the star product is not diagonal in this basis.
Computing the star product of two delta functions leads to a
kernel, which can be used to write an integral representation of
the product: 
\bea
\label{eq:starkernel} (f \star  g)(z) &= \int {\rm d}^d x\ {\rm
d}^d y\ K(x,y;z) f(x) g(y) ;\crt K(x,y;z) &= \delta(z-x) \star
\delta(z-y) \crt &= {1\over (2\pi)^d} \int {\rm d}^{d} k\
e^{ik(z-x)} \delta(z-y-\theta k) \crt &= {1\over{(2\pi)^d {\det
\theta}}} e^{i(z-x)\theta^{-1}(z-y)} . 
\eea 
In particular, the star product
\begin{equation}
\label{eq:stardelta} \delta(z) \star f(y) = {1\over (2\pi)^d \,
\det \theta} \int {\rm d}^d y\  e^{i \, y\theta^{-1}z} g(y)
\end{equation}
is a highly nonlocal operation:
it is the composition of a Fourier transform with the linear transformation
$z\rightarrow \theta^{-1}z$.

As in quantum mechanics,
one might expect the Gaussian to be a particularly nice basis
state, since it is simultaneously Gaussian in both conjugate
coordinates.  Let $\psi_{\BM,\Ba}$ be a
Gaussian with center $\Ba$, covariance $\BM$, and maximum $1$:
$$
\psi_{\BM,\Ba} =
 \exp -{(x^i-a^i) M_{ij} (x^j-a^j) }.
$$which satisfies $\int {\rm d}^d x\ |\psi|^2 = (\det
{{2M\over{\pi}}})^{-\half}$.

The star product of two Gaussians can be easily worked out using
\rfn{starkernel}. In particular, for concentric Gaussians
of width $a$ and $b$, we have
\begin{equation}
\label{eq:twogaussians} \psi_{{1\over a^2}\B1,0} \star
\psi_{{1\over b^2}\B1,0} = C(a,b)  \ \psi_{{1\over
\Delta(a,b)^2}\B1,0}
\end{equation}
with 
\bea \label{eq:productwidth} 
\Delta(a,b)^2 = {a^2b^2 +
\theta^2 \over a^2 + b^2} \crt C(a, b) = \left( 1 + {{\t}^2 \over
(ab)^2} \right)^{-{d\over 2}} 
\eea 
\ndt This result illustrates
the sense in which interactions in \nc\ theory obey a
position-space uncertainty principle. Formally, we can construct a
Gaussian configuration of arbitrarily small width in the \nc\
theory; its limit is the delta function we just discussed.  Unlike
commutative theory, however, multiplication by a Gaussian of width
$b^2<\theta$ does not concentrate a field configuration but
instead tends to disperse it. This is particularly clear for the
special case $a=b$ of \rfn{productwidth}.
More generally, the operation of multiplication by a Gaussian $\psi_{b\B1,0}$
(for any $b$) will cause the width to approach $\theta$,
decreasing ($\Delta<a$) if $a^2>\theta$ and increasing if $a^2<\theta$.

For some purposes,
one can think of a Gaussian as having a minimum ``effective
size'' $\max\{a,1/(\theta a)\}$.  This is a bit imprecise however.
For example, the result can be a Gaussian with
$\Delta(a,b)^2 < \theta$, which will be true if and only if
$(a^2-\theta)(b^2-\theta) < 0$.
A better picture is that star product with a small Gaussian is similar
to the Fourier transform \rfn{stardelta}.

The configuration with the minimum effective size is evidently the
Gaussian of width $a^2=\theta$.  One of its special features is that its
product with any Gaussian will be a Gaussian of width $\theta$, and thus
a basis can be defined consisting entirely of such Gaussians.  This
can be done using coherent states and we will return to this below.

\subsubsection{Fock space formalism}

A nice formal context which provides a basis including the minimal
Gaussian is to use as \nc\ coordinates creation and annihilation
operators acting on a Fock space. These are defined in terms of
the canonical coordinates of \rfn{ncm} in the usual way,
\bea \label{eq:defzzb} & {\ao}_\a = {q_\a + i p_\a
\over\sqrt{2{\t}_{\a}}}; \qquad {\co}_\a = {q_\a - i p_\a
\over\sqrt{2{\t}_{\a}}} \crt & z_{\a} = \sqrt{2\t_\a} {\ao}_{\a};
\qquad {\zb}_{\a} = {\sqrt{2\t_\a}} {\co}_{\a} \crt & \qquad
[{\ao}_\a ,{\co}_\b ] = \delta_{\a\b} . \eea

We can now identify elements of $\IR^d_\t$ with functions of the
$y_a$ valued in the space of operators acting in the Fock space
${\CH}_r$ of $r$ creation and annihilation operators.
The Fock space $\CH_r$ is the Hilbert space $\CH$ of our previous
discussion, and this basis makes the nature of $U(\CH)$ particularly
apparent: it is the group of all unitary operators on Hilbert space.
Explicitly,
\begin{eqnarray}
\label{eq:fock} & \qquad  {\CH}_r  = \bigoplus {\IC} \ket{ n_1,
\ldots, n_r } \crt &  {\ao}_{\a} \ket{\ldots,n_{\a},\ldots}  =
\sqrt{n_{\a}} \ket{\ldots,n_{\a}-1,\ldots } \crt  & {\co}_{\a}
\ket{\ldots,n_{\a},\ldots}
 = \sqrt{n_{\a}+1} \ket{\ldots,n_{\a}+1,\ldots} \crt &  \hat n_{\a}  =
{\co}_{\a} {\ao}_{\a} .
\end{eqnarray}
We have also introduced the number operator $\hat n_{\a}$. Real
functions of the original real coordinates correspond to the
hermitian operators.

In this language, the simplest basis we can use consists of the
elementary operators $\ket{\vec k}\bra{\vec l}$. These can be
expressed in terms of ${\co}_{\a}$ and ${\ao}_{\a}$ as 
$$
\ket{\vec k}\bra{\vec l} = \sum_{\vec n} \prod_{\a} (-1)^{n_{\a}}
{{{\co}_{\a}^{k_{\a} + n_{\a}} {\ao}_{\a}^{l_{\a} +
n_{\a}}}\over{n_{\a}! \sqrt{k_{\a}! l_{\a}!}}}
$$ 
Using
\rfn{rndercomm}, the derivatives can all be written as commutators
with the operators ${\ao}_{\a}, {\co}_{\a}$. The integral
\rfn{planeint} becomes the standard trace in this basis,
\bea
\int d^{2r}x \Tr \rightarrow \prod_\alpha (2\pi\theta_\alpha)  \Tr .
\eea

The operators $\ket{\vec k}\bra{\vec k}$, their sums and unitary
rotations of these, provide a large set of projections, operators $P$
satisfying $P^2=P$.  This is in stark contrast with the algebra $C(M)$
(where $M$ is a connected space) which would have had only two
projections, $0$ and $1$, and this is a key difference between
\nc\ and commutative algebras.  Physically, this will
lead to the existence of new solitonic solutions in \nc\ theories,
as we discuss in \rfs{sol}.

A variation on the projection
which is also useful in generating solutions
is the partial isometry, which by definition is any operator $R$ satisfying
\begin{equation}
\label{eq:partialisometry}
R R^\dagger R = R .
\end{equation}
Such an operator can be written as a product $R=PU$ of a
projection $P$ and a unitary $U$. The simplest example is the
shift operator $S^\dagger_\a$ with matrix elements
\begin{equation}
\label{eq:shift} S^\dagger_\a \ket{\ldots,n_\a,\ldots} =
\ket{\ldots,n_\a+1,\ldots } .
\end{equation}
It satisfies \bea \label{eq:shf} & S_\a S_\a^{\dagger} = 1, \quad
{\rm and} \crt & S_\a^{\dagger} S_\a = 1 - \sum_{n_\b; \b \ne
\a}\ket{n_\b,0_\a}\bra{n_\b,0_\a} \eea

\subsubsection{Translations between bases}

In this subsection we assume for simplicity that $d = 2r$.

The
translation between this basis and our previous descriptions
involving commutative functions and the star product, as we
commented above, can be done using the plane wave basis and
Fourier transform.

A standard tool from quantum mechanics which facilitates such
calculations is the coherent state basis  \cite{Klauder:1968a}.
Note that this is {\it not} a basis of $\IR^{d}_{\t}$ but
rather is a basis of $\CH_r$.
We recall their definition
\begin{equation}
\label{eq:defcoherent} \bbra{\xi} = \bra{0} e^{ \xi_\a {\ao}_\a }
, \qquad \kket{\eta} = e^{\eta_\a {\co}_\a} \ket{0} ,
\end{equation}
a useful formula for matrix elements of $\hat f$ in this basis,
\bea \label{eq:coherent}  \bbra{\xi} {\hat f} \kket{\eta}
 = \int f\left(z, {\zb} \right) {{{\rm d}^{r} z \,
 {\rm d}^{r} {\zb}}\over{\prod_\alpha \pi \t_\alpha}}
 \, e^{\xi\cdot\eta  - {1\over{\t}} ( \zb - \xi \sqrt{2\t} )\cdot
 (z - {\eta} \sqrt{2\t} )} \eea
 $$
( \zb - \xi \sqrt{2\t} )\cdot
 ( z - {\eta} \sqrt{2\t} ) = \sum_{\a} ( \zb - \xi_\a
 \sqrt{2\t_\a} ) ( z - {\eta}_{\a} \sqrt{2\t_\a}) $$
  and a formula which follows from this for matrix elements
between Fock basis states with vectors of occupation numbers $\vec
k$ and $\vec l$:
\begin{equation}
\label{eq:mtrx} {\langle }{\vec k} \vert {\hat f} \vert {\vec l}
\rangle = \prod_\a {1\over{\sqrt{{k_\a}!{l_\a}!}}}
{\p}^{k_\a}_{\xi_\a} {\p}^{l_\a}_{\eta_\a}\vert_{\xi = \eta = 0}
\bbra{\xi} {\hat f} \kket{\eta} .
\end{equation}

For example, let us consider the projection operator $\hat f_0
\equiv \ket{0}\bra{0}$. We have $\bbra{\eta}\hat f_0 \kket{\xi} =
1$, and using \rfn{mtrx} we can reproduce this with $f_0
(z, {\zb}) = 2^r\exp (-z\cdot {1\over{\t}} {\zb})$, so $\hat f_0$ is
precisely the minimal Gaussian we encountered earlier. One could
extend this to use
\begin{equation}
\label{eq:cohbasis}  {\hat f}_{\tilde\eta, \tilde\xi} =
e^{-\tilde\xi \cdot \tilde\eta} \vert{\tilde\eta} )(
{\tilde\xi}\vert \leftrightarrow f_{\tilde\eta, \tilde\xi} = 
2^r e^{ -(\zb - \tilde\xi \sqrt{2\t} ) \cdot {\t}^{-1}( z -
\tilde\eta\sqrt{2\t})}
\end{equation}
as an overcomplete basis for $\IR^{d}_{\t}$, consisting of minimal
Gaussians with centers $x = ({\tilde z}, {\tilde z}^*)$ (e.g. see
\refdipoleshift) multiplied by plane waves with momentum $k =
({\tilde\kappa}, {\tilde\kappa}^*)$, with $$ \tilde\kappa =
{1\over{2\t}} ({\tilde\xi} - {\tilde\eta}^*), \qquad {\tilde z} =
\sqrt{\t\over 2} ( {\tilde\xi} + {\tilde\eta}^*)$$ (where $^*$
denotes complex conjugation). As another example, the delta
function $\delta^{(d)}(x)$ has matrix elements (from
\rfn{coherent}) $\bbra{\eta} {\hat\delta} \kket{\xi} = e^{-\eta
\xi}$, leading to the expressions
\begin{equation}
\label{eq:dltor} \bra{\vec k} {\hat \d} \ket{\vec l} =
  {\d}_{{\vec k}, {\vec l}} (-1)^{\vert {\vec k} \vert}, \quad  \hat \d
= (-1)^{\vert \hat{\vec n} \vert}
\end{equation}
with $\vert {\vec k} \vert = \sum_\a k_\a, \vert \hat {\vec
n}\vert =\sum_\a \hat n_\a $. 

Let us now express the general
radially symmetric function in two \nc\ dimensions in the two
bases (we take $\theta=\half$).
These are functions of $r^2 = p^2+q^2 \sim z\zb = \hat n$,
so the general such function in the Fock basis is
\begin{equation}
\label{eq:radialsolution}
\sum_{n\ge 0} c_n \hat f_n = \sum_{n\ge 0} c_n \ket{n}\bra{n} .
\end{equation}
The corresponding symbols $f_n$ can be found
as solutions of the equations $z\zb * f_n = (n+\half) f_n$
\cite{Fairlie:1964a,Curtright:2000ux}.
A short route to the result is to
form a generating function from these operators,
\begin{equation}
\label{eq:genradial}
\hat f = \sum_{n\ge 0} u^n \ket{n}\bra{n} ,
\end{equation}
whose matrix elements are $({\eta}\vert \hat f \vert{\xi}) = \exp
u\eta\xi$.
This can be obtained from \rfn{coherent} by taking for $f$
the generating function $$ f(z\zb;u) = (\lambda-\half) \exp
\lambda z \zb $$ with $1-u=1/(\lambda-\half)$.  Substituting $u$
for $\lambda$ in this expression leads to a generating function
for the Laguerre polynomials $L_n(4r^2)$ (Bateman, 1953).

The final result for the symbol of \rfn{genradial} is 
$$ f = 2\sum_{n\ge 0} u^n  L_n(4r^2) e^{-2r^2} . $$

\subsubsection{Scalar Greens functions}

We now discuss the Greens function of the free \nc\ scalar 
field.  This is very simple in the plane wave basis,
in which the Klein-Gordon operator is diagonal:
$$
(-\p_i\p^i + m^2) e^{ik x} = (k^2 + m^2) e^{ik x} .
$$
The Greens function satisfying
\begin{equation}
\label{eq:nclaplace}
(-\sum_i {\p^2\over\p x_i^2} + m^2) G(x,y) = \delta(x,y)
\end{equation}
is then just
\begin{equation}
\label{eq:propagator}
G(k,k') = {\delta_{k,k'} \over k^2 + m^2} .
\end{equation}

This result can be easily transformed to
other bases using coherent states.
We set $\theta_\alpha=\half$ and start from
\begin{equation}
\label{eq:cohft} \bbra{\eta} e^{ik \zb + i\kb z} \kket{\xi} =
 e^{-\half k\kb + i\kb\eta + ik\xi + \eta\xi}
\end{equation}
which is derived from \rfn{coherent} by Gaussian
integration.

This allows us to derive the matrix elements of the Greens
function $\hat G$ by Fourier transform: $$ \bbra{\eta} \int {\rm
d}^{d} k\ {e^{ikx}\over k^2+i\epsilon} \kket{\xi} = $$ $$
\int {{\rm d}^r k\ {\rm d}^r \kb\ {\rm d}^{d-2r}p
 \over k\kb + p^2 +i\epsilon}\  %
e^{-\half k\kb + i\kb\eta + ik\xi + \eta\xi + i p y} . $$ It is
convenient to express this as a proper time integral. We can then
easily include commuting dimensions as well; let there be $2r$
noncommuting and $d-2r$ commuting dimensions, with momenta $k$ and
$p$ respectively.  We include a separation $y$ in the commuting
directions for purposes of comparison. We then have
\begin{eqnarray} &
\bbra{\eta} {\hat G} \kket{\xi} = \int_0^\infty {\rm d}t \int {\rm
d}^r k~ {\rm d}^r\kb~ {\rm d}^{d-2r}p\
 e^{-t p^2 +ipy} \crt
&  \times e^{-t k\kb -\half k\kb + i\kb\eta + ik\xi + \eta\xi} .
\end{eqnarray}
The integrals over momenta are again Gaussian:
\begin{equation}
\label{eq:greenproper}
 = (2\pi)^{d\over 2} \ e^{\eta\xi}
\int_{0}^{\infty}
 {{{\rm d}t}\over{(2t)^{{d-2r}\over 2}
(2t+1)^r}}
 e^{- {{2\eta\xi}\over{2t+ 1}} - {y^2 \over 4t}} .
\end{equation}
This is the result.  Let us try to compare its behavior in \nc\
position space, with that in commutative space. This comparison
can be based on \rfn{cohbasis}, which tells us that
$e^{-a^2} \bbra{\eta=a} {\hat G} \kket{\xi=a}$ evaluates the
Greens function on a Gaussian centered at $x=a$. The IR (long
distance) behavior is controlled by the limit $t\rightarrow\infty$
of the proper time integral, and we see that it has the same
dependence on $a$ as on $y$.  Thus, in this sense, the IR behavior
is the same as for the commutative Greens function.

The UV behavior is controlled by short times $t\rightarrow 0$, and
at first sight looks rather different from that of the commutative
Greens function, due to the shifts $t\rightarrow t+\half$.
However, this difference is only apparent and comes because the
\nc\ Greens function in effect contains a factor of the
delta function \rfn{dltor}, which is hiding the UV divergence.
To see this, we can compute $\Tr {\hat \delta} {\hat G}$,
which is the correct way to take the coincidence limit.

This can be done in the coherent state basis, but we instead make
a detour to expand the Greens function in the Fock basis.
This can be done using
\rfn{mtrx}; one finds that the matrix elements are nonzero
only on the diagonal  and are a function only of the sum of the
occupation numbers $\sum_a n_a$ -- this reflects rotational
invariance. Changing variables from $t$ to $\lambda=1/(t+\half)$,
setting $y=0$ and going to the Fock basis, we obtain
\begin{equation}
\label{eq:greensum}
 \bra{n} {\hat G} \ket{n} =
 \int_0^2 {\rm d}\lambda\ \lambda^{{d\over 2}-2}
    (2-\lambda)^{r -{d \over 2}}
    (1-\lambda)^n .
\end{equation}
This result makes it easy to answer the previous question:
the sum over modes $\Tr \hat\delta\hat G = \sum_n \bra{n}\hat G\ket{n}$
produces $(2-\lambda)^{-r}$, which is exactly
the UV divergent factor which was missing from \rfn{greensum}.

In effect, the coincidence limit of the Greens function diverges in
the same way in \nc\ theory as in \con\ theory.  This was
clear in the original plane wave basis, and will imply that many
loop amplitudes have exactly the same UV divergence structure as they
would have had in \con\ theory, as we will discuss in \rfs{quant}.

We will find ourselves discussing more general Greens functions
in the interacting theory.  Given a two-point function $G^{(2)}(k)$
in momentum space, the same procedure can be followed to convert it
to the coherent state basis and thus interpret it in \nc\ position
space.  We will not try to give general results but instead simply
assert that this allows one to verify
that the standard relations between asymptotic behavior in
momentum and position space are valid in \nc\ space.

First, the long distance behavior is controlled by the analytic
behavior in the upper half plane; if the closest pole to the real
axis is located at $\Im k = m$, the Euclidean Greens function will
fall off as $e^{-mr}$ in position space.

Similarly, the short distance behavior is controlled by the large
$k$ asymptotics, $G^{(2)}(k) \sim k^{-\alpha}$ implies $G(r) \sim
r^{\alpha-d}$.

\subsubsection{The membrane/hydrodynamic limit}

The formal relation between $\IR^d_\theta$ and the Heisenberg algebra of
quantum mechanics suggests that it should be interesting to consider
the analog of the $\hbar\rightarrow 0$ limit.  Since $\theta$ is not
$\hbar$, this is not a classical limit, but rather a limit in which
the \nc\ fields can be treated as functions rather than operators.
This limit helps provide some intuition for the \nc\ gauge symmetry,
and has diverse interpretations in the various physical realizations
of the theory.

We consider \nc\ gauge theory on ${\IR}^d_{\t}$ and take the
following scaling limit: let 
\bea \label{eq:hydrosc}
{\t} = {\ell}^2 {\t}_0; \ A = {\ell}^{-2} A_0; \ g^2 = g_{0}^2 {\ell}^{-4}
\eea
and take ${\ell} \to
0$, keeping all quantities with the $0$ subscript fixed.
This corresponds to weak noncommutativity with strong gauge
coupling, fixing the dimensionful combination ${\l}^2 = g {\t}$.

Assume that $\theta^{ij}$ is invertible, and define the
functions 
\bea \label{eq:bif} y^i = x^i + {\t}^{ij} A_j (x) 
\eea
In the limit, the Yang-Mills action \rfn{yangmills} becomes
\bea \label{eq:hydr} 
S_h = {1\over{{\l}^4}} \int {\rm d}^d x
\sum_{i,j} \left( \{ y^i, y^j \} - {\t}^{ij} \right)^2 
\eea 
where
$$ \{ f, g \} = {\t}_0^{ij} {\p}_i f {\p}_j g
$$
is the ordinary Poisson bracket on functions. 

Infinitesimal gauge transformations take the form \rfn{canontrans} in the
limit.  The corresponding finite gauge transformations are general
canonical transformations, also called symplectomorphisms, 
which are diffeomorphisms
$x \mapsto {\tilde x} (x)$ which preserve the symplectic form:
\be \label{eq:symplecto}
{\t}^{-1}_{ij} d {\tilde x}^i \wedge d
{\tilde x}^j = {\t}^{-1}_{ij} dx^i \wedge dx^j .
\ee
In general, not all symplectomorphisms are generated by Hamiltonians
as in \rfn{canon}; those which are not are the analogs of
``large'' gauge transformations.

The model \rfn{hydr} is a sigma model, in the sense that the fields
$y^i$ can be thought of as maps from a ``base space'' $\IR^d$ to
a ``target space,'' also $\IR^d$ in this example.  One can easily
generalize this to let $y$ be a map from one Poisson manifold $(M,\theta)$
to another Poisson manifold $(N,\pi)$, which must also have
a volume form ${\m}$ and a metric $\Vert \cdot \Vert^2$ on
the space of bi-vectors. The action \rfn{hydr} will read: 
$$ S =
\int_{N} {\m} \, \Vert y_{*} {\t} - \pi  \Vert^2 
$$ and will again have
the group $Diff_{\t} (M)$ of symplectomorphisms of $M$ as a
gauge group.  This generalization also appears very naturally;
for example if one starts with $d+k$-dimensional Yang-Mills theory,
dimensionally reduces to $d$ dimensions, takes these to be $\IR^d_\theta$
and takes the limit, one obtains $M=\IR^d$ and $N=\IR^{d+k}$.

If some of the coordinates are commuting, one gets a gauged sigma model.
The case of a single commuting time-like dimension is particularly simple
in the canonical formulation: one has canonically conjugate variables
$(p_i,y^i)$, a Hamiltonian $H = \sum_i p_i^2 + S_h$, and a constraint
$0 = J = \sum_i \{p_i,y^i\}$.
In this form, the construction we just described essentially appears
in \citeasnoun{Hoppe:1982a}, while its maximally supersymmetric
counterpart is precisely the light-cone gauge fixed supermembrane
action of \citeasnoun{deWit:1988ig}.  Thus one interpretation of the fields
$y^i$ is as the embedding of a $d$-dimensional membrane into
$d+k$-dimensional space.

Although this picture is a bit degenerate if $k=0$, the configuration
space of such maps is still nontrivial.  The related time dependent
theory describes flows of a ``fluid'' satisfying \rfn{symplecto}.
This is particularly natural in two dimensions, where 
$\theta_{ij} dx^i dx^j$ is the area form, and these are allowed
flows of an incompressible fluid.  This ``hydrodynamic picture'' has
also appeared in many works, for example 
\cite{Bordemann:1993ep},
and is also known in condensed matter theory.

We shall use this hydrodynamic picture in what follows to
illustrate solutions and constructions of the \nc\ gauge theory.
As an example, the hydrodynamic limit of the
Seiberg-Witten map \rfn{swmapthree} is \cite{Cornalba:1999ah}
\bea
\label{eq:hydra} F_{ij} + {\t}^{-1}_{ij}  = [ \{ y^k, y^l \}
]^{-1}_{ij} 
\eea 
As \citeasnoun{Susskind:2001fb} points out, for $d=2$ this is just the
translation between the Euler and Lagrange descriptions of fluid
dynamics.

Similarly, the hydrodynamic analogue of the straight Wilson line
is the Fourier transform of fluid density,
\bea \label{eq:wilspo} 
W[k] = \int {\rm d}^d x \ {\rm Pf} ({\t}^{-1}) e^{i k_i y^i (x)}
\eea

\subsubsection{Matrix representations}

In a formal sense, any explicit operator representation is a ``matrix
representation.''  In this subsection we discuss $\IT^d_\theta$ as a
large $N$ limit of a finite dimensional matrix algebra, and how this might
be used to formulate regulated \nc\ field theory.
We take $d=2$ for definiteness, but the ideas generalize.

One cannot of course realize \rfn{noncom} using finite dimensional
matrices.  One can realize \rfn{torusphase} for the special case
with $\theta^{12}=2\pi M/N$, 
for example by $U_1=\Gamma_1^M$ and $U_2=\Gamma_2$, where
$(\Gamma_1)_{m,n} \equiv \delta_{m,n} \exp 2\pi i {n\over N}$
is the ``clock'' matrix, and
$(\Gamma_2)_{m,n} \equiv \delta_{m-n,1 (\mod N)}$
is the shift matrix.

Products of these matrices form a basis for $\Mat_N(\IC)$
(we assume $\gcd(M,N)=1$)
and in a certain sense this allows us to regard 
$\Mat_N(\IC)$ as an approximation to $\IT^2_\theta$: namely, if we write
$$
\Phi[x] \equiv \sum_{0\le m_1,m_2<N} U_1^{m_1} U_2^{m_2} 
 e^{\pi i \theta^{ij} m_i m_j - 2\pi i m_i x^i} ,
$$
we find that the $\Phi[x]$ evaluated at lattice sites $x^i=n^i/N$
provide a basis in terms of which 
the multiplication law becomes \rfn{starkernel}, with the integrals
replaced by sums over lattice points.
This construction is sometimes referred to as a ``fuzzy $\IT^2$''
and was independently proposed in several works 
(\citeasnoun{deWit:1988ig,Hoppe:1989gk}; see also the references in
\citeasnoun{Bars:1999av})
as a starting point for regulated field theory on $\IT^2$, since we can regard
the bound $0\le m_i<N$ as a UV cutoff.

One can obtain $\IT^2_\theta$ with more general $\theta$ by taking the
limit $M,N\rightarrow\infty$ holding $\theta=2\pi M/N$ fixed, and
obtain $\IR^2_\theta$ by taking $g_{ij}$ (and thus the volume)
large in an obvious way.  Formulating this limit precisely enough to
make contact with our previous discussion requires some mathematical
sophistication, however, as there are many distinct algebras which can
be obtained from $\Mat_N(\IC)$ by taking different limits.  
For example, in \rfs{othernc} we will discuss a sense in which
the large $N$ limit of $\Mat_N(\IC)$ leads to functions on a sphere,
not a torus.  

For most physical purposes, the definitions of
$\IR^d_\theta$ we gave previously are easier to use.
On the other hand, making the limit explicit is a good starting point
for making a nonperturbative definition of quantum \nc\ field theory
and for a deeper understanding of the renormalization group.
Physically, one usually thinks of the continuum limit (at least at weak
coupling) as describing modes with low kinetic energy, so to decide which
algebra will emerge in the limit, we need to consider the derivatives.

Candidate derivatives for 
fuzzy $\IT^2$ are operators $D_i$ satisfying
$$
D_i \Phi[x] D_i^\dagger = \Phi[x + e_i]
$$
where $(e_i)^j = \delta_i^j/N$ is the lattice spacing; e.g. take
$D_1 = \Gamma_2^a$ with $a M = -1 \mod N$ and $D_2=\Gamma_1$.
Thus, a plausible regulated form of \rfn{genaction} might be
\begin{equation}
S = \frac{N^2}{2} \Tr \sum_{i} (D_i \Phi D_i^\dagger - \Phi)^2
  + \Tr V(\Phi) .
\end{equation}

The same approach could be followed for \nc\ gauge theory, leading to
a rather ugly action.  One aspires to the elegance of \rfn{matrixmodel},
in which the derivatives emerge from a choice of background
configuration.  It should be clear at this point (and will be made
more so in \rfs{opalg}) that one cannot simply take
$\CA=\Mat_N(\IC)$ in \rfn{matrixmodel}; the derivatives must be
specified somehow.  

Perhaps the best proposal along these lines at present is due to
\citeasnoun{Ambjorn:1999ts}.  We will just state this as a recipe; the
motivation behind it will become clearer upon reading
\rfs{nctorus}.

The starting point is the lattice twisted Eguchi-Kawai
model, a matrix model whose dynamical variables are $d$ unitary
matrices $U_i \in U(N)$, and the action
$$
S = -{1\over g^2} \sum_{i\ne j} Z_{ij}
 \Tr U_i U_j U_i^\dagger U_j^\dagger .
$$
This is Wilson's lattice gauge theory action restricted to a single
site, which is a natural nonperturbative analog of
\rfn{matrixmodel}, generalized by the ``twist factors'' $Z_{ij}$
which are constants satisfying $Z_{ij}=(Z_{ji})^*$.

This action is supplemented by the constraints
$$
\Omega_i U_j \Omega_i^\dagger =
 e^{2\pi i \delta_{ij} r_i/N} U_j
$$
which for suitable matrices $\Omega_i$ and constants $r_i$ can be
shown to admit $U_i = D_i \tilde U_i$ as solutions, where $D_i$ are
derivatives as above, and $\tilde U_i$ are unitary elements of $\IT^d_\theta$.

Following the ideas above, one can show that this reproduces
perturbative \nc\ gauge theory, and captures interesting
nonperturbative structure of the model, namely the Morita equivalence
described in \rfs{morita}, for rational $\theta$.  It would be
interesting to justify this further, for example by detailed analysis
in two dimensions.

\section{SOLITONS AND INSTANTONS}
\label{sec:sol}

Field theories and especially gauge theories admit many classical
solutions: solitons, instantons and branes,
which play important roles in nonperturbative physics.

In general, solutions of conventional field theory
carry over to the analogous \nc\ theory, but with any singularities
smoothed out, thanks to position-space uncertainty.
In particular, the \nc\ rank $1$ theory has
non-singular instanton, monopole and vortex solutions,
and in this respect (and many others)
is more like \con\ Yang-Mills, not Maxwell theory.

An even more striking feature of \nc\ theory is that
solitons can be stable when their \con\ counterparts
would not have been.  As we will see, noncommutativity provides a natural
mechanism for stabilizing objects of size $\sqrt{\t}$.

Another striking feature is how closely the properties of \nc\ gauge
theory solutions mirror the properties of corresponding
Dirichlet brane solutions of string theory.  We will discuss this
aspect in \rfs{strsol}.

\subsection{Large $\t$ solitons in scalar theories}

In commutative scalar field theories in two and more spatial dimensions,
there is a theorem which prohibits the existence of finite energy
classical solitons \cite{Derrick:1964a}.
This follows from a simple scaling argument: upon shrinking all length
scales as $L\rightarrow \lambda L$, both kinetic and potential energies
decrease, so no finite size minimum can exist.

This argument will obviously fail in the presence of a distinguished
length scale $\sqrt{\theta}$ and in fact one finds that for sufficiently
large $\theta$, stable solitons can exist in the \nc\ theory
\cite{Gopakumar:2000zd}.

The phenomenon can be exhibited in $2+1$ dimensions and we consider
such a field theory with action \refgenaction.
It is convenient to work with canonically commuting \nc\ coordinates,
defined as $x^1+ix^2=z\sqrt{\theta}$ and $x^1-ix^2=\zb\sqrt{\theta}$.
In terms of these, the energy becomes
$$
E = \int d^2z \half (\p\phi)^2 + \theta V(\phi) .
$$
In the limit of large $\theta V$, the potential energy dominates, and
we can look for an approximate solitonic solution by solving the
equation
\begin{equation}
\label{eq:criticalpt}
{\p V\over\p\phi} = 0 .
\end{equation}
For example, if we consider a cubic potential, this equation would be
\begin{equation}
\label{eq:cubicV}
V'(\phi) = g \phi^2 + \phi = 0 .
\end{equation}
While in commutative theory these equations would admit only constant
solutions, in \nc\ theory the story is rather more interesting.  It is
simplest in the Fock space basis, in which the field can be taken to be
an arbitrary (bounded) operator on $\CH_r$, and for which
the multiplication is just operator multiplication.
Since $\phi$ is self-adjoint it can be diagonalized, so we can immediately
write the general solution of \rfn{cubicV}:
$$
\phi = -{1\over g} U^\dagger P U
$$
where $U$ is a unitary and $P^2=P$ is a projection operator, characterized
up to unitary equivalence by $\Tr P$ (which must be finite for a finite
energy configuration) or equivalently the number of unit eigenvalues.
As discussed in \rfs{bases}.3, the diagonal operators correspond
to radially symmetric solutions, from which unitary rotations produce
all solutions.

The simplest solution of this type uses the operator
$P_0=\ket{0}\bra{0}$, of energy $2\pi\theta V(-{1/ g})=\pi\theta/3g^2$.
It is remarkable that this energy depends only on the value of the
potential at the critical point, and nothing else.

As we discussed in \rfs{bases}.2, this is a
Gaussian of width $\sqrt{\theta}$ which squares to itself under star
product.  We see that the scaling argument for instability is violated
because of the position-space uncertainty principle, which causes the
energy of smaller Gaussians to increase.

The general solution in this sector is $\phi=-{1\over
g}U^\dagger\ket{0}\bra{0}U$.  This includes Gaussians with arbitrary
centers, the higher modes discussed in \rfs{bases}.2, and various
``squeezed states.''  Neglecting the kinetic term, they are all
degenerate and are parameterized by an infinite dimensional moduli
space ``$\lim_{N\rightarrow\infty} U(N)/U(N-1)$.''  This infinite
degeneracy will however be lifted by the kinetic term.  The story is
similar for solutions with $\Tr P = n$; this moduli space has a limit
in which the solutions approach $n$ widely spaced $n=1$ solitons, with
exponential corrections.

This example and all of its qualitative features generalize immediately
to an arbitrary equation \rfn{criticalpt}.  Its
most general solution is
\begin{equation}
\label{eq:gensoliton}
\phi = U^\dagger \left(\sum_i \lambda_i P_i \right) U
\end{equation}
where the $\lambda_i$ are the critical points $V'(\lambda_i)=0$, and
the $P_i$ are a set of mutually orthogonal projections whose sum is the
identity.  The analysis also generalizes in an obvious way to higher
dimensional theories with full spatial noncommutativity ($d=2r+1$),
by tensoring projections in each Fock space factor.

Solutions for which all $\lambda_i$ are minima of $V(\lambda)$ are
clearly locally stable, if we neglect the kinetic energy term.
We now consider the full energy functional, which in the Fock basis becomes
$$
E = \Tr [z,\phi][\phi,\zb] + \theta \Tr V(\phi) .
$$
The kinetic term breaks $U(\CH_r)$ symmetry and might be expected to
destabilize most of the infinite dimensional space of solutions
\rfn{gensoliton}.  On the other hand, for low modes
($\ket{n}\bra{n}$ with $n\sim 1$), the kinetic energy will
be $O(1)$, so for sufficiently large $\theta$ a stable solution
should survive.  This was checked
by \citeasnoun{Gopakumar:2000zd} by an analysis of linearized stability,
with the result  (for the $n=1$ solutions in $d=2+1$)
that only the minimal Gaussian $\ket{0}\bra{0}$ and
its translates are stable.

The solution cannot exist at $\theta=0$ and it is interesting to ask
what controls the critical value $\theta_c$ at which it disappears
\cite{Zhou:2000xg,Durhuus:2000uz,Jackson:2001iy}.
One can easily see the rather surprising fact that this does not depend
directly on the barrier height.  This follows
because we can obtain a family of  equivalent problems with very
different barrier heights by the rescaling $\phi \rightarrow a\phi$
and $E \rightarrow E/a^2$, all with the same $\theta_c$.

Rather, the condition for \nc\ solitons to exist is that noncommutativity
be important at the scale set by the mass of the $\phi$ particle
in the asymptotic vacuum, i.e. $\theta V'' >> 1$.  Consider
the symmetric $\phi^4$ potential.  By the above argument, $\theta_c$ can
only depend on $V''$ at the minimum; numerical study leads to the result
\begin{equation}
\label{eq:criticaltheta}
\theta_c V''(0) = 13.92 .
\end{equation}
There is some theoretical understanding of this result, which suggests
that this critical value is roughly independent of the shape
of the potential.
It is plausible that only radially symmetric configurations are relevant
for stability and if one restricts attention to this sector,
the equation of motion reduces to a simple three-term recurrence relation
for the coefficients $c_n$ in \rfn{radialsolution},
\begin{equation}
\label{eq:recurrel}
(n+1)c_{n+1} - (2n+1)c_n + n c_{n-1} = {\theta\over 2} V'(c_n) ;
 \qquad n\ge 0.
\end{equation}
Suppose that $V(\phi)$ is bounded below, the vacuum is $\phi=0$,
and we seek a solution which approximates the $\theta = \infty$
one soliton solution $\left( c_0=\lambda; \, c_n =0, n > 0
\right)$. Finiteness of the energy requires
$\lim_{n\rightarrow\infty} c_n = 0$, and we can get the large $n$
asymptotics of such a solution by ignoring the nonlinear terms in
$V'(\lambda)$; this leads to
\begin{equation}
\label{eq:casympt}
c_n \sim n^{1/4} e^{-\sqrt{n \theta V''(0)}} .
\end{equation}
This shows that $c_n$ varies smoothly when
$
\theta V^{\prime\prime}(0) \ll 1,
$
and can be approximated by a solution of the differential equation analog of
\rfn{recurrel}.  However,
corresponding to the nonexistence of a solution in the commutative theory,
one can show on very general grounds that no solution of this differential
equation can have the boundary value $c(0)=\lambda$, even with the
nonlinear terms in $V'$ included.
A nontrivial soliton is possible only if this continuous approximation
breaks down, which requires the control parameter $\theta V''(0)$ to be large.

In particular, a nontrivial one soliton solution must have a discontinuity
$|c_0-c_1| >> 0$, and matching this on to \rfn{casympt} for $c_1$
provides a lower bound for $\theta V''(0)$.
This analysis leads to an estimate which is quite
close to \rfn{criticaltheta}.

Multi-soliton solutions have been studied recently in 
\cite{Lindstrom:2000kh,Gopakumar:2001yw,Hadasz:2001cn}.  
Solutions on $\IT^d_\theta$ have been studied
in \cite{Bars:2000yv}.

\subsection{Vortex solutions in gauge theories}

Derrick's theorem does not hold in gauge theories and as is well known,
the abelian Higgs model (Maxwell theory coupled to a complex scalar field)
has vortex solutions, which (among other applications)
describe the flux tubes in superconductors \cite{Nielsen:1973a}.

We proceed to discuss analogous solutions in the \nc\ gauge theory
\cite{Nekrasov:1998ss,Polychronakos:2000zm,Gross:2000ph,Aganagic:2000mh,%
Bak:2000ac}.
We work in $2+1$ dimensions with $\theta_{xy}=1$
and look for time-independent solutions in $A_0=0$ gauge.  
Using \refDCsub\ and the Fock basis, the energy is
\bea \label{eq:abelianhiggs}
E = {2\pi\over g^2}\int dt\Tr ~\half F^2
 + \sum_{i=1,2} D_i\phi~ D_i \phi^\dagger
 + V(\phi) .
\eea
Here 
$$
F = [C,\Cb]+1
$$
and $\phi$ is a complex scalar field (satisfying no hermiticity condition).

We first note that unlike Maxwell theory, even the pure rank
one \nc\ gauge theory admits finite energy solitonic solutions.
The pure gauge static equation of motion
is $$ 0 = [C,[C,\Cb]] $$ Of course, it is solved by the vacuum
configuration $C=\zb$ and $\Cb=z$, and gauge transformations of
this,
\begin{equation}
\label{eq:genvortex}
C = U^\dagger \zb U ; \qquad \Cb = U^\dagger z U .
\end{equation}
What is
amusing, is that $U$ does not need to be unitary in order for this
transformation to produce a solution \cite{Witten:2000nz,Harvey:2000jb}.
It needs to satisfy $U U^\dagger = 1$, but $U^\dagger U$ need not be
the identity.  This implies the partial isometry condition
\rfn{partialisometry} and is a bit stronger.

The simplest examples use the shift operator \rfn{shift}:
$U = S^m$.
To decide whether these are vortex solutions, we should compute the
magnetic flux. This is
$$
F = (S^\dagger)^{m} [\zb,z] S^{m} + 1 = 1 - (S^\dagger)^m S^m  =
\sum_{n=0}^{n-1} P_n
$$
where $P_n \equiv \ket{n}\bra{n}$.
The total flux is $\Tr F = m$.  Thus
\nc\ Maxwell theory allows nonsingular vortex solutions,
sometimes called ``fluxons,''
without needing a scalar field or the Higgs mechanism.

Physically, we might interpret this as an \nc\ analog of the
commutative gauge theory vortex $A_i = g^{-1} \p_i g$ with $g=e^{im\theta}$.
This is pure gauge except at the origin where it is singular, so we might
regard this as an example of \nc\ geometry smoothing out a singularity.

The soliton mass $M$ is proportional to $\Tr F^2 = m$.
To restore the dependence on the coupling constants, we first
note that (as for any classical soliton) the mass is proportional to
$1/g^2$.  This quantity has dimensions of length in $2+1$
dimensions, so on dimensional grounds the mass must be proportional
to $1/\theta$, consistent with the nonexistence of the fluxon
in the \con\ limit.
In the conventions of \rfn{abelianhiggs},
\bea \label{eq:stension}
M = \frac{\pi m}{g^2\theta} .
\eea

The most general solution is slightly more general than this; it is
\begin{eqnarray}
\label{eq:lift}
& C = (S^\dagger)^m \zb S^m + \sum_{n=0}^{m-1} c_{n}(x^0) \ket{n}\bra{n} \\
& \Cb = (S^\dagger)^m z S^m + \sum_{n=0}^{m-1} \cb_{n}(x^0) \ket{n}\bra{n}.
\end{eqnarray}
The $2m$ functions $c_n$,$\cb_n$ must satisfy $\p_0^2 c_n = \p_0^2 \cb_n = 0$
and can be seen to parameterize the world-lines of the $m$ fluxons.
This is particularly clear for $m=1$ by recalling \rfn{Cdefn}.

A peculiar feature to note is that fluxons exist with only one
sign of magnetic charge, $F$ aligned with $\theta$.  It is also
rather peculiar that they exert no force on one another; the
energy of the configuration is independent of their locations.

Even more peculiar, the equations $\p_0^2 c_n = \p_0^2 \cb_n = 0$
admit as solutions $c_n = x_n + v_n t$ and $\cb_n = \xb_n + \vb_n t$
with no upper bound on $v$.  In other words, the fluxons can move
faster than light \cite{Hashimoto:2000ys,Bak:2000im}.  Of course $\theta$
defines a preferred rest frame, and there is no immediate
contradiction with causality in this frame.

These peculiarities may be made more palatable by the realization
that all of these solutions (with no scalar field) are unstable,
even to linearized fluctuations.  For example, the $m=1$ solution
admits the fluctuation
\begin{equation}
\label{eq:fluxoninstab}
C + T = S^\dagger \zb S + t S^\dagger P_0
\end{equation}
where $t$ is a complex scalar parameterizing the fluctuation and
$S^\dagger P_0 = \ket{1}\bra{0}$.
One can straightforwardly compute
$$
[C+T,\Cb+\Tb] + 1 = P_0 + |t|^2 (P_1 - P_0) .
$$
The total flux $\Tr F$ is constant under this variation, while
the energy is proportional to
$$
\Tr F^2 = (1-|t|^2)^2 + |t|^4
$$
which exhibits the instability.  This is of course just like the
commutative case; the flux will tend to spread out over all of space
in the absence of any other effect to confine it.

There is no stable minimum in this topological sector.  This follows
if we grant that $\Tr F$ is ``topological'' and cannot change under
any allowed variation of the fields.  This of course depends on one's
definitions, but in \con\ gauge theory one can make a definition which
does not allow flux to disappear.  Starting with a localized configuration,
the flux may disperse or go to infinity, but one can always enlarge one's
region to include all of it, because of causality.  Similarly, energy
cannot be lost at infinity.

Since \nc\ theory is not causal and indeed a fluxon can move faster
than light, this argument must be reexamined.  The belief at present
is that flux and energy are also conserved in \nc\ gauge theory; they
are conserved locally and cannot not run off to infinity in finite
time.  In the case of the fluxon, a given solution will have some
finite velocity, and as it disperses, it would be expected to slow
down, so that once it has spread over length scales large compared to
$\sqrt{\theta}$ conventional causality will be restored.  This point
could certainly use more careful examination however.

This physical statement underlies the \con\ definition which
leads to topological sectors characterized by total flux; one only
considers variations of the fields which preserve a specified
falloff at infinity.  An analogous definition can be made in \nc\ theory,
as we will discuss in \rfs{opalg}.

One can use the same idea to
generate exact solutions to the \nc\ abelian Higgs theory,
even with a general scalar potential $V(\phi)$.
Now one starts with the scalar in a vacuum configuration $V'(\phi)=0$
and $D_i\phi=0$ (so $\phi \propto 1$), and again
applies an almost gauge transformation with
$U U^\dagger = 1$, but $U^\dagger U\ne 1$.
The same argument as above shows that this will be a solution for any $U$.
This ``solution generating technique'' has been used
to generate many exact solutions
\cite{Hamanaka:2000aq,Harvey:2000jb,Hashimoto:2000kq,%
Schnabl:2000cp,Tseng:2000sz,Bergman:2001kz}.

Of course the properties of these solutions, including stability,
depend on the specific form of the scalar potential and the
choice of matter representation; both ``adjoint'' matter with
$D_i\phi = [C_i,\phi]$, and fundamental matter as defined in
\rfs{kin}.D.3 have been studied.

A particularly nice choice of potential \cite{Jatkar:2000ei} is
$$
V = \half (\phi\phi^\dagger - m^2)^2 ,
$$
as in this case the energy can be written as the sum of squares and
a total derivative as in \cite{Bogomolny:1976de}, 
leading to a lower bound $E \ge |\Tr F|$.
Solutions saturating this bound are called BPS and are clearly stable.

For adjoint matter and $\theta m^2=2$,
the exact solution discussed above is BPS.
For fundamental matter, the bound can be attained in two ways.\footnote{
In the references, the two types of solution are sometimes referred to as
``self-dual'' and ``anti-self-dual,'' but not consistently.  It seems 
preferable to speak of positive flux ($F$ aligned with $\theta$) and
negative flux.}
One can have
$$
F = m^2 - \phi \phi^\dagger; \qquad \Db \phi = 0 ,
$$
which has positive flux solutions.  For $\theta m^2=1$ the exact solution
as above with $\phi=S^\dagger$ is BPS. \citeasnoun{Bak:2000im} show that
BPS solutions only exist for $\theta m^2\le 1$, and argue that
for $\theta m^2>1$ the exact solution is a stable non-BPS solution.

One can also have
$$
F = \phi \phi^\dagger - m^2; \qquad D \phi = 0 .
$$
This has been shown to have a negative flux
solution for any value of $\theta m^2$
\cite{Jatkar:2000ei,Lozano:2000qf},
which in the $\theta\rightarrow 0$ limit
reduces to the \con\ Nielsen-Olesen solution.

The vortex solutions admit a number of direct generalizations to higher
dimensional gauge theory.  If one has $2r$ noncommuting coordinates,
one can make simple direct products of the above structure to obtain
solutions localized to any $2(r-n)$-dimensional hyperplane.
One can also
introduce additional commuting coordinates, and it is not hard
to check that the parameters $c_n$, $\cb_n$ in the solution
\rfn{lift} must then obey the wave equation
$0=\eta^{ij} \p_i\p_j c$ (resp. $\cb$) in these coordinates.
This is as
expected on general grounds (they are Goldstone modes for space-time
translations and for the symmetries $\delta\phi=\epsilon$)
and fits in with the general philosophy that a soliton in $d+1$-dimensional
field theory which is localized in $d-p$ dimensions, should be regarded
as a ``$p$-brane,'' a dynamical object with a $p+1$-dimensional
world-volume which can be described by fields and a local effective action
on the world-volume.  While the vortex in $3+1$-dimensional gauge theory,
which is a string with $p=1$, may be the most familiar case,
the story in more dimensions is entirely parallel.

One can get a nontrivial fluxon solution in
the hydrodynamic limit (as in \rfs{bases}.5 with $k=0$)
by rescaling the magnetic charge as $m \sim \ell^{-d}$.
The solution becomes 
\bea \label{eq:dvortex} y^i = x^i \ \sqrt{
1 - {{L^{d}}\over{r^{d}}}}, &\qquad 
r > L, \crt y^i = 0, &\qquad r \leq L \crt r^2 =  \sum_i (x^i)^2 &
\eea 
The vortex charge $m \propto (L/\ell)^d$ is no longer quantized,
but it is still conserved.

\subsection{Instantons}

To obtain qualitatively new solutions of gauge theory,
we must move on to four Euclidean dimensions.  As is well known,
minima of the Euclidean action will be self-dual and anti-self-dual
configurations,
\def\asd{Eq. (\ref{eq:asd})}
\begin{eqnarray}
\label{eq:asd}
(P^{\mp})^{mn}_{kl} F_{mn} = 0 \\
\label{eq:sdproj}
(P^{\pm})^{mn}_{kl} \equiv
\half \delta^m_k \delta^n_l {\pm} {1\over{4}} {\ve}^{mn}_{\ \ kl} ,
\end{eqnarray}
where $P^+$ and $P^-$ are the projectors on self-dual
and anti-self-dual tensors.
These solutions are classified topologically by the instanton charge:
\begin{equation}
\label{eq:inch}
N = -
{1\over{8\pi^2}} {\int}{\Tr} F \wedge F
\end{equation}
Note that $\theta$ breaks parity symmetry,
and the two types of solution will have different properties.
Let $\theta$ be self-dual; then
one can obtain self-dual solutions by the direct product construction
mentioned in the previous subsection, while the anti-self-dual solutions
turn out to be the \nc\ versions of the Yang-Mills instantons.

Instantons play a central role in the nonperturbative
physics of Yang-Mills theory
\cite{Schafer:1998wv}
and have been studied from many points of view.
The most powerful approach to constructing
explicit solutions and their moduli space is
the so-called ADHM construction \cite{Atiyah:1978a}, which reduces this
problem to auxiliary problems involving simple algebraic equations.
Although when it was first proposed,
this construction was considered rather recherch\'e by physicists
\cite{Coleman:1979a}, modern
developments in string theory starting with
\cite{Witten:1996gx} have placed it in a more physical context, and
in recent years
it has formed the basis for many practical computations in nonperturbative
gauge theory, e.g. see \cite{Dorey:2000ww}.  We will explain
the stringy origins of the construction in \rfs{strsol}.

It turns out that the ADHM construction can be adapted very readily to
the \nc\ case \cite{Nekrasov:1998ss}.
Let us quote the result for anti-self-dual gauge fields, $P^+ F = 0$,
referring to \citeasnoun{Nekrasov:2000ih} for proofs, further explanations
and generalizations.  See also \citeasnoun{Furuuchi:2000vc}.

To construct charge $N$ instantons in the $U(k)$ gauge theory we must
solve the following auxiliary problem involving the following
finite-dimensional matrix data.
Let $X^i, i=1,2,3,4$, be a set
of $N \times N$ Hermitian matrices,
transforming as a vector under $SO(4)$ space rotations and
in the adjoint of a `dual' gauge group $U(N)$.
Let ${\l}_{\a}$,
${\a} =1,2$ be a Weyl spinor of $SO(4)$,
transforming in $(N,k)$ of $U(N) \times U(k)$.
Instanton solutions will then be in correspondence with
solutions of the following set of equations:
\begin{equation}
\label{eq:mmnt}
0 = (P^+)^{kl}_{ij} \left( [X^i, X^j]
 + {\bar \l} {\s}^i \bar\s^j {\l} - {\t}^{ij} {\bf 1}_{N \times N} \right).
\end{equation}
These equations admit a $U(N)$ symmetry
$(X^i, {\l}_{\a}) \mapsto (g_N X^i g_N^{-1},
{\l}_{\a} g_{N}^{-1} )$, and two solutions related by this symmetry
lead to the same instanton solution.  Thus the moduli space of
instantons is the non-linear space of solutions of \rfn{mmnt}
modulo $U(N)$ transformations.
By counting parameters, one finds that it
has dimension $4Nk$.

The only difference with the \con\ case is the shift by $\theta$.
This eliminates solutions with $\lambda=\bar\lambda=0$,
which would have led to a singularity in the moduli space.
In \con\ Yang-Mills theory, instantons have a scale size parameter
which can be arbitrary (the classical theory has conformal invariance),
and the singularity is associated with the zero size limit.
The scale size is essentially $\rho \sim \sqrt{\bar\lambda\lambda}$,
and in the \nc\ theory is bounded below at $\sqrt{\t}$, again
illustrating that position-space uncertainty leads to a minimal size
for classical solutions.

To find the explicit instanton configuration corresponding to a
solution of Eq. (\ref{eq:mmnt}), one must solve an
auxiliary linear problem: find a pair
$({\psi}_{\a}, {\xi})$, with the Weyl spinor ${\psi}_{\a}$ taking
values in $N \times k$ matrices over ${\CA}_{\t}$, and ${\xi}$
being a $k \times k$ ${\CA}_{\t}$-valued matrix, such that
\begin{eqnarray}
\label{eq:drc}
X^i {\s}_i {\psi} + {\l} {\xi} = 0 \crt
{\psi}^{\dagger} {\psi} + {\xi}^{\dagger} {\xi} = {\bf 1}_{k
\times k}
\end{eqnarray}
Then the instanton gauge field is given by
\begin{eqnarray}
\label{eq:inst}
& A_i = {\psi}^{\dagger} {\p}_i
{\psi} + {\xi}^{\dagger} {\p}_i {\xi} \\
& C_i = -i {\t}_{ij}
\left( {\psi}^{\dagger} x^j {\psi} + {\xi}^{\dagger} x^j {\xi}
\right) .
\end{eqnarray}

One can make these formulae more explicit in the $k=N=1$ case.
We follow the conventions of Eq. (\ref{eq:canon}).
Let
$$
{\Lambda} = {{\sum_{\a} {\t}_{\a} ( {\hat n}_{\a} + 1
)}\over{\sum_{\a} {\t}_{\a} {\hat n}_{\a}}} ;
\, D_{1} = C_1 + i C_2; \, D_2 = C_3 + i C_4.
$$
Then
$$
D_{\a} =
{1\over{\sqrt{{\t}_{\a}}}} S {\Lambda}^{-{1\over 2}} {\zb}_{\a}
{\Lambda}^{{1\over 2}} S^{\dagger},
$$ where the operator $S$ is the shift operator in Eq. (\ref{eq:shift}).

This solution is nonsingular with size $\sqrt{\t}$.  One can take the
limit $\t\rightarrow 0$ to return to \con\ Maxwell theory, obtaining a
configuration which is pure gauge except for a singularity at the
origin.  Whether or not this counts as an instanton depends on the
underlying definition of the theory; in string theory we will see later
that it is.

\subsection{Monopoles and monopole strings}

One might at this point suspect that in general \nc\ gauge theory solitons
look quite similar to their Yang-Mills counterparts, if perhaps less
singular.  The monopole will prove an exception to this rule.

We consider static field configurations
in $3+1$ gauge theory with an adjoint
scalar field $\phi$, and the energy functional
\begin{equation}
\label{eq:lagint} E =  {1\over{4g_{\rm YM}^2}}
\int\Tr \  \sum_{ 1\le i < j\le 3}  (F_{ij})^2
 + \sum_{i=1}^3  (\nabla_i\phi)^2 .
\end{equation}
Just as in the \con\ theory,
one can rewrite this as a sum of a total
square and a total derivative:
\begin{eqnarray}
\label{eq:splt}
E =
{1\over{4g_{\rm YM}^2}}
\int d^3x
 {\Tr} \left( {\nabla}_{i} {\phi} \pm B_i \right)^2 \mp Z \\
\label{eq:moncharge}
Z = {1\over{4g_{\rm YM}^2}}
\int d^3x
\Tr {D}_i \left( B_i \star {\phi} + {\phi} \star B_i  \right)
\end{eqnarray}
The total derivative term $Z$ depends only on the boundary conditions
and in the \con\ case would have been proportional to
the magnetic charge of the soliton.
Minimizing the energy with fixed $Z$ leads to $E=|Z|$ and
the equations \cite{Bogomolny:1976de}
\begin{equation}
\label{eq:bgmlni}
\nabla_i \phi = \pm B_i.
\end{equation}

We are only going to consider nonsingular configurations,
and the first observation to make is that this implies that
the diagonal part of the total magnetic charge is zero,
$$
0 = Q = \int d^3x \Tr [D_i, B^i] ,
$$
by the Bianchi identity, analogous to the \con\ theory.

This would seem to rule out the possibility of a nonsingular
rank $1$ monopole solution.  Nevertheless, it turns out that such
a solution exists, as was discovered by explicit construction
\cite{Gross:2000wc}
making use of the Nahm equations \cite{Nahm:1980yw}.  These equations
are very analogous to the ADHM equations and
admit an equally direct \nc\ generalization.
They are ordinary
differential equations and their analysis is somewhat intricate;
we refer to \cite{Nekrasov:2000ih} for this and approach the solution in
a more elementary way.

We start with the fact that in three spatial dimensions (we assume
only spatial noncommutativity), since the Poisson tensor has at most
rank two, there will be a commutative direction.  Call its coordinate
$x^3$.  This theory will admit the vortex solution of \ref{sec:sol}.B,
a string extending to infinity along $x^3$, say
$$
B_3 = \ket{0}\bra{0} .
$$
This can be used to find
a simple rank 1 solution of \rfn{bgmlni} by postulating $A_3=0$ and
$$
\phi = x^3 \ket{0}\bra{0} .
$$

Now, we might not be tempted to call this solution a monopole.
However it is only one point in a moduli space of solutions. The
linearized equation of motions around the solution take the form
$$ D^i D_i \delta\phi = 0, \quad D^i D_i {\d} A_j = 0, \quad D_j
{\d} A_j + [ {\phi}, {\d} \phi] = 0. $$ Besides variations of the
location of the vortex in the $(x^1,x^2)$ plane and constant
shifts of $\phi$, these also include 
$$ {\d}{\phi} \propto e^{-(x^3)^2/{\t}} \ \vert 0 \rangle\langle 1
\vert $$ Turning on this mode corresponds to splitting the string
in two, as one can see by looking at the eigenvalues of the
operator ${\phi} + {\d} {\phi}$.

This linearized variation extends to a finite variation of the
solution which can be used to send one half of the string off to
infinity.  The remaining half is a rank $1$ monopole attached to a
{\it physical} analog of the Dirac string, the flux tube of
\ref{sec:sol}.B stretching from the monopole to infinity carrying
magnetic flux and energy, and cancelling the monopole magnetic
flux at infinity.  Its energy diverges, but precisely as the
string tension \rfn{stension} times the length of the string.
Thus the \nc\ gauge theory has found a clever way to produce a
solution despite the absence of $U(1)$ magnetic charge in three
dimensions.

One can express the solution in closed form in terms of error
functions \cite{Gross:2000wc}.  Self-dual solutions to
gauge theory generically admit closed form expressions; the deep
reason for this is the integrable structure of these equations
\cite{Ablowitz:1991xb}.  The \nc\ deformation respects
this; for example the \nc\ Bogomolny equations for axially
symmetric monopoles are equivalent to nonabelian Toda lattice
equations.

Multimonopoles also exist, and as for the vortices, this moduli space
will have limits in which it breaks up into $U(1)$ monopoles.  In
particular, the \nc\ analog of the 't Hooft-Polyakov monopole of the
$U(2)$ theory is better thought of as a multimonopole solution, with
two centers associated to the two $U(1)$ subgroups, connected by a
string.

A simple observation following from the form of the noncommutative
Nahm equations is that (in contrast to the instantons) this
multimonopole moduli space is the same as that of the \con\
theory.  Since the \con\ solutions and moduli space were already
nonsingular, this fits with the general picture of desingularization
of solitons by noncommutativity, but is still non-trivial (it was not
forced by symmetry).

There are pretty string theory explanations for all of this, discussed
in \rfs{strsol}.  In string theory, monopoles
turn out naturally to be associated with ``D-strings'' extending in a
higher dimension.  In commutative gauge theory, the string extends
perpendicular to the original dimensions, and its projection on these
is point-like.  Noncommutativity tilts the string in the extra
dimensions, leading to a projection onto the original dimensions which
is itself a string.

\section{NONCOMMUTATIVE QUANTUM FIELD THEORY}
\label{sec:quant}

In this section we discuss the properties of quantized \nc\
field theories.  We start by developing the Feynman rules and explain
the role of planarity in organizing the perturbation theory
\cite{Filk:1996dm}.

We then give examples illustrating the basic structure of
renormalization: the UV properties are controlled by the planar
diagrams, while nonplanar diagrams generally lead through what is
called ``UV/IR mixing'' \cite{Minwalla:1999px} to new IR phenomena.
The limit $\theta\rightarrow 0$ in these theories is non-analytic.
We cite a range of examples in scalar and gauge theory which illustrate
the possibilities; there is some physical understanding of its
consequences.
We also discuss the question of gauge invariance of
the effective action.

We then discuss a variety of results for physical observables,
including finite temperature behavior and
a universal high energy behavior of the Wilson line operator.
We also outline the Hamiltonian treatment of \nc\ gauge theory on a
torus, and conclude with miscellaneous other results.

\subsection{Feynman rules and planarity}

We start by considering the theory of a single scalar field $\phi$
with the action $$ S = \int\Tr \half (\p\phi)^2 - V(\phi) ;\qquad
V(\phi) \equiv {m^2\over 2}\phi^2 + \sum_{n>2} {g_n\over n!}
\phi^n . $$ Functional integral quantization can be done in the
standard way and leads to Feynman rules
which are almost the same as for \con\ scalar field theory.
In particular the propagator is the same thanks to Eq.
(\ref{eq:quadraticint}). The only difference is an additional
momentum dependence in the interaction vertices following from
\refmoyalproduct: a $\phi^n$ vertex with successive incoming momenta
$k_{\m}$ has the phase factor
\begin{equation}
\label{eq:cyclicvertex}
 \exp -{i\over 2} \sum_{1\le{\m}<{\n}<n} k_{\m}\times k_{\n}
\end{equation}
\def\refcyclicvertex{Eq. (\ref{eq:cyclicvertex})}
where (as in \refmoyalproduct) $k\times k'\equiv \theta^{ij} k_i
k'_j$.  The same holds in the presence of derivative interactions,
multiple fields and so on.

\begin{figure}
\epsfxsize=3in
\epsfbox{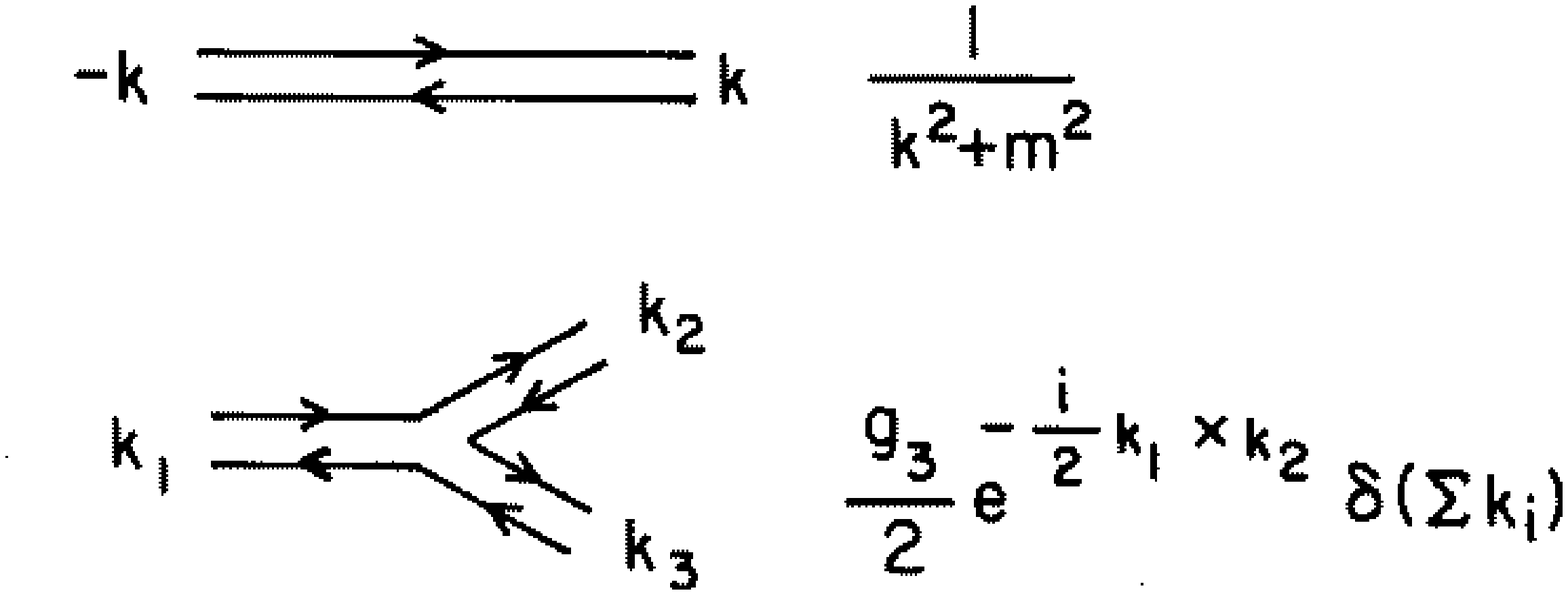}
\caption{Double line notation and phase factors.}
\end{figure}

The factor \refcyclicvertex\ is not permutation symmetric but is
only cyclically symmetric.  This is just as for a matrix field
theory, and the same double line notation can be used to represent the
choice of ordering used in a specific diagram
\cite{'tHooft:1974jz,Coleman:1979a}.  In this notation, propagators are
represented as double lines, while the ordering of fields in a vertex
is represented by connecting pairs of lines from successive
propagators.

This additional structure allows defining ``faces'' in a diagram as
closed single lines and thus the Euler character of a graph can be
defined as $\chi=V-E+F$, with $V$, $E$ and $F$ the numbers of
vertices, edges and faces respectively.
For connected diagrams, $\chi\le 2$ with the maximum attained for
planar diagrams, those which can be drawn in a plane without crossing
lines.  A diagram with $\chi=2-2g$ for $g\ge 1$ is nonplanar of genus
$g$ and can be drawn on a surface of genus $g$ without crossings.

In matrix field theory, summing over indices provides a factor
$N^F$, while including appropriate $N$ dependence in the action (an
overall prefactor of $N$ will do it) completes this to a factor
$N^\chi$.  This is the basis for the famous 't Hooft limit in which
the dynamics of certain field theories is believed to reduce to that
of a free string theory: as $N\rightarrow \infty$, planar diagrams
dominate.

In \nc\ field theory, planarity plays an important role in organizing
the phase factors \refcyclicvertex.  The basic result is that a planar
diagram in \nc\ theory has the same amplitude as that of \con\ theory
multiplied by an overall phase factor \refcyclicvertex, where the
momenta $k_i$ are the ordered external momenta.  This can be seen by
checking that the diagrams in figures 3 and 4 can each be replaced by a
single vertex while preserving the phase factor, and that these
operations can be used to reduce any planar diagram with $n$
external legs to the single vertex $\Tr \phi^n$.

\begin{figure}
\epsfxsize=3in
\epsfbox{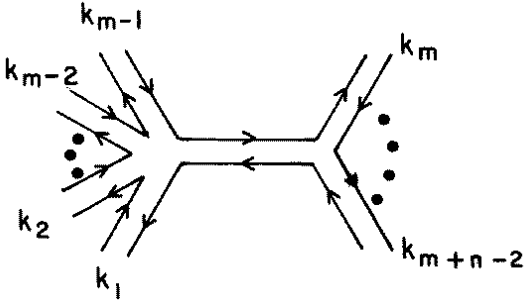}
\caption{The contraction
$(\Tr \phi^m)(\Tr \phi^n) \leftrightarrow \Tr \phi^{m+n-2}$.}
\end{figure}

\begin{figure}
\epsfxsize=3in
\epsfbox{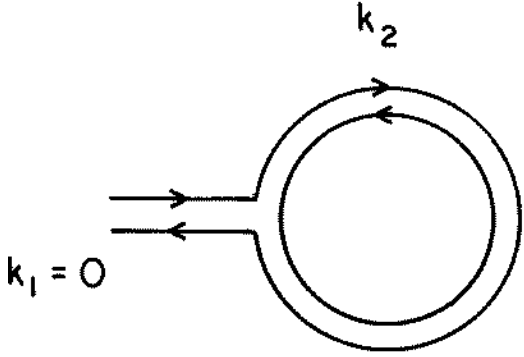}
\caption{Tadpoles come with no phase factor.}
\end{figure}

Since this additional phase factor is completely independent of the
internal structure of the diagram, the contribution of a planar diagram to
the effective action will be {\it the same} in \nc\ theory and in the
corresponding $\theta=0$ theory
(the \nc\ phase factor will be reproduced in the course of evaluating
the vertex in the effective action).
In particular, divergences and renormalization will be the same for
the planar subsector as in a \con\ theory.

We can obtain the phase factor for a nonplanar diagram by
comparison with the planar case.  Given a specific diagram, we
choose a way to draw it in the plane, now with crossing
propagators. Let $C^{\m\n}$ be the intersection matrix,
whose $\m\n$ matrix element counts the number of times the $\m$'th
(internal or external) line crosses the $\n$'th line (with sign; a
crossing is positive as in the figure).  Comparing this diagram
with the corresponding planar diagram obtained by replacing each
crossing with a vertex, we find that it carries the additional
phase factor
\begin{equation}
\label{eq:nonplanarphase} \exp {-{i\over 2}\sum_{{\m},{\n}}
C^{\m\n} k_{\m} \times k_{\n}}.
\end{equation}
Although the matrix $C^{{\m\n}}$ is not uniquely determined by the
diagram, the result \rfn{nonplanarphase} is.

\begin{figure}
\epsfxsize=3in
\epsfbox{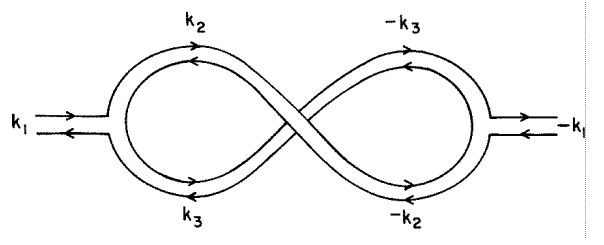}
\caption{A non-planar diagram.}
\end{figure}

Since the phase factor \rfn{nonplanarphase} depends on the
internal structure of the diagram, nonplanar diagrams can have very
different behavior from their $\theta=0$ counterparts.
Since the additional factor is a phase, one would expect it to improve
convergence of integrals, leading to better UV behavior.
This expectation will be borne out below, leading to the principle
that the leading UV divergences (in particular the beta function)
come from the planar diagrams (no matter what the rank)
and are thus the same as for the large $N$ limit of a matrix field theory.

In practice, this principle might be obscured by the presence of other
divergences.  A more precise statement one can make is that in the
limit $\theta^{ij}\rightarrow \infty$, with fixed external momenta, UV
and IR cutoffs, integrating over this phase factor would cause the
general nonplanar diagram to vanish.  Thus this limit would keep only
planar diagrams.  This observation goes back to the early works on the
twisted Eguchi-Kawai models, as mentioned in the introduction, and
indeed was the main focus of interest in these works.

\subsection{Calculation of nonplanar diagrams}

The most important features are already visible in the one-loop
renormalization of the scalar field theory propagator.  We
consider two examples, the $\phi^4$ theory and the $\phi^3$ theory
in $d$ Euclidean \nc\ dimensions.  We will not be careful about
$\CO(1)$ numerical factors in our discussion and often substitute
generic positive real constants $c$, $c'$, etc., to better make
the qualitative points.  We will try to keep track of signs.

The $\phi^4$ theory has two one-loop self-energy diagrams, 
one planar and one nonplanar.  They contribute
\begin{eqnarray}
\Gamma^{(2)}_{1 \; planar} &={g_4 \over 3(2\pi)^d} \int {{\rm d}^d
k \over k^2+m^2} \\ \Gamma^{(2)}_{1 \; nonplanar} &={g_4 \over
6(2\pi)^d} \int {{\rm d}^d k \over k^2+m^2} e^{ik \times p} .
\end{eqnarray}
The planar contribution is proportional to the one-loop diagram of
the $\theta=0$ theory, and is divergent for $d\ge 2$.  If we
introduce a momentum space cutoff $\Lambda$, we will find $$
\Gamma^{(2)}_{1 \; planar} =
 {g_4\over 3(4\pi^2)^{d\over 2}}\left(\Lambda^{d-2} + \cdots\right),
$$
a mass renormalization which can be treated by standard
renormalization theory.

Compared to this, the nonplanar contribution has an oscillatory
factor $e^{ik \times p}$, arising from the \nc\ nature of space-time,
which will play the role of a cutoff.  On general grounds,
this cutoff will come in for momenta $k$ such that
$d/dk(k\times p) \sim \CO(1)$, i.e.
$k^i \sim \Lambda_p \equiv |\theta^{ij} p_j|^{-1}$, and produce
$$
\Gamma^{(2)}_{1 \; nonplanar}
 \sim {g_4} (c\Lambda_p^{d-2} + \cdots).
$$
This cutoff goes
to infinity as we take the limit $p\rightarrow 0$ of the external
momentum.  To get a finite result in this limit, we will need to keep
the original cutoff $\Lambda$ in the calculation as well.
This can be done in many ways; the usual approach is a proper
time cutoff $e^{-1/\Lambda^2 t}$ in the integrand of
\rfn{greenproper} which leads to the result
\begin{eqnarray}
\Lambda_{p}^2 \equiv
 {1 \over {1/\Lambda^2 + p \circ p}} ; \\
\label{eq:defcirc}
p \circ p \equiv p_i \theta^{ik} \theta^{jk} p_j
\end{eqnarray}
(in fact this is the simplest function with the
qualitative behavior we want).
The notation $p \circ p$ for the quantity controlling the cutoff is
standard; it is positive definite if $\theta^{0i}=0$
(purely spatial noncommutativity).  If $\theta^{ij}$ has maximal rank
and all its eigenvalues are $\pm\theta$, then $p\circ p=\theta^2 p^2$.
If there are commutative directions as well, they will not enter into
this quantity.

Adding the classical and one-loop contributions to the two-point
function leads to a 1PI effective action \bea
\label{eq:selfenergy} & \Gamma[\phi] =\int {\rm d}^d p\
\half\phi(p) \phi(-p) \Gamma^{(2)}(p) + \ldots\\
\label{eq:phifouronePI} & \Gamma^{(2)}(p) = p^2 + M^2 + { c g_4
\over (p \circ p+ {1/\Lambda^2})^{d-2}}
 + \cdots
\eea where $M^2 = m^2 + {c' g_4 \Lambda^{d-2}}$ is the
renormalized mass.  At $d=2$ the powerlike divergences become
logarithmic in the usual way.

The novel feature of this result is that the limit
$\Lambda\rightarrow \infty$ is finite, and the UV divergence of the
nonplanar diagram has been eliminated, but only if we stay away from
the IR regime $p\rightarrow 0$.  The limiting theory has a new IR
divergence, arising from the UV region of the momentum integration.
This type of phenomenon is referred to as ``UV/IR mixing'' and is very
typical of string theory, but is not possible in local field theory.
However, it is possible in \nc\ theory, thanks to the nonlocality.

One way to see that this is not possible in local field theory is
to observe that it contradicts the standard dogma of the
renormalization group.  Let us phrase this in terms
of a Wilsonian effective action, defined by integrating out all
modes at momentum scales above a cutoff $\Lambda$.  The result is
an effective action \bea & S_{Wilson} [ {\phi};  \Lambda ] = \crt
& \qquad \int {\rm d}^4 x \, {Z(\Lambda) \over 2} \left(
(\p\phi)^2 + m^2(\Lambda) \phi^2 \right) +
{g^2(\Lambda)Z^2(\Lambda) \over 4!} \phi^4. \eea 
Renormalizability
in this framework means that one can choose the functions
$Z(\Lambda)$, $m(\Lambda)$ and $g^2(\Lambda)$ in such a way that
correlation functions computed with this Lagrangian have a
limit as $\Lambda \rightarrow \infty $, and
correlation functions computed at finite $\Lambda$ differ from
their limiting values by terms of order ${1\over \Lambda}$ for all values
of momenta.

However, one sees from \rfn{phifouronePI} that this is manifestly
untrue of \nc\ scalar $\phi^4$ theory, and indeed the
generic \nc\ field theory with UV divergences.
The two point function at any finite value of $\Lambda$ differs
drastically from its $\Lambda\rightarrow\infty$ value, for small
enough momenta ($(p\theta)^2  << 1/\Lambda^2$).

The arguments above
lead to the general expectation that a nonplanar diagram with
UV cutoff dependence $f(\Lambda^2)$, will obtain an IR divergence
$f(1/p\circ p)$ in the \nc\ theory.  
One might further expect that,
if the \con\ amplitude had no IR divergences, the leading
IR divergence of the \nc\ theory would take precisely this form.
We will refer to this as ``the standard UV/IR relation.''
It is typical but not universal and in particular is violated
in gauge theory.

Another example is the one-loop self-energy diagram in $\phi^3$ theory
(figure 5).
The expectations we just stated are indeed valid
\cite{VanRaamsdonk:2000rr} and we find
\begin{equation}
\label{eq:phithreeonePI}
\Gamma^{(2)}(p) =
 p^2 + M^2 + {c g_3^2 M^2 \over
    (p\circ p + 1/\Lambda^2)^{d-4}} + \ldots
\end{equation}
where $M^2$ is again the renormalized mass.  At $d=4$ this becomes
a logarithmic divergence.

The effects we have seen so far drastically change the IR behavior and
we refer to this case as ``strong UV/IR mixing.''  There are also
theories, particularly supersymmetric theories, in which the IR
behavior is not qualitatively different from \con\ theory, and we can
speak of ``weak UV/IR mixing.''  This would include models where
renormalization is logarithmic and only affects the kinetic terms
(assuming the standard UV/IR relation), such as the $d=4$ Wess-Zumino
model \cite{Girotti:2000gc}.

In proceeding to higher loop orders, one faces the danger that the new
IR divergences will mix with other divergences in an uncontrollable way.
In fact the IR divergences
under discussion are rather similar to those appearing in thermal
field theory, in that they give a large effective mass to low
momentum modes (we will discuss this point further below),
and can be addressed by similar techniques
\cite{Fischler:2000fv,Gubser:2000cd,Griguolo:2001wg}.

One easy way to do this is to add and subtract a counterterm to
the bare action which represents the leading divergence, so that
it can be taken into account in the bare propagator. This leads to
the next correction to \rfn{phifouronePI} being 
$$
\Gamma^{(2,1)}(p) = c' g_4 \int {{{\rm d}^{d} k \ \left(  1 +
e^{ik\times p} \right)} \over k^2 + M^2 + c g_4 (p \circ p)^{2-d}}
- { c g_4 \over (p \circ p)^{d-2}} . 
$$ 
with the IR divergence explicitly subtracted.

The main feature of this result is that the IR divergent term in
the propagator (for $d\ge 2$) causes it to vanish at low momenta,
and thus the resummed loop corrections are controlled in this
region.

In fact, at weak coupling
these corrections to $\Gamma^{(2)}(p)$ are small for all momenta.
They do exhibit nonanalyticity in the coupling similar to that
in finite temperature field theory \cite{Griguolo:2001wg},
beginning at $\CO(g_4^3 \log g_4)$ in the massive theory and
$\CO(g_4^{3/2})$ in the massless theory.  This brings new difficulties
into the perturbation theory; one could try to work
at finite coupling by using a self-consistent Hartree equation,
or any of the many other approximation methods available for field theory.

In any case, this discussion appears to justify considering Eqs.
(\ref{eq:phifouronePI}) and (\ref{eq:phithreeonePI}) as valid
approximations to the Greens function at weak coupling, and we
discuss physics following from this idea in the next subsection.
It falls short of a proof, which would require discussing (at the
least) all Greens functions which obtain large UV renormalizations
in \con\ theory, such as the four-point function in $\phi^4$
theory in $d\le 4$.  The argument we just gave generalizes to some
extent to these diagrams as their IR corrections will be
controlled by the propagator in the same way.  This might fail for
special values of external momenta however, and a real proof of
these ideas has not yet appeared.  Relevant work includes
\cite{Chepelev:1999tt}.  

See also \cite{Chen:2001qg,Li:2000ig,Kinar:2001yk}.

\subsection{Physics of UV/IR mixing}

The UV/IR mixing observed by \citeasnoun{Minwalla:1999px} appears to
be the main qualitative difference between \con\ and \nc\ perturbation
theory.  One cannot say that its full significance, and the issue of
whether or not it spoils renormalizability or leads to other
inconsistencies, is well understood at present.  On the other hand,
there is an emerging picture which we will outline.

A reason for caution at this point is that,
for the reasons we just discussed, we cannot blindly assume that the
general framework of the renormalization group, which underlies most
of our understanding in the \con\ case, is applicable.
This is not to say that it is obviously inapplicable, but that to
justify it one must show how to clearly separate UV and IR phenomena,
taking into account the higher loop subtleties mentioned above.

If this can be done, justifying the idea that the high energy behavior
is controlled by the planar diagrams, then the question of whether a
\nc\ theory is renormalizable will have the same answer as for the
corresponding \con\ theory in the planar limit (normally this limit
has the same UV behavior as the finite $N$ theory).  
It has recently been argued by \citeasnoun{Griguolo:2001ez} that this
can be done in the Wilsonian RG approach to proving renormalizability
\cite{Polchinski:1984gv}.
There are also arguments against this in certain cases 
\cite{Akhmedov:2000uz}, which we return to below.

We assume for sake of discussion that this is 
indeed true and now discuss the physics of strong UV/IR mixing.
Before we start, we note that the supersymmetric 
theories of primary interest to string theorists 
have weak or no UV/IR mixing, and the considerations we are about to
discuss, except for those regarding unitarity,
have not played a role in this context so far.
On the other hand they are likely
to be important in other applications.

Once one observes new IR effects at the quantum level, the first
question one must ask is whether the original perturbative vacuum
$\phi=0$ (the disordered phase, in the language
of statistical field theory)
is stable, or whether these effects are a signal that we are
expanding around the wrong vacuum.  This question could be answered
if we knew the exact quantum effective action; in particular the
perturbative vacuum will be at least metastable if the
inverse propagator $\Gamma^{(2)}(p)$ in
Eq. (\ref{eq:selfenergy}) is positive at
all momenta.  This includes the usual condition on the effective
potential $V''(0) > 0$ but since the effects we are discussing modify
the dispersion relation so drastically, we need to entertain the
possibility that a phase transition could be driven by modes with
non-zero momentum, perhaps leading to a ``stripe'' phase as one finds
in certain condensed matter systems.

This is a real possibility here, as is clear from
Eqs. (\ref{eq:phifouronePI}) and (\ref{eq:phithreeonePI}).
One expects and it might be possible to prove that with certain
hypotheses, the standard UV/IR relation will hold for the exact
(cutoff) quantum effective action.  This would mean that
$\Gamma^{(2)}(p) \sim \Delta M^2(1/p\theta)$
in theories whose planar limit has a UV divergent
mass renormalization $\Delta M^2(\Lambda)$.

Then the relevant question is whether this mass renormalization is
positive or negative (for large $\Lambda$). If it is negative, the
perturbative vacuum is clearly unstable to condensing $p\rightarrow 0$
modes. On the other hand, if it is positive, the resulting dispersion
relation will make the low $p$ modes ``stiff'' and there might or
might not be a phase transition, but if there is, it will be driven by
a mode with $p\ne 0$.  \citeasnoun{Gubser:2000cd} have argued that
such a phase transition is indeed expected (in scalar field theory in
$d>2$) and will be first order.

If we grant that the perturbative vacuum is stable, we can go on to
ask about the meaning of the propagator $\Gamma^{(2)}(p)^{-1}$.
This discussion will depend on whether the noncommutativity
is purely spatial ($\theta_{0i}=0$) or has a time-like component.

It is not hard to see that timelike noncommutativity combined with
UV/IR mixing leads to
violations of perturbative unitarity.  This shows up in unphysical
branch cut singularities in loop diagrams \cite{Gomis:2000zz} and
can also be seen from the behavior of the propagator and commutators
of fields \cite{Alvarez-Gaume:2001ka}; see also \cite{Seiberg:2000gc}.
We now give a simple argument along these lines.

A theory with timelike noncommutativity will be invariant under
boosts along the conjugate spatial momentum (e.g. $p_i \propto
\theta_{01}$) and therefore, if we set the other momenta to zero,
the propagator will be a function of the $1+1$ Lorentz invariant
$p^2=-E^2+(p_i)^2$. Thus we can write a spectral representation $$
{1 \over \Gamma^{(2)}(p)} = \int_0^\infty {{\rm d}m^2 \over p^2 +
m^2} \rho(m^2) $$ with $\rho(m^2)\ge 0$ in a unitary theory.
Unitarity requires
\begin{equation}
\label{eq:unitarity} \lim_{p^2\rightarrow 0} {1 \over
\Gamma^{(2)}(p)} = \int_0^\infty {{\rm d}m^2 \over m^2} \rho(m^2)
> 0
\end{equation}
assuming this limit exists (it might also diverge).
On the other hand, the general behavior produced by UV/IR mixing is
$$
\lim_{p^2\rightarrow 0} \Gamma^{(2)}(p) = +\infty ,
$$
which is incompatible with Eq. (\ref{eq:unitarity}).

We turn to consider purely spatial noncommutativity.  This will lead
to a dispersion relation of the form
$$
E^2 = \vec p^2 + m^2 + \Delta M^2({1\over p\theta}) .
$$
If $\Delta M^2(\Lambda)$ grows at infinity, we again find low momentum
modes are stiff.

We first discuss the Euclidean theory.
The Greens function $G^{(2)}(p)=\Gamma^{(2)}(p)^{-1}$
can be transformed to position space following our discussion in 
\rfs{bases}.4.
As discussed there, the long distance behavior is controlled by the
pole in the upper half plane nearest the real axis, as in \con\ field
theory.
Let us consider $\Delta M^2={1\over 4}g^2 \Lambda^2$ for definiteness, then
the poles in $G^{(2)}(p)$ can be found by solving a quadratic equation,
leading to the following two limiting behaviors.
If $g << \theta m^2$ (weak coupling or strong noncommutativity),
the closer root will be at $p=i g/4\theta m$.  This can be interpreted
as a new mode with mass $m_2 = g/4\theta m$, and the precondition
is equivalent to $m_2 << m$, i.e. this is the limit in which the new
mode dominates the long distance behavior.
In the other limit, there are two poles at equal distance
$(g/8\theta)^{1/2}$, leading to oscillatory-exponential behavior.

Thus, in either case, the IR effects appear to have a sensible
description at finite $g$, in terms of a new mode.
One can go on to ask whether the new mode can be described by adding
an additional field to the effective Lagrangian.
This question is best discussed in the context of the connection to
string theory and we postpone it for \rfs{strq}.

We now discuss the quantum \nc\ field theory in Minkowski space.
The primary question is whether it is unitary.
Formally, the main thing to check is that the
$(\p_0\phi)^2$ term in the effective action
is positive for all spatial momenta (we also assume that
higher time derivative terms are not significant).  This is necessary
for perturbative unitarity of the effective action but almost certainly
will be true if the analog \con\ theory was renormalizable and unitary.
The cutting rules can also be checked and appear compatible with
unitarity in this case \cite{Gomis:2000zz}.  Since these theories
admit a (cutoff) Hamiltonian formulation, any problem with unitarity
would almost certainly be tied to instability as well.

However, the position space Greens function in this case is quite
bizarre at low momenta.  Its main features can be understood by
considering wave propagation in a medium with an index of refraction
$n(\omega)=k(\omega)/\omega$ which grows as $\omega\rightarrow 0$
faster than $1/\omega$.\footnote{
Strictly speaking, we are discussing spatial dispersion
\cite{Landau:1984a}, but the difference is not relevant for us.}
In our example, $\omega \sim g/\theta k$.
This leads to a negative group velocity $v_g=\p\omega(k)/\p k$.
One can still proceed formally to derive a position space Greens function,
which exhibits superliminal propagation.  
In a non Lorentz invariant
theory, this might not be considered a major surprise; however
its short time propagation is dominated by long wavelengths and
it does not satisfy the usual defining property
$\lim_{t\rightarrow 0} G(x,t) = \delta(x)$.
This certainly looks unphysical and at present further 
interpretation would seem to be a purely academic exercise.
On the other hand, perhaps the usual position space interpretation is
inappropriate and there is something deep here yet to be discovered.

Finally, we return to our original assumption of the validity of the
Wilsonian RG, and discuss the work of
\cite{Akhmedov:2000uz,Akhmedov:2001fd,Girotti:2001gs}
on the \nc\ Gross-Neveu model.
This is a model of $N$ flavors of interacting massless fermions whose
perturbative vacuum is unstable.  In the $1+1$ \con\ theory a
nonperturbative condensate forms, as can be shown exactly in the large
$N$ limit by solving a gap equation, leading to a vacuum with massive
fermionic excitations.  This theory exhibits dimensional
transmutation; the bare coupling $g$ can be eliminated in favor of the
mass gap $M \sim \Lambda\exp -1/g^2$.

Following the procedure we just outlined for the \nc\ theory leads to
$\Gamma^{(2)}(p) < 0$ at low $p$ (for the condensate) and instability.
However, Akhmedov \etal\ argue that instead of following the standard
Wilsonian approach and defining the continuum theory as a flow out of
a UV fixed point, one should enforce the gap equation.  This leads to
a stable vacuum but spoils dimensional transmutation and results in a
trivial continuum limit.  Akhmedov \etal\ also point out the
interesting possibility of a nontrivial ``double scaling limit'' with
fixed $\Lambda M \theta$.

One cannot {\it a priori} say that one of these definitions is
``right;'' this depends on the underlying microscopic physics and the
application.  What we would insist on is that one can choose to define
the theory as a flow from a UV fixed point and that in this sense the
theory is renormalizable; one however does not know (at present)
whether it has a stable vacuum.

Furthermore, we feel that even the alternate definition can be fit
within the conventional RG picture, allowing for IR physics to
determine a condensate which will then react back on the UV physics,
determining the couplings through dynamics at various scales.  In
itself, this phenomenon is not new but has fairly direct analogs in
\con\ field theory, particularly supersymmetric field theory where IR
physics can lift a flat direction, leading to large scalar vevs and
large masses for other degrees of freedom.

However, one has opened Pandora's box as the larger point seems to be
that strong UV/IR mixing can lead to very different physics for
condensates, which is not understood.  For example, if the \nc\
Gross-Neveu model defined as a UV fixed point has a stable vacuum, it
might not be translationally invariant and perhaps not even be
describable by a classical condensate.

The tentative conclusion we will draw from all of this is that quantum
field theories with spatial noncommutativity and strong UV/IR mixing
can be consistent, often with rather different physics from their
\con\ analogs.

\subsection{Gauge theories}

The usual discussion of perturbative gauge fixing and the Fadeev-Popov
procedure go through essentially unchanged.  The Feynman rules for
\nc\ gauge theory are the standard ones for Yang-Mills theory, with
the Lie algebra structure constants augmented by the phase factors
\refcyclicvertex.

Gauge theories can have IR divergences which are stronger than the
standard UV/IR relation would suggest \cite{Hayakawa:1999zf,Matusis:2000jf}.
Consider for example the one-loop contribution to the NCQED photon
self-energy due to a massless fermion.  Standard considerations
lead to the amplitude
$$ i \Pi_{ij}(p) = -{4g^2\over(2\pi)^4} \int {\rm d}^4 k
 \ {(2k_i k_j-\eta_{ij} k^2) \over k^4} \left( 1 -  e^{i k\times p} \right).
$$
The nonplanar part of this is
\begin{eqnarray}
i \Pi^{(np)}_{ij}(p) &= {4 i g^2\over(2\pi)^4}
 (\p_i\p_j - \eta_{ij} \p^2) \log \left(p\circ p+1/\Lambda^2\right) \\
 &= -{16 i g^2\over(2\pi)^4}
    {\theta_{ik} p^k \theta_{jl} p^l \over (p\circ p+1/\Lambda^2)^2} .
\end{eqnarray}
If we remove the cutoff, this diverges as $1/\theta^2 p^2$ as $p\rightarrow 0$.

Physically, this term would lead to different dispersion relations for
the two photon polarizations.  Peculiar as this is, it is consistent
with gauge invariance in this non-Lorentz invariant theory.
A similar one-loop contribution can be found to the three-point function.
It behaves as $(\theta p)_i (\theta p)_j (\theta p)_k/(p\circ p)^2$
at low momenta and diverges as $1/\theta p$ as $p\rightarrow 0$.
It is not clear at present what the 
significance of these effects might be.  

In any case, it has been observed that these effects vanish in
supersymmetric theories.  The logarithmic effects expected from our
previous discussion will still be present.  A detailed discussion at
one loop is given by \citeasnoun{Khoze:2000sy};
see also \cite{Zanon:2000nq,Ruiz:2000hu}.

All of the UV and IR divergences found in any computation done to date
are gauge invariant, and it is generally believed that these theories
are renormalizable.  It has also been argued that spontaneously broken
\nc\ gauge theories are renormalizable \cite{Petriello:2001mp}.\footnote{
There is some controversy about this and about the renormalizability
of the related
nonlinear sigma model, possibly related to issues discussed at the
end of \rfs{quant}.C;
see \cite{Campbell:2001ek,Sarkar:2001fn,Girotti:2001gs}.}
Unitarity has recently been discussed in \cite{Bassetto:2001vf}.

Thus we turn our attention to
the structure of the \nc\ effective action.
It turns out to be somewhat subtle to write this
in a gauge invariant form \cite{Liu:2000ad,Zanon:2000gy} (this was
also seen in related string computations by \citeasnoun{Garousi:1999ch}).
One cannot use the
usual rule for Yang-Mills theory, according to which one can infer
higher-point functions by completing an operator
$\p_i A_j-\p_j A_i$ to the gauge-invariant $F_{ij}$.
The basic problem is
apparent from the analogy with \con\ large $N$ field theory.  The
rules of \ref{sec:quant}.A tell us that nonplanar diagrams will
produce contributions of the form
$$
\int \prod_i d^dk_i\ \Tr O_1(k_1) \ldots \Tr O_n(k_n)
G(k_1,\cdots,k_n) .
$$
Such terms do not follow the rule from \rfs{ft}
of associating a single trace with each integral and thus we cannot
apply the arguments made there.

Clearly this problem is related to the problems with
defining local gauge invariant operators discussed in \rfs{gt}.2, 
and it is believed that the effective action will be gauge invariant
if expressed in terms of the the open Wilson loop operator
\cite{Liu:2000mj,Mehen:2000vs}.
This has been checked in examples \cite{Pernici:2000va}
and one can also make strong arguments for it from string theory
(see \rfs{strq}).

The basic gauge invariant local Greens function then is the two-point
function of two open Wilson loops, $\vev{ W(k) W(-k) }$.
A particularly interesting question is the high energy behavior
of this quantity, which was found by \citeasnoun{Gross:2000ba}.
Our general arguments that high energy behavior reduced to planar
diagrams applied to naive local observables, but do not apply to
the open Wilson line, because its extent grows with energy.

At high energy, the open Wilson loops will become very long and
the computation becomes identical to that
for the expectation value of a single rectangular
Wilson loop, a well known result in the \con\ theory.
Since the leading large momentum behavior
is given by the sum of planar ladder diagrams, the result
can be applied directly.
For a rectangular Wilson loop of sides $L$ and $r$ (with $L>>r$), it is
$\exp -(E+M)L$ with the Coulomb interaction energy $E= -g^2 N/4\pi r$
and $M$ a nonuniversal mass renormalization.
This result
can then be Fourier transformed, leading to a growing exponential
in momentum space, $\exp +\sqrt{{g^2 N|k|L\over 4\pi}}$.

The main adaptation required to compute the open Wilson loop Greens
function is just to use Eq. (\ref{eq:dipolelength}) and take
$L=|\theta k|$.  This produces
\begin{equation}
\label{eq:wilsontwo}
\vev{ W(k) W(-k) } \sim
 \exp +\sqrt{{g^2 N|k\theta||k|\over 4\pi}}
\end{equation}
This exponential growth with momentum is universal and
applies to correlators
of Wilson loops with operator insertions $W[O](k)$ as well.

This result cannot be properly interpreted without knowing something about
the higher point correlators, as the overall normalization of the operator
is not a matter of great interest.  Thus we consider a
ratio of correlation functions in which this cancels,
$$
\vev{ W(k_1) W(k_2) \ldots W(k_n) } \over
\sqrt{ \vev{ W(k_1) W(-k_1) } \ldots \vev{ W(k_n) W(-k_n) } } .
$$
\citeasnoun{Gross:2000ba} argue that generic
higher point functions do not share the exponential growth,
essentially because more than two Wilson lines will not generally
be parallel and thus the dependence on $L$ will not exponentiate.

Thus, Eq. (\ref{eq:wilsontwo}) should be interpreted as leading to a
universal exponential falloff of correlation functions at large
momentum.  This is another strong analogy between \nc\ gauge theory
and string theory, as it is a very characteristic feature of string
theory (however, see the caution in \citeasnoun{Gross:2000ba} about
this analogy).

\citeasnoun{Rozali:2000np} have further argued that in a multipoint
function in which a pair of momenta become antiparallel (within a
critical angle $\phi < \phi_0 \sim 1/|k_i||\theta k_j|$), the
exponential growth is restored, which would again be consistent with
field theory behavior.  See also \cite{Dhar:2000ht}.

The main motivation for this computation was to check it against
the AdS/CFT dual supergravity theory, and we discuss this aspect
in \rfs{ads}.

Loop equations governing these correlators are discussed in 
\cite{Abou-Zeid:2000qi,Dhar:2000nj}.

\subsection{Finite temperature}

Another limit which will probe noncommutativity is the high
temperature limit $\theta T^2 >> 1$.  We now discuss this regime,
following \cite{Arcioni:1999hw,Fischler:2000fv,Fischler:2000bp}, 
and using standard
techniques of finite temperature field theory \cite{Kapusta:1990a}.
The temperature is $T$ and $\beta=1/T$, and we take
$\theta_{12}=\theta$ and $x^3$ and time commutative.

The leading nonplanar diagram contributing to the finite
temperature free energy of $\phi^4$ theory is the two loop diagram
obtained by contracting the external legs in figure 5.
This contributes (for $d=4$) 
$$ -g^2 T^2
\sum_{n,l=-\infty}^\infty
 \int {{\rm d}^3 p\over (2\pi)^3} {{\rm d}^3k\over (2\pi)^3} \, 
e^{i p \times k} 
$$ 
$$
 {1\over
  ({4\pi^2 \over\beta^2}n^2 + p^2 + m^2)
  ({4\pi^2 \over\beta^2}l^2 + k^2 + m^2)} .
$$
If we neglect the mass (appropriate if $m << T$),
this can be reduced using standard techniques to
$$
- g^2 T  \int {d^3p\over (2\pi)^3}
 {1 + 2 n_\beta(|p|) \over 2|p|}
 {1 + 2 n_{1/\beta}(2\pi|\theta p|) \over 4\pi|\theta p|}
$$
where $n_\beta(|p|) = (e^{\beta|p|} - 1)^{-1}$ is a Bose distribution
at temperature $\beta=1/T$.

This expression shows intriguing similarities to string theory, and
\citeasnoun{Fischler:2000fv} interpret the second thermal distribution
as describing ``winding states;'' see also
\citeasnoun{Arcioni:2000bz}.  It also leads to an IR divergence which
must be addressed, either along the lines of the resummation discussed
above, or by considering supersymmetric theories with better
convergence properties.

In either case, one finds the following fascinating behavior
for the the nonplanar contribution to the free energy.
While the regime $\theta T^2 << 1$ is essentially as for \con\ field
theory, results for the regime $\theta T^2 >> 1$ are very much as if
the theory had many fewer degrees of freedom in the UV than \con\ field
theory.  For $\phi^4$ theory, supersymmetric $\phi^4$ theory
(the Wess-Zumino model) and $N=4$ NCSYM (all in $D=4$), one finds
(at two loops)
\begin{equation}
\label{eq:phifour}
{F_{nc}\over V} \sim -{g^2\over\theta^2} T^2\theta \log T^2\theta .
\end{equation}
In $\phi^3$ theory in $D=6$, one finds
\begin{equation}
\label{eq:phisix}
{F_{nc}\over V} \sim -{g^2\over\theta^2} T^2\theta .
\end{equation}
This latter result is even more amusing as it can be derived from
{\it classical} statistical mechanics.  This is essentially the
same as quantum field theory in one lower dimension and leads to
an integral $$ F \sim T^2 \int d^5 k d^5 p {e^{ik\times p}\over
k^2 p^2 (k+p)^2} . $$ While normally such integrals are badly UV
divergent (the famous ``ultraviolet catastrophe'' of classical
physics), this one actually converges and reproduces
\rfn{phisix}.  This would seem a very concrete demonstration
of the idea that the number of UV degrees of freedom has been
drastically reduced, presumably to a finite number per
$\theta^{d/2}$ volume or ``Moyal cell'' in \nc\ space.

However, the more general situation is that the corresponding
classical integrals are still UV divergent.  In particular, this is
true in the cases summarized in \rfn{phifour}.  It must also be
remembered that the planar contributions are still present and
dominate the contributions we are discussing.  Thus the full import
of these rather striking results is not clear.

See also \citeasnoun{Landsteiner:2001ky}, who argue for a finite
temperature phase transition in \nc\ gauge theory.

\subsection{Canonical formulation}

In general terms, canonical quantization of \nc\ field theory and
gauge theory with purely spatial noncommutativity can be done by
following standard procedures, which we assume are familiar.  This
leads to commutation relations between free fields, say $\phi$ and a
conjugate momentum $\pi$, which in momentum space take the same form
as \con\ theory.  One furthermore has standard expressions for energy
and momentum operators associated to commutative dimensions.
Possible quantum corrections to their
Ward identities have not been much studied.

One can define momentum operators
$P_i$ for the \nc\ dimensions by using the restricted energy-momentum
tensor $T^0 = i[\phi,\pi]$, as in \rfn{momentum}.
In \nc\ gauge theory, this makes sense for the reasons discussed in
\rfs{gstress}.

A point where interesting differences from the \con\ case appear
is in discussing the action of gauge transformations and charge
quantization.  Consider $2+1$ Maxwell theory on a square torus
$T^2$ defined by the identifications $x^i \cong x^i + 2\pi$. The
total electric flux has two components $\int {\rm d}^2 x\ E^{i}$,
each conjugate to a Wilson loop $W_i = \exp i\int dx^i A_i$:
$$[E^i, W_j] = \delta^i_{j} W_i. $$ This theory admits large gauge
transformations, acting on a charge $1$ matter field $\psi$ as
$\psi \rightarrow e^{in_i x^i} \psi$, with $n_i\in\IZ$ to make the
transformation single valued. Thus the zero mode of $A_i$ is a
periodic variable, and eigenvalues of $E^i$ must be integrally
quantized.

A similar argument can be made in the rank 1 \nc\ theory on
$\IT_\theta^d$, but now the corresponding gauge transformations
$\psi \rightarrow U_i \psi U_i^\dagger$ (notation as in II.B.2)
act on the nonzero modes of $\psi$ as well as in
\rfn{dipoleshift}, taking $\psi(x^j) \to \psi(x^j-\theta^{ji})$.
Thus the wave function must be invariant under a simultaneous
shift of the zero mode of $A_i$ and an overall translation
\cite{Hofman:1998iv},
leading instead to a quantization law for the quantity 
\be \label{eq:qelec}
\int \Tr\ E^i - \theta^{ij} P_{j} . 
\ee
One also finds modified quantization laws for magnetic charges,
which we will discuss in \rfs{math}.D.

Taking into account these considerations, one can find the quantum
Hamiltonian and its perturbative spectrum along standard lines.
We quote as an example the ground state energy in $2+1$ \nc\ Yang-Mills
on $\Mat_p(\IT^2_\theta)$
at leading order in perturbation theory (this can be shown to be the
exact result given enough supersymmetry).  Sectors of this theory are
labeled by conserved integral quantum numbers $n^i$ (related to
electric flux as in \rfn{qelec}), $q$
(magnetic flux), and $m^i$ (momentum).  Finally, we write 
$N=\Tr 1=p-\theta q$; as we explain in \rfs{math}, the
naive expression $\Tr 1=p$ is not valid on $\IT_\theta$.

One then has \cite{Connes:1998cr,Brace:1998ai,Hofman:1998iv,%
Konechny:1998wv}
\bea \label{eq:torusenergy}
E = {g_{YM}^2\over 2\sqrt{g}N}
g_{ij}(n^i + \theta^{ik} m_k)g_{ij}(n^j + \theta^{jl} m_l) \crt
+ {\pi^2\over 2g_{YM}^2 \sqrt{g}N} q^2
+ {2\pi\over N} |m_i p - q \epsilon_{ij} n^j| .
\eea
The first two terms are the energies associated to zero modes, while
the last term is the energy $E=|p|$ associated to a state with
massless excitations carrying momentum $m_i$ as well as
a contribution from $E\times B$.

\subsection{Other results}

Unfortunately, space does not permit us to discuss all of the
interesting results obtained in quantum 
\nc\ field theory, but let us mention a few more.

The Seiberg-Witten solution of $\CN=2$ supersymmetric Yang-Mills
\cite{Seiberg:1994rs}, giving a prepotential which encodes the
dependence of BPS masses and low energy couplings on the choice of
vacuum, is a benchmark for nonperturbative studies.  It turns out that
the solution is the same for \nc\ as for \con\ theory; this can be
seen from instanton computations \cite{Hollowood:2001ng} and M theory
considerations \cite{Armoni:2001br}.  See also \cite{Bellisai:2000aw}.

One may wonder what happened to the UV/IR mixing.  The \nc\ theory
necessarily includes a ``$U(1)$ sector,'' and one can show that only
this sector is affected.  It is not visible in the prepotential,
which is independent of the diagonal component of the scalar vev.
This sector (and the rank $1$ theory) is nontrivial, and does not reduce
to Maxwell theory as $\theta\rightarrow 0$ \cite{Armoni:2000xr}.
It is not understood at present.

\Nc\ sigma models are discussed by \citeasnoun{Lee:2000ey}.
The \nc\ WZW model is discussed in
\cite{Moreno:2000kt,Moreno:2000xu,Lugo:2000mf}.

Anomalies have been studied in
\cite{Ardalan:2000qk,Gracia-Bondia:2000pz,Martin:2000wc,%
Bonora:2000he}.
Gauge anomalies appear to be directly analogous to those of \con\
theory, consistent with their topological origin, and can be described
using fairly straightforward
\nc\ generalizations of the Wess-Zumino consistency conditions, descent
equations, etc.
Somewhat surprisingly, however, these formulas appear to lead to more
restrictive conditions for anomaly cancellation, because more
invariants appear.  For example, while the \con\ $d=4$ triangle
anomaly only involves the invariant $d^{abc}=\tr t^a\{t^b,t^c\}$,
which can cancel between different representations, its \nc\ generalization
involves $\tr t^a t^b t^c$, which cannot.  
This apparent contradiction to the general similarity we have seen
between \nc\ and \con\ quantum effects as well as to the topological
nature of the anomaly deserves further study.

\section{APPLICATIONS TO THE QUANTUM HALL EFFECT}
\label{sec:qhe}

A physical context leading directly to noncommutativity in the
Fock space basis is the dynamics of electrons in a constant
magnetic field $\vec B$, projected to the lowest Landau level (LLL).
This is well known in the theory of the quantum Hall effect (QHE)
\cite{Girvin:1987a,Girvin:1999}, and we summarize the basic idea here.

More recently, it has been proposed that a good description of the
fractional quantum Hall effect (FQHE) can be obtained using
\nc\ rank $1$ Chern-Simons theory
\cite{Susskind:2001fb,Polychronakos:2001mi},
and we give a brief introduction to these ideas as well.

It seems fair to say that so far, the \nc\ framework has mainly provided
a new language for previously known results.  However, it also seems
fair to say that this formulation connects these problems to a very
large body of field theory results whose potential relevance had not
been realized, and one can hope that these connections will prove
fruitful.

\subsection{The lowest Landau level}

The Lagrangian of a system of interacting electrons in two
dimensions, subject to a perpendicular magnetic field, is
\bea \label{eq:landau} 
L = \sum_{{\m}=1}^{N_e} {\half}
m_e {\dot{\vec x}}^2_{\m} - {i e B \over 2c} {\ve}_{ij} x^i_{\m}
{\dot x}^j_{\m} + V(\vec x_\m) \\
+ \sum_{{\m} < {\n}} U ( {\vec x}_{\m} - {\vec
x}_{\n} ) 
\eea 
Defining a projection operator $P$ to the first Landau level for
each electron, one finds that the projected coordinates $P x^i_\mu P$
do not commute, but instead satisfy
\bea \label{eq:nclandau} 
[x^i_{\m}, x^j_{\n}] =
i{\d}_{\m\n} {\ve}^{ij} {{\hbar} c \over {eB}} \equiv
i{\d}_{\m\n} {\theta}^{ij} .
\eea 
Heuristically, this is because in the limit of strong magnetic field one can
neglect the kinetic term, i.e. formally put $m_e = 0$. The resulting
Lagrangian is first order in time derivatives, turning the original
coordinate space into an effective phase space defined by \rfn{nclandau}.
A more precise argument can be made by taking coordinates $\z,\bar z$ as
in \rfs{ncflat} and showing that 
\be \label{eq:defbarz}
P\bar z P = \theta\p/\p z + z
\ee
acting on LLL states
\be \label{eq:LLL}
\ket{k} = \frac{1}{\sqrt{k!}}z^k e^{-z z^*/2\theta} .
\ee
The resulting single particle Hamiltonian has been much studied and
we refer to \cite{Bellissard:1994a} for a discussion of the uses of
\nc\ geometry in this context.

To obtain a field theory, one introduces electron
creation and annihilation operators  $\psi^\dagger(x)$ and $\psi(x)$,
in terms of which the electron density is
\bea \label{eq:density} 
{\rho} ({\vec x}) &=
{1\over{\sqrt{N_e}}} \sum_{\m} {\d}^{2} ( {\vec x}_{\m} - {\vec x}
)\crt & \equiv \psi^\dagger(x) \psi(x) .
\eea
The single particle and pairwise interaction Hamiltonians then become
\bea
H_V &= \int dx\ V(x)\ {\rho}(x); \\
H_U &= \int dx\ dx'\ U(x-x')\ {\rho}(x) {\rho}(x') .
\eea
The effects of truncation to the LLL are now expressed by noncommutativity,
and more specifically enter when we use the star product
to compute the commutators of density operators.  In momentum space,
$$
{\rho}({\vec q}) = \int e^{iqx} \rho(x) ,
$$
we have
$$ 
[{\rho} ({\vec q}), {\rho} ({\vec q}^{\prime})] = 
\sin {i {\vec q} \times_{\t} {\vec q}^{\prime}} 
{\rho}({\vec q} + {\vec q}^{\prime}) 
$$
which leads to deformed equations of motion, etc.

This type of description has been used by many authors; we cite
\cite{Girvin:2000} on localization in quantum Hall states, and
\cite{Zee:2000} on its classical limit, as the tip of a large iceberg.

\subsection{The fractional quantum Hall effect}

The generally accepted explanation of the FQHE is that interactions
lead to a state similar to the filled LLL but allowing fractionally
charged quasi-particle excitations.  
A good microscopic description of such a state
is provided by the $N$-electron wavefunction 
(Laughlin, 1983) 
\bea \label{eq:laughlin}
\Psi = \prod_{\mu<\nu} (z_\mu - z_\nu)^m e^{-\sum_\mu z_\mu z_\mu^*/2\theta} .
\eea
where $m$ is an odd integer.  This state has charge density $1/m$ that
of a filled Landau level and was argued by Laughlin to be a ground state
of \rfn{landau}, at least for small $m$.

States with quasiparticles are obtained by simple modifications of this.
An operator creating a quasi-hole at $z_0$ acts as multiplication by
$\prod_i (z_i-z_0)$, while the conjugate operator (which can create
quasi-particles) is $\prod_i (\bar z_i-z_0^*)$ with $\bar z_i$ as in
\rfn{defbarz}.  Acting $m$ times with one of these operators adds or
subtracts a particle, so the quasiparticles have charge $1/m$.

A more subtle property of the quasiparticles is that they
satisfy fractional statistics; a $2\pi$ rotation of a state of two
quasiparticles produces a phase $\exp 2\pi i/m$, as can be seen by 
a Berry phase argument (Arovas \etal, 1984).

The low energy excitations of this ground state can also be described
by a Landau-Ginzburg theory of a superfluid density $\phi$ and a
fictitious vector potential $A$, such as \rfn{abelianhiggs} 
\cite{Kallin:1984,Laughlin:1985,Girvin:1985,Fradkin:1991}.  
In this picture,
the original quasi-particles are vortex solutions, and their fractional
statistics is reproduced by an abelian Chern-Simons term in the action,
\be \label{eq:cs}
S_{CS} = {i m \over 2\pi} \int \epsilon^{ijk} A_i \p_j A_k .
\ee
With this term, a vortex with unit magnetic charge 
will also carry electric charge $1/m$, so the Aharonov-Bohm effect
will lead to fractional statistics \cite{Wilczek:1982du}.

Recently \citeasnoun{Susskind:2001fb} has proposed that noncommutative
Chern-Simons theory is a better description of fractional quantum Hall
states, which can reproduce the detailed properties of these
quasiparticles.  Perhaps the simplest argument one could give for this
is simply to combine the arguments leading to LG theory with the
arguments leading to noncommutativity in a magnetic field.

One way to make this claim precise has been proposed by
\citeasnoun{Polychronakos:2001mi}.  
One first observes that \rfn{laughlin} 
(considered as a function of one dimensional positions $x_\mu=\Re z_\mu$)
is the ground state for
a Calogero model, defined by the quantum mechanical Hamiltonian
\be \label{eq:calogero}
H = \sum_\mu \half p_\mu^2 + \frac{1}{2\theta^2} x_\mu^2
+ \half\sum_{\mu< \lambda} {m(m+1)\over (x_\mu-x_\lambda)^2} .
\ee
One can continue and identify all the excited Calogero
states with excited Laughlin wavefunctions \cite{Hellerman:2001rj}.

It is furthermore known \cite{Olshanetsky:1976a}
that this model can be obtained from
a matrix-vector $U(N)$ gauged quantum mechanics, with action
\cite{Polychronakos:1991bx}
\be \label{eq:matvec}
S = \int dt\ \Tr \left( \epsilon_{ij} X^i D_0 X^j - \frac{1}{2\theta^2} X_i^2 
+ 2 m A_0 \right)
+ \psi^\dagger D_0 \psi
\ee
where $X^i$ are Hermitian matrices with $i=1,2$ and $\psi$ is a complex
vector.  The gauge field in this action is nondynamical but enforces a
constraint which selects the sector with $\psi$ charge $m$, for which
the Hamiltonian is \rfn{calogero}.

Reversing the procedure which led to \rfn{matrixmodel}, 
we can regard $X^i$ as covariant derivative operators, to
obtain an \nc\ gauge theory action, whose kinetic term (coming from
the first term of \rfn{matvec}) is precisely \rfn{cs}.
The other terms are secondary: the
$X^2$ term localizes the state in space, while $\psi$, although
required for consistency at finite $N$, plays no dynamical role.

In this sense, one has a precise \nc\ field theory description of the
fractional Hall state.  In particular, the quasiparticles are
well-defined excitations of the \nc\ gauge field; for example the
quasihole is rather similar to the fluxon \rfn{genvortex}.  We refer
to the cited references for more details.  

\section{MATHEMATICAL ASPECTS}
\label{sec:math}

As we mentioned in the introduction, \nc\ gauge theory was first
clearly formulated by mathematicians to address questions in \nc\
geometry.  Limitations on length would not permit more than the most
cursory introduction to this subject here, and since so many
introductions are already available, starting with the excellent
\cite{Connes:1994a}, much of which is quite readable by physicists, and
including 
\cite{Connes:1995mw,Connes:2000by,Connes:2000ti,%
Gracia-Bondia:2001tr,Douglas:1999ge} 
as well as the reviews cited in the introduction, we will content
ourselves with a definition:

\Nc\ geometry is a branch of mathematics
which attempts to generalize the notions of geometry, broadly defined,
from spaces $M$ whose function algebras $C(M)$ are commutative, to
``spaces'' associated to general algebras.  The word space is in
quotes here to emphasize that there is no {\it a priori}
assumption that these spaces are similar to manifolds;
all of their attributes emerge through the course of
formulating and studying these geometric notions.

The remainder of this section provides an introduction to other
examples of \nc\ spaces, and other topics for which a more
mathematical point of view is advantageous.  We will only be able to
discuss a few aspects of \nc\ geometry with direct relevance for the
physics we described; there are many others for which such a role may
await, or for which we simply lack the expertise to do them justice.
Among them are cyclic cohomology and the related index theorems
\cite{Connes:1982}, and the concept of spectral triple
\cite{Connes:1996gi}.

\subsection{Operator algebraic aspects}
\label{sec:opalg}

The origins of \nc\ geometry are in the theory of operator algebras,
which grew out of functional analysis.  Certain issues in \nc\ field
theory, especially the analogs of topological questions in \con\ field
theory, cannot be understood without these ideas.  The following is
loosely inspired by a discussion of the meaning of the instanton
charge in \cite{Schwarz:2001ru}; we also discuss a proposed definition
of the \nc\ gauge group \cite{Harvey:2001pd}.

A good example of a topological quantity is the total magnetic flux in
two spatial dimensions, $\int \Tr F$.  In commutative theory, this is
the integral of a total derivative $F = dA$.  In a pure gauge
background, this integral will be quantized; furthermore, it cannot
change under variations $A+\delta A$ by functions $\delta A$ which are
continuous, single-valued and fall off faster than $1/r$ at infinity.
Thus one can argue with hardly any dynamical input that the total flux
in a sufficiently large region must be conserved.

Let us return to the question raised in \rfs{sol}.B, of whether
this argument can be generalized to \nc\ theory.
The magnetic flux has the same formal expression.
In operator language, it is the trace
of a commutator,
\begin{equation}
\label{eq:traceF}
\Tr F = \Tr [C,\Cb]+1.
\end{equation}
We need to understand in what sense
this is a ``boundary term,'' which is preserved under continuous
variation of the fields with a suitable falloff condition.

If one does not try to define these conditions,
one can easily exhibit counterexamples,
such as the path
\begin{equation}
\label{eq:paradox}
\Cb = \lambda z + (1-\lambda) S^\dagger z S
\end{equation}
which purports to continuously interpolate between the fluxon and
the vacuum.
Is this a continuous variation?  We have to decide;
there is no single best definition of continuous in this context.

The minimal criterion we could use for a continuous variation of the
fields is to only allow variations by bounded operators.  A bounded
operator has finite operator norm $|A|$ (this is the largest eigenvalue of
$(A^\dagger A)^{1/2}$).  A small variation $C'$ of the connection $C$
then has small $|C'-C|$, so will be a bounded operator times a small
coefficient.  Boundedness is mathematically a very natural condition
as it is the weakest condition we can impose on a class of operators
which guarantees that the product of any two will exist, and thus the
definition of an operator algebra normally includes this condition.

However, boundedness is not a strong enough condition to force traces
of commutators to vanish: one can have $\Tr [A,B]\ne 0$ even
if both $A$ and $B$ are bounded operators (e.g. consider $\Tr
[S^\dagger,S]$ where $S$ is the shift operator).  The operator
$O=z - S^\dagger z S$ appearing in \rfn{paradox} is bounded.

A condition which does guarantee $\Tr [A,B]=0$ is for $A$ to be
bounded and $B$ to be trace class, roughly meaning that its
eigenvalues form an absolutely convergent series.  More precisely, $A$
is trace class if $|A|_1$ is finite, where $|A|_p = (\Tr(A^\dagger
A)^{p/2})^{1/p}$ is the Schatten $p$-norm.  More generally, the
$p$-summable operators are those for which $|A|_p$ is finite.  This is
more or less the direct analog of the \con\ condition that a function
(or some power of it) be integrable.  A related condition which also
expresses ``falloff at infinity'' is for $A$ to be compact, meaning
that the sequence of eigenvalues of $A^\dagger A$ has the limit zero.

Although these are important conditions, they do not solve the problem at
hand, because they are not preserved by the derivatives
\rfn{rnderivatives}, so do not give strong constraints on
\rfn{traceF}.  In particular, one can have $\Tr [z,K]\ne 0$ (and even
ill-defined) for $K$ in any of the classes above (e.g. try $\Tr[O,\zb]$).

Another approach is to adapt a falloff condition on functions on
$\IR^n$ to $\IR^n_\theta$ by placing the same falloff condition on the
symbol \rfn{weyl}.  An important point to realize about the \nc\
case is that one cannot separately define ``growth at large radius''
from ``growth with large momentum.''  The only obvious criterion one
can use is the asymptotics of matrix elements at large mode number,
which does not distinguish the two.  Physically, as we saw in our
considerations of Gaussians in II.2.C, interactions easily convert one
into the other.  Thus questions about whether configurations disperse
or go off to large distance, are hard to distinguish from questions
about the possibility of forming singularities.

Thus, useful falloff conditions must apply to both position and
momenta.  Such conditions are standard in the theory of
pseudodifferential operators, and can be used to define various
algebras which are closed under the star product \cite{Schwarz:2001ru}.
Perhaps the simplest is the class of operators $S(\IR^d_\theta)$,
which can be obtained as the transform Eq. (\ref{eq:weyl}) of smooth
functions with rapid decrease, i.e. which fall off both in position
and momentum space faster than any power.  This condition is very much
stronger than boundedness (roughly, it requires matrix elements in the
Fock basis to fall off exponentially) and is preserved under
differentiation, so $\Tr F$ will certainly be preserved by this type
of perturbation.

As we discussed in III.B, the question of physical flux conservation
is whether time evolution allows flux to get to infinity at finite
time.  This is a dynamical question, but might be addressed by finding
conditions on the fields which are preserved under time evolution.
For example, in the \con\ case one can take fields which are
pure gauge outside a radius $r=R$, and then causality guarantees that
they will stay pure gauge outside of $r=R+t$.  Since $S(\IR^d_\theta)$
is preserved by both the product and the Laplacian, it is a good
candidate for an analogous class of \nc\ fields which is preserved
under time evolution.  The fluxon and its perturbations by finitely
many modes fall into this class, so an argument along these lines
should show that flux is conserved, and thus the decay of the fluxon
does not lead in finite time to a stable ground state.  Of course
conservation of $\Tr F$ (and energy $\Tr F^2$) is surely true for a
larger class of initial conditions.

A related question, discussed by \citeasnoun{Harvey:2001pd}, is the
precise definition of the gauge group $U(\CH)$.  In \con\ gauge
theory, the topology of the group $\CG_0$ of gauge transformations
which approach the identity at infinity, is directly related to that
of the configuration space of gauge fields (connections modulo gauge
transformations) by a standard argument: since the space of
connections $\CA$ is contractible, we have $\pi_n(\CA/\CG_0) \cong
\pi_{n-1}(\CG_0)$.  One might expect the topology of $U(\CH)$ to play
an analogous role and in fairly direct correspondence to the
considerations above, this would imply that we cannot identify $U(\CH)$
with the group of unitary bounded operators on Hilbert space; this
group is contractable \cite{Kuiper:1965a}.  
One can get the expected nontrivial topology,
$\pi_k(G)=\lim_{N\rightarrow\infty} \pi_k(U(N)) = \IZ$ for
$k=2n+1$, by using the unitaries $1+K$ with $K$ compact or $p$-summable
\cite{Palais:1965a}.  \citeasnoun{Harvey:2001pd} suggests using $K$
compact, as the largest of these groups.  Alternatively, one might try
to use a smaller group defined by imposing conditions involving the
derivatives.

A similar open question about which less is known at present is
that of what class of fields to integrate over in the functional
integral (the functional measure).  As we commented in \rfs{gt}.1,
formally all pure bosonic \nc\ gauge theories have the same action;
however we expect that different quantum theories exist, distinguished
by the choice of measure.  In particular, this subsumes the choice of
the dimension of space-time.

Again, the minimal proposal is that one uses an action such as
\rfn{matrixmodel}, expands around a pure gauge
configuration $C_i=\zb^i$ and $\Cb_i=z^i$ for $1\le i\le r$, and
integrates over all variations of this by bounded operators.  This
type of prescription does make sense in the context of the
perturbation theory of \rfs{quant} with IR and UV cutoffs, and would
specify the dimension of space-time there.  This dimension can also be
inferred from the nature of one-loop divergences; see
\cite{Connes:1994a,Varilly:1998gq}.

For the reasons we discussed above, this prescription probably does
not make sense beyond perturbation theory.  One might address this by
proposing a smaller class of variations to integrate over.  At our
present level of understanding, however, it may be better to work
with an explicit cutoff, such as that provided by the large $N$ limit
of matrix approximations as discussed in \rfs{bases}.6.  We described
there the proposal of \cite{Ambjorn:1999ts} along these lines; while
concrete, this appears to share the problems of \con\ lattice
definitions of breaking the symmetries of the continuum theory, and of
not admitting supersymmetric generalizations.  Thus the problem of
finding the best regulated form of \nc\ field theory remains open.

Given such a regulated theory, one then wants to study the continuum
limit.  Of course the analogous questions in \con\ quantum field
theory are not trivial, and their proper understanding requires the
ideas of the renormalization group.  We already made such a discussion
in \ref{sec:quant}, following the usual paradigm in which the momentum
space behavior of Greens functions is central.  It is not yet clear
that this is the best paradigm for \nc\ quantum theory; perhaps other
classes of fields such as discussed here will turn out to be equally
or more useful.

\subsection{Other noncommutative spaces}
\label{sec:othernc}

Our discussion so far was limited to \nc\ field theory on
$\IR_\theta^{d}$, mostly because it and $\IT_\theta^{d}$ are the
only examples for which field theoretic physics has been explored
sufficiently at present to make it worth writing a review. There
are many other examples which allow defining \nc\ field theories,
whose physics has been less well explored.

For example, let us try to define a \nc\ sphere $S^d_\theta$.  
This should be a space associated to an algebra of $d+1$ operators
$x^i$, $1\le i\le d+1$, satisfying the relation
\begin{equation}
\label{eq:ncsphere}
\sum_{i=1}^{d+1} (x^i)^2 = R^2~ {\bf 1}.
\end{equation}

We next need to postulate some analog of \rfn{noncom}.
Although one can define algebras without imposing commutation
relations for each pair of variables, these will be much ``larger''
than the algebra of functions on $S^d$.
For $d=2$, there is a natural choice to make,
$$ 
[ x^i, x^j ] = i\theta\epsilon^{ijk} x^k , 
$$ 
which preserves $SO(3)$ symmetry.

In fact there is a very simple way to define such an algebra,
called the ``fuzzy two-sphere,''
discussed in depth by \citeasnoun{Madore:1992bw}, in
\cite{Madore:1995a} and many other works.
It is to consider the $2j+1$-dimensional irreducible
representation of the $SU(2)$ Lie algebra, defined by hermitian
generators $t^i$ satisfying the commutation relations 
$ [ t^i,t^j ] = i\epsilon^{ijk} t^k .$ 
One can then set $$ x^i = {R\over
\sqrt{j(j+1)}} t^i $$ from which \rfn{ncsphere} follows easily,
and one finds $\theta = R/\sqrt{j(j+1)}$.

This algebra can serve as another starting point for defining \nc\
field theory.  One cannot use \rfn{rnderivatives} to define the
derivatives, however, as this is inconsistent with \rfn{ncsphere}.
A simple choice which works is to define a linearly dependent set
of derivatives $\p_i f = [X^i, f]$ and use $g_{ij}=\delta_{ij}$ in
this basis as a metric.  Since these derivatives do not commute,
the natural definition of curvature becomes $$ F_{ij} = i[\p_i +
A_i,\p_j + A_j] - i[\p_i,\p_j] , $$ (as in \rfn{matrixfield}), in
terms of which one can again use the Yang-Mills action.  This and
related theories have been discussed by \citeasnoun{Grosse:1996ar}.

Unlike $\IR^d_{\t}$ and ${\IT}^d_{\t}$, this algebra is finite
dimensional. If we base our theory on it directly, it will have a
finite number of degrees of freedom, and one might question the use of
the term ``field theory'' to describe it.

One can certainly take the limit $j\rightarrow \infty$, but if we 
take $R\propto j \rightarrow\infty$ as well to keep $\theta$
finite, we lose the curvature of the sphere, and end up with $\IR^2_\theta$.

Finally, if we keep $R$ fixed and take $j\rightarrow\infty$, it is a
theorem that (with suitable definitions) this algebra goes over in the
limit to the algebra of continuous functions on ordinary $S^2$, so we
obtain a \con\ field theory.  This feature of restoring commutativity
in the limit applies to a wide class of constructions, as we will
discuss further in \rfs{defq}.
It should be said that this theorem is classical, and there might be a
way to quantize the theory which does not commute with the large $j$
limit, leading to a nontrivial \nc\ quantum field theory.

One can still argue that for large but finite $j$, a theory based on
this algebra deserves the name ``\nc\ field theory.''  We would
suggest that the nomenclature be based, first, on the extent to which
a theory displays physical characteristics similar to those we have
seen for theories on $\IR_\theta^d$, and second on the extent to which
it shows universality ({\it e.g.} has finitely many parameters)
analogous to field theory; this is not usually the case for
constructions with a finite number of degrees of freedom.  These
questions have not been settled and we will not take a position on
this here.

We move on and discuss other \nc\ algebras comparable to function
algebras which can clearly serve to define field theories.  The
simplest possibility was of course $\Mat_n(C(M))$.  Interesting
variations on this can be obtained by imposing further conditions
which respect the product.  For example, one could consider an algebra
and a $\IZ_2$ automorphism preserving the metric, call it $g$.
A simple example would be to take $\IR^d_\theta$
and let $g$ be the reflection about some hyperplane.
One could then impose $a^\dagger = g(a)$ where $g$ acts on each matrix
element.  This idea has been used to propose \nc\ gauge theories with
analogy to the other classical ($SO$ and $Sp$) Lie algebras
\cite{Bonora:2000td,Bars:2001iq}.

One can go on in this vein, using non-simple finite algebras and more
complicated automorphisms.  Indeed, one can obtain the complete action
for the standard model by choosing the appropriate algebra
\cite{Connes:1991qp}.  Obviously the significance of this observation
is for the future to judge, though any example of a formalism which
only describes a subset of all possible gauge theories but can lead to
the standard model probably has something important to teach us.

\subsection{Group algebras and noncommutative quotients}
\label{sec:grpalg}

A large class of ``more \nc'' algebras are provided by the group algebras.
Given a group $G$, we define $\CA_G$ to be the algebra of all linear
combinations of elements of $G$, with multiplication law inherited from $G$.
For example, consider $G=\IZ_2$ with elements $1$ and $g$ satisfying
$g^2=1$.  The general element of $\CA_G$ is $a + b g$, and
$(a + b g)(c + d g)=(ac+bd)+(bc+ad)g$.

It is well known that for a finite group $G$, $\CA_G$ is a direct sum
of matrix algebras, one for each irreducible representation of $G$.
One goal of representation theory is to try to make analogous
statements for infinite groups.  This requires being more precise
about the particular linear combinations allowed, and leads one deep
into the theory of operator algebras.  We will not go into details,
but these algebras are clearly a good source of \nc\ field theories,
as the original definitions of \citeasnoun{Connes:1987a} can be
applied directly to this case.  One of the main problems in trying to
define \nc\ field theories on more general spaces is to either define
a concept of metric, or get away without one.  Since group spaces are
homogeneous spaces, this problem becomes very much simpler.

A variation on this construction is the twisted group algebra
$\CA_{G,\epsilon}$, which can be defined if $H^2(G,U(1))\ne 0$.
This allows for nontrivial projective representations $\gamma$
characterized by a two-cocycle $\epsilon$, 
\bea \label{eq:twist}
\gamma(g_1) \gamma(g_2) =
\epsilon(g_1,g_2) \gamma(g_1 g_2), 
\eea
and $\CA_{G,\epsilon}$ is just
the group algebra with this multiplication law.  Since the phases
\rfn{moyalphase} are a two-cocycle for $\IZ^d$, $\IT_\theta^d$
itself is an example.

A very important source of noncommutative algebras is the
``crossed product'' construction.  One starts with an algebra
$\CA$, with a group $G$ acting on it, say on the left: $$ a
\rightarrow U(g) a . $$ One then chooses a representation $R$ of
$G$ acting on a Hilbert space $H$, $g \mapsto {\g} (g) \in
{\CB}(H)$, considers the tensor product $\CA \otimes {\CB}(H)$,
and imposes the condition
\begin{equation}
\label{eq:crossproduct}
\gamma^{-1}(g)~ a ~\gamma(g) = U(g) a .
\end{equation}

The simplest example of this construction would be to take $\CA$
finite dimensional.  If $G$ is abelian, we can even take $\CA=\IC$,
and the general solution for $a$ will be some particular solution
added to an arbitrary element of $\CA_G$.  As we discuss below,
the particular solution has the interpretation of a connection, and
this construction leads directly to gauge theory on $\CA_G$.

Suppose we had started with $\CA=C(M)$.  If we take $\gamma(g)$ to be
the trivial representation, \rfn{crossproduct} defines the algebra of
functions on the quotient space $C(M/G)$.  More general choices of
$\gamma(g)$ thus lead to a generalized concept of quotient.  The
striking feature of this is that it provides a way to define quotients
by ``bad'' group actions, those for which the quotient space $M/G$ is
pathological, as is discussed in Connes (1994).  This definition
of quotient also follows from the standard string theory definition of
orbifolds, as discussed in 
\cite{Douglas:1999ge,Konechny:1999zz,Martinec:2001hh}.

A natural generalization of this construction is
the twisted cross product, whose definition is precisely the same
except that we take $\gamma$ to be a projective representation
as in \rfn{twist}.
This leads to gauge theory on $\CA_{G,\epsilon}$; we will discuss
the toroidal case below.

Another construction with a related geometric picture, the foliation
algebra, is discussed in \cite{Connes:1994a}.

\subsection{Gauge theory and topology}
\label{sec:gaugetop}

A good understanding of the topology of \con\ gauge field
configurations requires introducing the notions of principal bundle
and vector bundle.  We recall that \con\ gauge fields are connections
in some principal $G$-bundle, while matter fields are sections of
vector bundles with structure group $G$.  On a compact space such as a
torus, the topological classication of these bundles has direct
physical implications.

The \nc\ analog of these ideas is a central part of \nc\ geometry.  We
refer to \citeasnoun{Konechny:2000dp} for a detailed discussion
focusing on the example of the \nc\ torus, but we give the basic
definitions here.  See also \cite{Harvey:2001yn} for a related
discussion.

One aspect of a \con\ vector bundle is that one can multiply a section
by a function to get another section.  This will be taken as the
defining feature of a \nc\ vector bundle associated to the algebra
$\CA$; one considers the (typically infinite dimensional) linear space
of sections and requires it to be a module over the algebra $\CA$.

A module $E$ over $\CA$ is a linear space admitting a multiplicative
action of $\CA$ which is bilinear and satisfies the rule
\begin{equation}
\label{eq:moduledef}
a\cdot(b\cdot v) = (ab)\cdot v; \qquad (a,b \in \CA;\ v \in E),
\end{equation}
and carrying whatever other structure $\CA$ has (e.g. continuity,
smoothness, etc.)  The simplest example is just a $N$-component vector
with elements in $\CA$, which is called a free module of rank $N$.
Sections of a trivial rank $N$ \con\ vector bundle form the free module
$C(M)^N$, and this is the obvious generalization.

We chose the multiplication in Eq. (\ref{eq:moduledef}) to act on the
left, defining a left module, but one can equally well let it act on
the right, defining a right module, or postulate independent
multiplication laws on both sides, defining a bimodule.
One can also speak of a $(\CA,\CB)$-bimodule, which admits an action
of $\CA$ on the left and another algebra $\CB$ on the right.

A simple example of a bimodule would be the space of $M\times N$
matrices of elements of $\CA$, with $\CA \otimes {\bf 1}_M$
acting on the left and $\CA \otimes {\bf 1}_N$ on the right.
These sit inside left and right actions of $\Mat_M(\CA)$ and
$\Mat_N(\CA)$, respectively.
An obvious but important point is that these two actions commute,
i.e. $(a\cdot v)\cdot b = a\cdot (v\cdot b)$.

We can regard the free module $\CA^N$ as a bimodule (the $1\times N$
matrices) and this comment shows that it admits a right action of
$\Mat_N(\CA)$ which commutes with the action Eq. (\ref{eq:moduledef}).
There is a general term for the linear maps acting on a module $E$
which commute with \rfn{moduledef}; they are the
endomorphisms of the module $E$, and the space of these
is denoted $\End_\CA E$.  In fact $\End_\CA (\CA^N) \cong \Mat_N(\CA)$
so we know all endomorphisms of the free module.  For more general
$E$, $\End_\CA E$ will always be an algebra, but need not be a
matrix algebra.

We can now obtain more examples by starting with free modules and
applying a projection.  We use the fact that the left module $\CA^N$
admits a right action of $\Mat_N(\CA)$.  Given a projection $P\in
\Mat_N(\CA)$, the space of solutions of $v\cdot P=v$ is a
module with multiplication law $a\times v = a\cdot v$.
These examples are known as finitely
generated projective modules and in fact it is only these modules
which are natural generalizations of vector bundles, so one normally
restricts attention to them.
The endomorphisms of these modules also admit a simple description:
they are the elements in $\Mat_N(\CA)$ of the form $PaP$.

We are now prepared to make a more general definition of connection
\cite{Connes:1980ji}.
So far we have been taking a connection to be a set of operators
$D_i = \p_i + A_i$ where the components $A_i$ are taken to be elements
of $\Mat_N(\CA)$, in direct analogy to \con\ Yang-Mills theory.
We used this in several ways; as \rfn{adjointderiv} acting on
fields in $\Mat_N(\CA)$ (the ``adjoint action''),
in the curvatures \rfn{matrixfield}, and finally acting on
``fundamental'' matter in \rfs{gt}.3.  The last of these
is the point where vector bundles enter the \con\ discussion, and where
our generalization will apply most directly.

More generally, a connection on the module $E$ could be
any set of linear operators $D_i$
which act on $E$ and satisfy the Leibniz rule,
\bea \label{eq:defconn}
D_i(a \cdot v) = \p_i(a) \cdot v + a \cdot D_i(v) .
\eea
To see the relation to our previous definition, we first note that
the difference between two connections $A_i\equiv D_i - D'_i$
will commute with the action of $\CA$, and is thus an
endomorphism of $E$.
Thus we can also describe the general connection of this class
by choosing a fiducial connection $D^{(0)}_i$ and writing
$$
D_i = D^{(0)}_i + A_i; \qquad A_i \in \End_\CA E.
$$

Let us consider $E=\CA^N$ so we can compare with \rfs{gt}.3.
One can clearly take $D^{(0)}_i=\p_i$ for the free module, and as
we discussed, the endomorphisms are just $\Mat_N(\CA)$, so we see
that in this case the new and old definitions of connection agree.

All of the modules constructed from projections as above also admit
a natural candidate for $D^{(0)}_i$, namely we apply $\p_i$ and
project back:
$$
D^{(0)}_i(v) = (\p_i v) \cdot P.
$$
Thus connections on $E$ can be identified with elements of $\End_\CA E$,
which as we discussed will be some subalgebra of matrices $\Mat_N(\CA)$,
but not itself a matrix algebra.

Finally, the gauge theory action uses one more ingredient, the trace.
This can also be defined in terms of the projection $P$; an endomorphism
which can be written $PaP$ as above has trace $\Tr|_{\Mat_N(\CA)} PaP$.
In particular, we can define the dimension of the module as
$$
\Tr_E 1 = \Tr_{\Mat_N(\CA)} P.
$$

These definitions can be used to obtain all the ``\nc\ vector
bundles,'' and largely reduces the classification problem to
classifying the projections in $\Mat_N(\CA)$.  The next step in such a
classification is to find invariants which tell us when two
projections are related by a continuous deformation.  A natural guess
is that one wants to generalize the Chern classes of the \con\ theory,
and indeed we have been implicitly doing this in claiming that
quantities like Eq. (\ref{eq:traceF}) are topological quantum numbers.
This turns out to be true for the \nc\ torus and indeed there is a
well-developed formalism generalizing this to arbitrary algebras,
based on cyclic cohomology \cite{Connes:1982}.

\subsection{The \nc\ torus}
\label{sec:nctorus}

The \nc\ torus and its associated modules can be obtained using almost
any of the constructions we cited, so besides its physical relevance
it serves as a good illustration.  We follow the definition of
$\IT^d_\theta$ made in \cite{Connes:1980ji}, as the algebra of linear
combinations of products $\prod_i U_i^{n_i}$ of generators with the
relations \rfn{torusphase}, with coefficients decreasing faster than
any power of $|n|$.  Our discussion will mostly stick to
$\IT^2_\theta$.

One can regard commutative $C(\IT^2)$ as the group algebra $G_{\IZ^2}$,
i.e. all linear combinations of products of two commuting
generators $U_1=\gamma(g_1)$ and $U_2=\gamma(g_2)$.  We can think of
$\gamma$ as the regular representation; it decomposes into the direct
sum of all irreducible representations of $\IZ^2$, which can be written
$U_i = \exp i\sigma_i$, parameterized by $\sigma_i$ coordinates on $\IT^2$.
Each such representation is a one dimensional
module over $G_{\IZ^2}$.

Taking instead a projective representation \rfn{twist}
with $\epsilon(g_1,g_2)=e^{-i\theta/2}$ leads directly to \rfn{torusphase}.
The regular representation can be written explicitly as
\bea \label{eq:twistedrep}
U_1 = e^{i \sigma_1} S_{\sigma_2,-\theta}; \qquad
U_2 = e^{i \sigma_2}
\eea
where $S_{\sigma,a} f(\sigma) \equiv f(\sigma+a) S_{\sigma,a}$.
This can again be decomposed into irreducibles, each labelled
by a fixed value of $\sigma_1$.  All of these are equivalent
however (if we take this to act on functions on $\IR^2$), under
conjugation by $e^{i\alpha \sigma_2}$.

This gives us an example of an projective module $E$ over $\IT^2_\theta$:
to repeat, we take for $E$ the smooth
functions on a real line $\CS(\IR)$, and define the action of the operators
$U_i$ on them as
\begin{equation}
\label{eq:torusone}
(U_1 f)(\sigma) = f(\sigma - \theta); \qquad
(U_2 f)(\sigma) = e^{i \sigma} f(\sigma).
\end{equation}

In terms of the general construction of the previous subsection,
we obtained $E$ from $\CA^1$ by a projection, which should allow
us to compute $\dim E = \Tr_E 1$.  Naively $P=\delta(\sigma_1)$;
this is too naive as this operator is not bounded.  A correct
projector can be found using the ansatz
$$
P = U_2^\dagger g(U_1)^\dagger + f(U_1) + g(U_1) U_2
$$
for functions $f$ and $g$ chosen to satisfy $P^2=P$ 
\cite{Connes:1980ji,Rieffel:1981,Bars:2000yv},
and this can be used to compute $\dim E = \theta/2\pi$.

Rather than continue by following the general theory, 
in this case it is easier to write
explicit results.  For example, the endomorphisms of $E$ are
\begin{equation}
\label{eq:torustwo}
(Z_1 f)(\sigma) = f(\sigma + 2\pi); \qquad
(Z_2 f)(\sigma) = e^{2 \pi i \sigma/\theta} f(\sigma).
\end{equation}
These operators also satisfy the defining relations of a \nc\ torus,
but one with $\theta'=4\pi^2/\theta$.  

The fact that two dual tori are involved may seem
counterintuitive.  However, there is a sense in which the tori
$\IT^2_\theta$ and $\IT^2_{\theta'}$ are equivalent, called Morita
equivalence, based on the observation that $E$ is also a module
for $\IT^2_{\theta'}$ (since we have an action of the $Z_i$).
We return to this below.

A reference gauge connection $D_i^{(0)}$ on $E$ satisfying 
$D_i^{(0)}(U_j) = 2\pi i\delta_{i,j} U_j$
can now be defined as
\bea \label{eq:ncconn}
D_1^{(0)} f =  - {2\pi i\sigma\over\theta} f; \qquad
D_2^{(0)} f = 2\pi {\p f\over\p \sigma}
\eea
The general connection is a sum $D_i^{(0)} + A_i$ with
$A_i \in \End_{\IT^2_\theta} E$.  In other words, the vector potential
naturally lives on a ``dual'' \nc\ torus $\IT^2_{\theta'}$.
Note that ${1\over 2\pi i}\Tr [D_1,D_2] = 1$ 
is integrally quantized, even though $\dim E$ was not.

This construction can be generalized to produce all modules over
$\IT^2_\theta$.  These modules $E_{p,q}$ are characterized by two
integers $p$ and $q$, and can be produced by tensoring a
representation with $\theta_1=p/q$ constructed in \rfs{bases}.6,
with \rfn{torusone} with $\theta_2$ chosen to satisfy
$\theta=\theta_1+\theta_2$.  The computations above then generalize to
\bea \label{eq:ttwochern}
\dim E_{p,q} &= |p-q\theta/2\pi|; \\
\Tr_{E_{p,q}} F &= 2\pi i q .
\eea
The example $E$ of \rfn{torusone} is $E_{0,1}$, while regarded as a
module on $\IT^2_{\theta'}$ it turns out to be $E_{1,0}$.

The same construction generalizes to produce all modules on
$\IT^d_{\theta}$ (at least for irrational $\theta$) 
\cite{Connes:1980ji,Rieffel:1988a}
and thus we can get the topological classification of gauge field
configurations on $\IT^d_\theta$.  This turns out to be the same, in a
sense, as for the commutative torus.  In both cases, the topological
class of a connection is determined by its Chern classes
\cite{Connes:1980ji}, which can be
defined as $\int\Tr 1$, $\int\Tr F_{i_1i_2}$ $\int\Tr F_{i_1i_2}
\wedge F_{i_3i_4}$, etc. and which obey additive quantization rules
corresponding to the group $\IZ^{2^{d-1}}$.  This is reasonable as a
continuous variation of $\theta$ should not change a topological
property.

Many properties of these modules are easier to see from other
constructions.  In particular, one can also regard $\IT^2$ as the
quotient $\IR^2/\IZ^2$, and then define this quotient using the
crossed product.  Let $n^i\in\IZ^2$ act on $\IR^2$ as $x^j \rightarrow
x^j + n^i (e_i)^j$, and take $X^i \in \IR^2\otimes B(\CH)$, then
\rfn{crossproduct} becomes

\bea \label{eq:tquotient} 
U_i^{-1} X^j U_i = X^j + (e_i)^j .  
\eea 
If we take the regular representation as above, and $(e_i)^j=\delta_i^j$
for simplicity, these equations are solved by
\bea \label{eq:concovder}
X^j = -i{\p\over\p \sigma^j} + A_j
\eea
where $A_j$ commute with the $U_i$; i.e. are general functions on $\IT^2$.
It is no coincidence that these solutions look like covariant derivatives;
we can rewrite \rfn{tquotient} as
\bea \label{eq:tconn}
X^j U_i = U_i X^j + (e_i)^j U_i
\eea
which is precisely \rfn{defconn} with $X^j = D_j$.

To get $\IT_\theta^2$, we instead take the twisted crossed product,
which amounts to solving \rfn{tquotient} with $U_i$ satisfying
\rfn{torusphase}.  Using the representation \rfn{twistedrep} for this, 
the solutions become 
\bea \label{eq:tcovder}
X^i = D_i^{(0)} + A_i
\eea
with $D_i^{(0)}$ as in \rfn{ncconn}, and $A_i$ general elements of
$\IT^2_{\theta'}$.

This generalizes to a procedure for deriving connections on
$\IT^d_\theta$, which by taking more general $U_i$ produces all
constant curvature connections.  It corresponds directly to the
string and M(atrix) theory definition of quotient space, and thus
we will find in \rfs{ncmth} that $\IT_\theta$ appears naturally
in this context.

Although we started with the continuum definition \rfn{twistedrep},
one could also obtain this as an explicit limit of matrix representations.
Using this in the constructions we just discussed
leads to the regulated gauge theory discussed in \rfs{bases}.6.

\subsection{Morita equivalence}
\label{sec:morita}

Two algebras $\CA$ and $\hat\CA$ are Morita equivalent if there is a
natural one-to-one map carrying all modules and their associated
structures from either algebra into the other.  On the face of it,
this is quite a strong relation, which for commutative algebras
$C(M)$ and $C(\hat M)$ would certainly imply that $M\cong \hat M$.

We first note that if such a map exists, we can derive the general map
just knowing the counterpart in $\hat\CA$ of $\CA^1$, the free module of
rank $1$.  Call this $P$; it is an $(\hat\CA,\CA)$-bimodule, because we know
that $\CA^1$ also admits a right action of $\CA$.  There is then a general
construction (the mathematical notion of tensor product) which produces
the map:
\be \label{eq:morita}
E \rightarrow \hat E = P \otimes_\CA E .
\ee
An example is that $\hat \CA = \Mat_N(\CA)$ is Morita equivalent to $\CA$,
by taking $P$ to be $\CA^N$.  Thus we do not want to think of Morita
equivalent algebras as literally equivalent; however their K theory and
many other properties are the same.

A more striking example of a Morita equivalence is that $\IT^2_\theta$
is Morita equivalent to $\IT^2_{\theta'}$ as above.  The module $E$ we
constructed in Eqs. (\ref{eq:torusone}), (\ref{eq:torustwo}) is the
bimodule which provides this equivalence.  Thus every module $E_{p,q}$
on $\IT^2_\theta$ is associated by \rfn{morita} to a module $E_{-q,p}$
on $\IT^2_{\theta'}$.

There is a second, simpler equivalence of this type obtained by taking
$\theta\rightarrow\theta+2\pi$, which manifestly produces the same
algebra.  Considering \rfn{ttwochern} shows that this acts on the modules
as $E_{p,q} \rightarrow E_{p+1,q}$.

These two transformations generate the group $SL(2,\IZ)$ of $2\times
2$ matrices $\left(\matrix{a&b\cr d&d}\right)$ with integer entries
and determinant $1$, acting on $\tau\equiv\theta/2\pi$ as
$\tau\rightarrow (a\tau+b)/(c\tau+d)$ and on $(p,q)$ as a vector.
This is an example of a ``duality group'' which is a candidate
equivalence between gauge theories based on the pair of Morita
equivalent modules.  This equivalence can be further strengthened by
introducing a stronger notion of ``gauge Morita equivalence''
\cite{Schwarz:1998qj}, which produces a map between the spaces of
connections on the two modules which preserves the Yang-Mills action.

Although somewhat abstract, in this example these equivalences have a
simple geometric origin, which can be understood in terms of the
constructions of the previous subsection
\cite{Connes:1994a,Douglas:1999ge}.  It is that $\IT^2_\theta$ can be
obtained as a quotient of $\IR$ by the identifications $x \sim x+2\pi
\sim x+\theta$, in other words by a two dimensional lattice, and as
such its moduli space will admit the $SL(2,\IZ)$ symmetry of
redefinitions of the lattice, just as in the construction of $\IT^2$ as
$\IR^2/\IZ_2$.  Indeed, as this suggests, $\IT^2_\theta$ can be regarded
is a zero volume limit of $\IT^2$, a picture which will reappear in the
string theory discussion.

Morita equivalence between higher dimensional \nc\ tori is
also understood (for irrational $\theta$ at least).  The
basic result is the following \cite{Rieffel:1998a}: two tori
$\IT^d_\theta$ and $\IT^d_{\theta'}$ are Morita equivalent 
if they are related as
$$
\theta' = (A \theta + B)(C \theta + D)^{-1} .
$$
where $\left(\matrix{A&B\cr C&D}\right)$ is a $2d\times 2d$
matrix belonging to the group $SO(d,d;\IZ)$.  The simplest example is
$A=D=0$, $B=-1$, $C=1$ which corresponds to the
Morita equivalence on $\IT^2_\theta$ we described above.
The corresponding transformation on the modules is also given by
a linear action of $SO(d,d)$ on the Chern class data, which can
be regarded as a spinor of this group.

These transformations agree precisely with the action of T-duality in
toroidal compactifications of string theory (in the limit of a zero
volume torus) and this is how they were first conjectured.
Conversely, the mathematical proof of these equivalences is part of
a new argument for these dualities in M theory.

\subsection{Deformation quantization}
\label{sec:defq}

Deformation quantization \cite{Bayen:1978ha,Sternheimer:1998} 
is a reformulation of
the problem of quantizing a classical mechanical system as follows.
One first considers the algebra of observables $\CA$ of the classical
problem; if one starts with a phase space $M$ (perhaps the cotangent
space to some configuration space, but of course it can be more
general), this will be the algebra of functions $C(M)$.
One then finds a deformation of this algebra
in the sense of \rfs{ncflat}.3, i.e.
a family of algebras ${\CA}_{h}$ depending on a parameter $h$ which
reduces to $C(M)$ as $h\rightarrow 0$, and for which the leading term
in the star commutator in a power series expansion in $h$ is the
Poisson bracket,
\begin{equation}
\label{eq:defquant}
f * g - g * f = ih \{f,g\} + \ldots
\end{equation}
One can then reinterpret Heisenberg picture equations of motion such as
$$
{\p f\over\p t} = {i\over h}[H,f]
$$
as equations for observables in $\CA_h$ involving star commutators.

As discussed in \rfs{ncflat}.3, the Moyal product is precisely such
a deformation of the multiplication law of functions, and deformation
quantization would appear to be a very direct way to generalize the
construction of \nc\ field theory
on $\IR_\theta$ and $\IT_\theta$ to general \nc\ spaces.
Not only is it more general, it can be formulated geometrically,
without recourse to a specific coordinate system.
The primary input, the
Poisson bracket, can be specified by the choice of an antisymmetric
bi-vector field $\theta^{ij}$,
$$
\{f,g\} = \theta^{ij} \p_i f \p_j g
$$
such that $\{f,g\}$ satisfies the Jacobi identity.

For the further discussion, it will be useful to recall a bit of the
canonical formulation of classical mechanics and its underlying
mathematics.  A simplifying assumption which often holds is that
$\theta^{ij}$ is everywhere nondegenerate, in which case one speaks of
a manifold with symplectic structure.  The symplectic form is
$\omega_{ij}=(\theta^{ij})^{-1}$ (the matrix inverse at each point),
and the Jacobi identity for the Poisson bracket is equivalent to
$d\omega=0$.  More generally, $\theta^{ij}$ can degenerate at points or
be of less than full rank, in which case one speaks of a manifold
with a Poisson structure.

The simplest symplectic structure
is constant $\theta^{ij}$, in which case by a linear
transformation we can go to canonical coordinates as in
Eq. (\ref{eq:canon}).  Indeed, one can make a coordinate
transformation to canonical coordinates in any contractable
 region
in which $\theta^{ij}$ is everywhere nondegenerate (Darboux's
theorem).  One then distinguishes a special class of coordinate
transformations which preserve $\theta$, the canonical
transformations, or symplectomorphisms as they are called in
mathematics.  In infinitesimal form, these are defined by $\delta x =
\{S,x\}$ for some generating function $S$ (if $\pi_1(M)\ne 0$, one
must allow multivalued generating functions).  One can regard these
transformations as generating an infinite dimensional Lie algebra, for
which the symplectomorphisms are the corresponding Lie group.

Having defined the deformations of interest,
we can now discuss the question of whether they exist.
Constructing one involves
postulating additional terms in Eq. (\ref{eq:defquant})
at all higher orders in
$h$ to make the star product associative, $(f*g)*h = f*(g*h)$.
One might wonder if this procedure requires more data than just
the Poisson bracket, and a little reflection shows that it surely
will.  After all, we know of a valid star product for constant
$\theta^{ij}$, the Moyal product \refmoyalproduct, and if this were
the output of a general prescription depending only on $\theta^{ij}$,
we would conclude that the Moyal product intertwines with canonical
transformations;
i.e. if $f \rightarrow T(f)$ is a canonical transformation, we would have
$$
T(f * g) = T(f) * T(g) .
$$
However, a little experimentation with nonlinear canonical transformations
should convince the reader that this is false.  Thus the question
arises of what is the additional data required to define a star
product, and what is the relation between these different products.

Deformation quantization has been fairly well understood by
mathematicians and we briefly summarize the main results, referring to
\cite{Fedosov:1996,Kontsevich:1997vb}
for more information.  First,
deformation quantization always exists.  In the symplectic case this
was shown by \citeasnoun{DeWilde:1983a} and by
\citeasnoun{Fedosov:1994a}, who also constructed a trace.  For more
general Poisson manifolds, it was shown by
\citeasnoun{Kontsevich:1997vb}.  As we will discuss in \rfs{ncstring},
Kontsevich's construction is in terms of a topological string theory,
and has been rather influential in the physical developments already.

In the symplectic case, it is known that
the additional data required is a choice of symplectic connection,
analogous to the connection of Riemannian geometry but preserving the
symplectic form, $\nabla\omega=0$.  Unlike Riemannian geometry, this
compatibility condition does not determine the connection uniquely,
and acting with a canonical transformation will change the connection.
In this language, the special role of the canonical coordinates is
analogous to that of normal coordinates;
they parameterize geodesics of the symplectic connection, $\nabla x^i = 0$.

Although the precise form of the star product depends on the connection,
different choices lead to ``gauge equivalent'' star products, in the
sense that one can make a transformation
$$
G(f * g) = G(f) *' G(g)
$$
of the form
$$
G(f) = f + h g_1^i \p_i f + h^2 g_2^{ij} \p_i \p_j f + \ldots
$$
which relates the two.  Thus the two algebras are (formally) equivalent.

We now ask what is the relation between deformation quantization and
more conventional physical ideas of quantization, or other
mathematical approaches such as geometric quantization.  Now the usual
intuition is that the dimension of the quantum Hilbert space should be
the volume of phase space in units of $(2\pi\hbar)^d$, so a
finite volume phase space should lead to a finite number of states.

This fits with our earlier discussion of fuzzy $S^2$ but leads us to
wonder how we obtained field theory on $\IT_\theta$.  This appears to
be connected to the fact that $\pi_1(T^2)=\IZ^2\ne 0$, and a nice
string theory explanation of how this changes the problem can be found
in \citeasnoun{Seiberg:1999vs}; because a string can wind about
$\pi_1$, the phase space of an open string on $T^2$ is in fact
noncompact.

In any case, deformation quantization gets around all of these
questions in a rather peculiar way: the series expansions in $h$ one
obtains are usually ``formal'' in the sense that they do not converge,
not just when applied to badly behaved functions but for any
sufficiently large class of functions.  Thus they typically (i.e. for
generic values of $h$) do not define algebras of bounded operators,
and do not even admit representations on Hilbert space of the sort
which explicitly or implicitly lay behind many of our considerations
\cite{Fedosov:1996,Rieffel:1998cf}.

Given some understanding of this point, in the general case one also
needs to define a \nc\ metric $g_{ij}(x)$ to make sense of
\rfn{yangmills}; other field theory actions require even more
structure.

At this writing, the question of whether and when deformation
quantization can be used to define \nc\ field theory is completely
open.  Some suggestions in this direction have been made in
\cite{Asakawa:2000bh,Jurco:2001rq}.

\section{RELATIONS TO STRING AND M THEORY}
\label{sec:string}

We now discuss how \nc\ field theories arise from string theory and
M theory, and how they fit into the framework of duality.

Historically, the first use of noncommutative geometry in string
theory was in the formulation of open string field theory due to
\citeasnoun{Witten:1986cc}, which uses the Chern-Simons action in a
formal setup much like that of \rfs{kin}, with an algebra
$\CA$ defined using conformal field theory techniques, whose elements
are string loop functionals.  Noncommutativity is natural in open
string theory just because an open string has two ends, and an
interaction which involves two strings joining at their end points
shares all the formal similarities to matrix multplication which we
took advantage of in \rfs{ft}.

Although these deceptively simple but deep observations combined with
the existence of the string field framework strongly suggested that
\nc\ geometry has a deep underlying significance in string theory, it was
hard to guess just from this formalism what it might be.  Further
progress in this direction awaited the discovery of the Dirichlet
brane \cite{Dai:1989ua,Polchinski:1995mt}, which gave open strings a
much more central place in the theory, and allowed making geometric
interpretations of much of their physics.

Now contact between string theory and conventional geometry, as
epitomized by the emergence of general relativity from string
dynamics, relies to a large extent on the curvatures and field
strengths in the background being small compared to the string length
$l_s$.  Conversely, when these quantities become large in string
units, one may (but is not guaranteed to) find some stringy
generalization of geometry.

The simplest context in which \nc\ field theory as we described it
arises, and by far the best understood, is in a limit in which a large
background antisymmetric tensor potential dominates the background
metric.  In this limit, the world-volume theories of Dirichlet branes
become \nc\ \cite{Connes:1998cr,Douglas:1998fm}.  This can be seen
from many different formal starting points, as elucidated in
subsequent work, and it provides very concrete pictures for much of
the physics we discussed in sections \ref{sec:sol} and
\ref{sec:quant}.  It will also lead to new theories: \nc\ string
theories, and even more exotic theories such as open membrane (OM)
theory.

After reviewing a range of arguments which lead to \nc\ gauge theory,
we focus on its origins from the string world-sheet, following
\citeasnoun{Seiberg:1999vs}, who were the first to precisely state the
limits involved.  We also describe related arguments in topological
string theory, originating in work of \citeasnoun{Kontsevich:1997vb}.
We then give the string theory pictures for the solitonic physics of
\rfs{sol}, and other contacts with duality such as the
AdS/CFT proposal \cite{Maldacena:1998re}.  Finally, we discuss some of
the limits which have been proposed to lead to new \nc\ theories.

\subsection{Lightning overview of M theory}
\label{sec:mth}

Obviously space does not permit a real introduction to this subject,
but it is possible to summarize the definition of M theory so as to
provide a definite starting point for our discussion.  Details of the
following arguments can be found in
\cite{Polchinski:1996nb,Polchinski:1996na}.  Throughout this section,
we will try to state the central ideas at the start of each
subsection, though as the discussion progresses we will reach a point
where we must assume prior familiarity on the part of the reader.

A unified way to arrive at the various theories which are now
considered part of M theory is to start with one of the various
supergravity actions with maximal supersymmetry ($32$ supercharges in
flat spacetime), compactify some dimensions consistent with this
supersymmetry (the simplest choice is to compactify $n$ dimensions on
the torus $T^n$), find the classical solutions preserving as much
supersymmetry as possible ($16$ supercharges), make arguments using
supersymmetry that these are exact solutions of the quantum theory,
and then claim that in a given background (say, with specified sizes
and shapes of the torus), the lightest object must be fundamental, so
some fundamental formulation should exist based on that object.

Thanks to the remarkable uniqueness properties of actions with maximal
supersymmetry, in each case only one candidate theory survives even
the simplest consistency checks on these ideas, and thus one can be
surprisingly specific about these fundamental formulations and see
some rather non-trivial properties of the theory even in the absence
of detailed dynamical understanding.

The simplest and most symmetric starting point is eleven-dimensional
supergravity \cite{Cremmer:1978km}, a theory with no free parameters
and a single preferred scale of length, the Planck length $l_p$.  Its
fields are the metric, a spin $3/2$ gravitino, and a third rank
antisymmetric tensor potential, traditionally denoted $C_{ijk}$.  Such
a potential can minimally couple to a $2+1$-dimensional extended
object, and indeed a solution exists corresponding to the background
fields around such an object and preserving $16$ supersymmetries.  One
can also find a supersymmetric solution with magnetic charge emanating
from a $5+1$ dimensional hypersurface.  We can thus define M theory as
the well-defined quantum theory of gravity with the low energy
spectrum of this supergravity, containing solitonic ``branes,'' the
$2$-brane (or supermembrane) and $5$-brane, whose long range fields
(at distances large compared to $l_p$) agree with the solutions just
discussed.  These branes have tensions (energy per unit volume)
$c_2/l_p^3$ and $c_5/l_p^6$ as is clear by dimensional analysis;
arguments using supersymmetry and charge quantization determine
the constants $c_n$.

This is not a constructive definition and indeed one might doubt
that such a theory exists at all, were it not for its
connections to superstring theory.  The simplest connection is to
consider a compactification of the theory on $\IR^{9,1}\times S^1$
with a flat metric, circumference $2\pi R$ for the $S^1$, and no other
background fields.  One can derive the resulting ten-dimensional supergravity
by standard Kaluza-Klein reduction and find \IIa\ supergravity, a theory
which can be independently obtained by quantizing the \IIa\ superstring.
Indeed, the superstring itself can be identified with a $2$-brane with
one spatial dimension identified with (or ``wrapped on'') the $S^1$;
in the small $R$ limit this will look like and has precisely the action
for a string in ten dimensions.  This relation determines the
string length $l_s$, the average extent of a string, and the string
tension, $1/l_s^2 = c R/l_p^3$ (again with a known coefficient).

The claim is now that this compactification of M theory exhibits two limits,
one with $R >> l_p$ which can be understood as eleven-dimensional
supergravity with quantum corrections, and one with $R << l_p$ which can
be understood as a weakly coupled superstring theory.  Considerations
involving supersymmetry as well as many nontrivial consistency checks have
established this claim beyond reasonable doubt, as part of a much larger
``web of dualities'' involving all of the known string theories
and many of their compactifications.

The string theory limits are still much better understood than the
others, because the string is by far the most tractable fundamental
object.  One can use them to make a microscopic definition of certain
branes, the Dirichlet branes.  A Dirichlet brane is simply an allowed
end point for open strings.  The crucial generalization beyond the
original definition of open string theory is that one allows Dirichlet
boundary conditions for some of the world-sheet coordinates and this
fixes the endpoint to live on a submanifold in space-time.  For a
simple choice of submanifold such as a hyperplane, the world-sheet
theory is still free, so the physics can be worked out in great
detail.

The central result in this direction and the starting point for most
further considerations is that the quantization of open strings ending
on a set of $N$ D$p$-branes, occupying coincident hyperplanes in
ten-dimensional Minkowski spacetime,  leads to $p+1$-dimensional
$U(N)$ MSYM.
Its field content is a vector field, $9-p$ adjoint scalars and their
supersymmetry partner fermions, and its action is
the dimensional reduction of Eq. (\ref{eq:msym}).

A crucial point is the interpretation of the adjoint scalars.  Let us
denote them as hermitian matrices $X^i$; the action contains a potential
$$
V = -\sum_{i<j} \Tr [X^i,X^j]^2
$$
(the sign is there for positivity).
A zero energy configuration satisfies $[X^i,X^j]=0$ and is thus given by
a set of $9-p$ diagonal matrices (up to gauge equivalence).

The point
now is that the $N$ vectors of eigenvalues $X_{nn}^i$ must be identified
as the positions of the $N$ branes in the $9-p$ transverse dimensions.
This identification is behind most of the geometric pictures arising
from D-brane physics, and the promotion of space-time coordinates to
matrices is at the heart of the noncommutativity of open string theory.

As the simplest illustration of this, a generic configuration of
adjoint scalars will break $U(N)$ gauge symmetry to $U(1)^N$ by the
Higgs effect, giving masses $|X_{mm}-X_{nn}|$ to the $(m,n)$
off-diagonal matrix elements of the fields.  This corresponds to the
mass of a string stretched between two branes at the positions
$X_{mm}$ and $X_{nn}$ and we see that the Higgs effect has a simple
picture in terms of the geometry of an extra dimension.

This picture shows some resemblance to pictures from \nc\ gauge theory
appearing in \cite{Connes:1994a}, and this observation 
\cite{Douglas:1996vj,Ho:1997jr} led to a search for a
more direct connection between \nc\ gauge theory and D-brane physics.

\subsection{Noncommutativity in M(atrix) theory}
\label{sec:ncmth}

As we commented in \rfs{gt}.1, the simplest derivation of \nc\
gauge theory from a more familiar physical theory is to start with a
dimensionally reduced gauge theory (or ``matrix model'') action such
as \rfn{matrixmodel}, and find a situation in which the
connection operators $C_i$ of (\ref{eq:Cdefn}) obey the defining
relations of a connection on a module over a \nc\ algebra as given in
\rfs{gaugetop}, perhaps after specifying appropriate boundary
conditions or background fields.

A particularly significant theory of this type is maximally
supersymmetric quantum mechanics, the $p=0$ case of MSYM, with action:
\begin{equation}
\label{eq:maxqm}
S = \int dt\ \Tr\ \sum_{i=1}^9 (D_t X)^2 - \sum_{i<j} [X^i,X^j]^2 +
\chi^\dagger (D_t + \Gamma_i X^i) \chi .
\end{equation}
Here $D_t = \p/\p t + iA_0$, and varying $A_0$ leads to the constraint
that physical states be invariant under the action of $U(\CH)$

This action first entered M theory as a regulated form of the action
for the supermembrane, which as we discussed one might try to use as a
fundamental definition of the theory \cite{deWit:1988ig}.  How this
might work was not properly understood until the work of
\citeasnoun{Banks:1997vh}, who argued that a simpler and equally valid
way to obtain (\ref{eq:maxqm}) from string theory was to take the
theory of D$0$-branes in \IIa\ string theory and boost it along the
$x^{11}$ dimension to the infinite momentum frame.  Bound states of
these D$0$-branes would be interpreted as the supergravity spectrum,
while the original membrane configuration could also be obtained as a
non-trivial background.  An important feature of this interpretation
is that the compact eleventh dimension of our previous subsection does
not disappear in taking this limit; one should think of the resulting
theory as M theory compactified on a light-like circle.
See \cite{Taylor:2001vb} for a recent review.

In this framework, compactification on the torus $T^n$ is quite simple
to understand, in more than one way.  One can first compactify the
\IIa\ string and take a similar limit to obtain $n+1$-dimensional
MSYM, with a similar interpretation.  The case $n=1$ reproduces the
original string theory in a slightly subtle but convincing way 
\cite{Motl:1997th,Dijkgraaf:1997vv}
and this is one of the main pieces of confirming evidence for the
proposal.

Another approach, spelled out by \citeasnoun{Taylor:1997ik},
is to define toroidal compactification using the general theory
of D-branes on quotient spaces discussed in \rfs{grpalg}.
Letting $U_i=\gamma(g_i)$ for a set of generators of $\IZ^n$,
and taking $\CA=\Mat_n(\IC)$, this leads to the equations \rfn{tquotient},
$$
U_i^{-1} X^j U_i = X^j + \delta_i^j 2\pi R_i .
$$
These are solved by the connection \rfn{concovder},
and substituting into \rfn{maxqm} leads to MSYM on $T^n\times\IR$.

This construction admits a natural generalization, 
namely one can impose the relations
$$U_i U_j = e^{i\theta_{ij}} U_j U_i.$$
Again as discussed in \rfs{grpalg}, \rfn{tquotient} now defines a
twisted crossed product, and its solutions \rfn{tcovder}
are connections on the \nc\ torus.
Substituting these into \rfn{maxqm} leads directly to \nc\ gauge theory.  
This was how noncommutativity was first
introduced in M theory by \citeasnoun{Connes:1998cr}.

Having seen this possibility, one next must find the physical
interpretation of this noncommutativity.  Since M theory has no
dimensionless parameters, one is not allowed to make arbitrary
modifications to its definition but rather must identify all choices
made in a particular construction as the values of background fields.
Although $\theta$ had not been previously noticed as such a choice, it
appears naturally given the interpretation of M(atrix) theory as M
theory on a light-like circle, as a background constant value for the
components $C_{ij-}$ of the three-form potential, where $-$ denotes
the compact light-like direction.  This interpretation was supported
by comparing the expected duality properties of the \nc\ gauge theory
and of M theory in this background, a subject we return to below.

Having seen how M(atrix) theory can lead to noncommutativity and then
to a string, one wants to close the circle and show that
string theory can lead to noncommutativity on brane world-volumes,
from which \nc\ M(atrix) theory can be derived.  The \IIa\ string
interpretation of $C_{ij-}$ is as the ``NS $B$-field,''
a field which minimally couples to the string world-sheet as in
the action
\begin{equation}
\label{eq:wsa}
S = {1\over{4{\pi}{l_s^2}}} \int_{\Sigma}
\left( g_{ij} {\p}_{a} x^i {\p}^a x^j - 2{\pi}i{\ap} B_{ij}
{\e}^{ab} {\p}_a x^i {\p}_b x^j \right)
\end{equation}
We consider space-time $\IR^{9-n,1} \times T^n$ where the torus has
metric $g_{ij}=R^2\delta_{ij}$
and constant Neveu-Schwarz $B$-field $B_{ij}$.
In this case,
the term in Eq. (\ref{eq:wsa}) involving $B$ is an integral
of a total derivative, and will be nonzero either because of the nontrivial
topology of the torus or in the presence of a world-sheet boundary.

Contact with M(atrix) theory suggests that we
study the physics of D$0$-branes in this theory.
One way to proceed \cite{Douglas:1998fm}
is to apply a T-duality along one axis
(say $x^1$) of the torus, which one can show turns the D$0$-branes
into D1-branes extending along the $x^1$ axis, and the $T^2$ into
another $T^2$ with $B=0$ and metric defined
by the identifications $(0,0)\sim(l_s,0)\sim(\theta=B,\epsilon=V/l_s)$.

\begin{figure}
\epsfysize=3in
\epsfbox{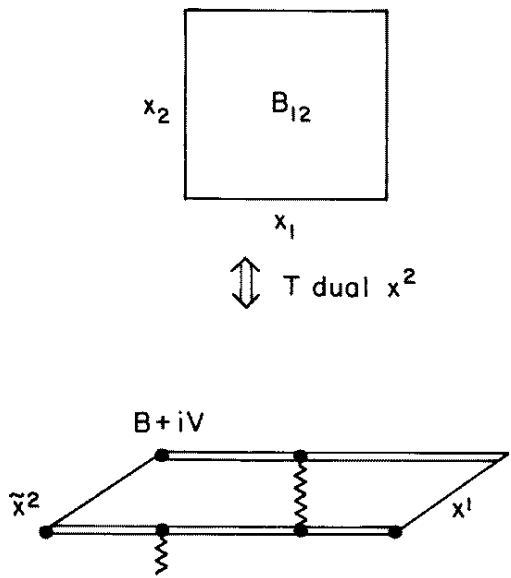}
\caption{T-duality to an anisotropic torus.}
\end{figure}

This gets rid of the $B$ field and thus one must be able to understand
the physics in conventional geometric terms.  The point is that while
general arguments lead to a $1+1$-dimensional gauge theory on the
world-volume of this brane, in the limit of small $\epsilon$, open
strings which wind about the $x^2$ dimension will also become light
and must be included in the action; their winding number $w^2$ becomes
a new component of momentum in $2+1$-dimensional gauge theory.  In
this anisotropic geometry, the two ends of a winding open string will
have different locations in $x^1$, with separation $\theta w^2$.  Thus
the fundamental objects turn out to be dipoles in exactly the sense
described in \rfs{ncflat}.2, with the corresponding \nc\ %
interactions.  This construction directly reproduces the quotient
construction of $\IT_\theta$ as $C(S^1)/\IZ$.

A world-sheet argument treating D$0$-branes on the original torus
leads to the same conclusion \cite{Cheung:1998nr}.
Now one must take the size $R$ of both axes of the torus small, and
keep winding strings in both directions.  The point now is that
since the term $\int B$ in Eq. (\ref{eq:wsa}) is a total derivative,
it contributes to the interaction
of two strings with winding numbers $(w_1,w_2)$ and $(w'_1,w'_2)$
about the two axes of the torus
by a phase proportional to the product $w_1 w'_2 - w_2 w'_1$,
directly producing \rfn{moyalphase}.  This argument can be
generalized to other string theory situations involving similar phases,
such as discrete torsion on orbifolds \cite{Vafa:1986wx}, 
and leads to the twisted crossed product with finite groups
\cite{Ho:1998xh,Douglas:1998xa}.

These derivations arose naturally in the consideration of M(atrix)
theory, and in this context there is a rather striking test one can
make.  The original derivation led to an identification of M theory
compactified on $T^d$ as a large $N$ limit of $d+1$-dimensional MSYM,
and the most basic prediction of this identification is that the two
theories share the same duality properties.  
\citeasnoun{Connes:1998cr} discussed compactification on $T^3$,
which according to M theory considerations must be invariant under
an action of the U-duality
group $SL(3,\IZ)\times SL(2,\IZ)$ on the moduli and brane spectrum.
The $SL(3,\IZ)$ can easily be identified with large diffeomorphisms of
the $T^3$, but the $SL(2,\IZ)$ symmetry is a prediction: in fact it is
just the $SL(2,\IZ)$ duality of $N=4$ SYM proposed by 
\citeasnoun{Montonen:1977sn}.

The generalized compactifications with $C_{-ij}\ne 0$ allow accessing
more dualities, namely those which can be seen in the limit of
compactification on a $T^{d+1}$ containing a light-like dimension.
For example, the \nc\ $2+1$ MSYM is predicted to have
$SL(2,\IZ)\times SL(2,\IZ)$ duality symmetry, the subgroup of
$SL(3,\IZ)\times SL(2,\IZ)$ preserving the choice of light-like direction.
Now the second, nonclassical
duality is manifest: it is the $SL(2,\IZ)$ Morita duality group of
$\IT^2_\theta$, and one can verify that the spectrum predicted in
\rfn{torusenergy} is invariant under this change of parameters.

This type of argument has been extended in many directions, and an
in-depth treatment would require another full length review, which
happily already exists \cite{Konechny:2000dp}.
In the M(atrix) theory context, it has been argued \cite{Verlinde:2001}
that by careful treatment
of the limit, one can extend the duality to a general $T^{d+1}$,
not necessarily preserving the light-like dimension.
In the string theory context, as we discuss below, one can see that
the nonclassical duality arises from T-duality, leading to the prediction
that \nc\ gauge theory on $T^d$ will have $SO(d,d;\IZ)$ duality,
which was the motivation behind the theorem of 
\citeasnoun{Rieffel:1998a} discussed in \rfs{math}.E.

\subsection{Noncommutativity in string theory}
\label{sec:ncstring}

The arguments we just gave established that limits of M theory and
string theory compactified on a torus will lead to \nc\ gauge theory,
realizing the intrinsic noncommutativity of open strings.  However the
torus is not the simplest \nc\ space; one is led to ask whether
noncommutativity can also emerge without compactification, leading to
gauge theory on $\IR^d_\theta$.  This question gained particular focus
after the discovery of \nc\ instantons on $\IR^d_\theta$ 
\cite{Nekrasov:1998ss}.

While the arguments of the previous section lead directly to this
result (to obtain $\IR^d_\theta$, one just adjusts the parameters to
make the string modes light, which takes the volume of $\IT^d_\theta$ to
infinity), many other string theory computations on $\IR^d$ with
general background $B$ had been done previously and noncommutativity
had not been seen.  Indeed, there are general arguments which lead
from open strings to \con\ gauge theory, making the new claim appear
paradoxical.

On the other hand, \citeasnoun{Kontsevich:1997vb} had argued that
deformation quantization and the Moyal product in particular could
come from ``open string theory,'' at least in a mathematical sense.
This was turned into an argument in topological string theory by
\citeasnoun{Cattaneo:1999fm} and in physical string theory by
\citeasnoun{Schomerus:1999ug}, proving that $\IR^d_\theta$ could indeed
emerge directly from world-sheet considerations.

The paradox was resolved by \cite{Seiberg:1999vs} who explained
how both \con\ and \nc\ descriptions could be correct, along the
general lines we already indicated in \rfs{kin}.D.4.  Their careful
treatment of the limit leading to \nc\ field theory has spurred
numerous further developments.

The first point to make is that on $\IR^d$ and unlike $T^d$,
a constant background potential $B_{ij}$ is pure gauge
and thus cannot affect closed string physics.
However, this is not true
in the presence of a Dirichlet brane extending along the directions $i,j$,
because the gauge transformation also acts on
its world-volume gauge potential $A$ as
\begin{equation}
\label{eq:branegt}
\delta B^{(2)} = d\lambda(x) ;\qquad \delta A = \lambda(x) .
\end{equation}
This can be seen by rewriting the total derivative term in
Eq. (\ref{eq:wsa}) as a boundary term, as is appropriate
in the presence of open strings, and adding the gauge coupling:
\begin{equation}
\label{eq:boundact}
S_b =
i \int_{{\p}{\Sigma}} \left(A_j(x) - B_{ij} x^{i}\right) {\p}_t x^j ,
\end{equation}
where ${\p}_t$ and ${\p}_{n}$
denote the tangential and normal derivatives along the boundary
${\p}{\Sigma}$ of the worldsheet ${\Sigma}$.  The transformation
Eq. (\ref{eq:branegt}) can be undone by an integration by parts.

Eq. (\ref{eq:branegt}) implies that the open string effective action
can only depend on the combination $F+B$, not $F$ or $B$ separately.
In particular, one can gauge $B$ to zero, replacing it by a
background magnetic field.  However, because of stringy effects,
in the limit $l_s^2 B >> 1$, this could lead to physics quite different
than that of a magnetic field in the usual Yang-Mills action.

To proceed,
we will need to make use of the standard relation between world-sheet
correlation functions of vertex operators, the S-matrix for string
scattering, and effective actions which can reproduce this physics
\cite{Green:1986,Polchinski:1998}.
The basic relation is that each local world-sheet operator $V_n(z)$ corresponds
to a space-time field $\Phi_n$.  Operators in the bulk of the world-sheet
correspond to closed strings, while operators on the boundary correspond
to open strings and thus fields which
propagate on the worldvolume of a D-brane.
A term in the effective Lagrangian
\begin{equation}
\label{eq:effac}
\int d^{p+1}x \sqrt{{\det}G} \Tr {\Phi}_1 ~ {\Phi}_2 \ldots {\Phi}_n
\end{equation}
is obtained as a correlation function
\begin{equation}
\label{eq:wscor}
\vev{ \int dz_1  V_1(z_1)\ \int dz_2  V_2(z_2)\ \ldots
\int dz_n  V_n(z_n) }.
\end{equation}
on a world-sheet $\Sigma$ with the disk topology, with operators $V_i$
at successive points $z_i$ on the boundary $\p\Sigma$,
integrated over all $z_i$ satisfying the same ordering as in \rfn{effac}.

Taking only vertex operators for the massless fields, one finds that
the leading $l_s\rightarrow 0$ limit of the S-matrix is reproduced by the
MSYM effective action.
It turns out that these considerations also lead to a simple universal
effective action which describes the physics of a D-brane with arbitrarily
large but slowly varying field strength,
the Nambu-Born-Infeld (NBI) action
\begin{equation}
\label{eq:nbi}
S_{NBI} = {1\over g_s l_s (2\pi l_s)^p} \int d^{p+1}x\ %
 \sqrt{\det(g+2\pi l_s^2(B+F))} .
\end{equation}
Here $g_s$ is the string coupling, and $g$ is the induced metric on
the brane world-volume.  See \citeasnoun{Schwarz:2001ps} for its
supersymmetrization, and other references on this topic.

The original string computations leading to the Born-Infeld action
were done by \citeasnoun{Fradkin:1985qd} and \citeasnoun{Abouelsaood:1987gd}
in the context of type I open string theory, i.e. the case $p=9$.  For
$p<9$, the induced metric $g$ contains the information about the
embedding of the brane, and substituting $F=0$, \rfn{nbi} reduces
to the Nambu action governing its dynamics.  Finally, the prefactor is
the D-brane tension as computed from string theory.  Besides detailed
computation, there are simple physical arguments for the Born-Infeld
form \cite{Bachas:1996kx,Polchinski:1998}.

This action summarizes essentially all weakly coupled and ``weakly
stringy'' physics of a single D-brane and is even valid in the
large $l_s^2(B+F)$ limit, so one might at first hypothesize that
it is valid in the limit of large $B$ we just discussed, without
need of noncommutativity.  However, it is not in general valid for
rapidly varying field strengths $l_s \p F \sim 1$, nor is its
nonabelian generalization understood.  Thus we can
reconcile our earlier arguments for noncommutativity with the
NBI action if $\theta < l_s^2$, as all of the new physics
we observed would be associated with length scales at which the
NBI action broke down.

\subsubsection{Deformation quantization from the world-sheet}

The key point in arguing that string theory will lead to \nc\ field
theory is to see that a correlation function \rfn{wscor} will obtain
the phase factors \refcyclicvertex.
Since this is a product of terms for each successive pair of fields,
it should also be visible as a phase \rfn{moyalphase}\
in the operator
product expansion of two generic boundary operators carrying
momenta $k$ and $k'$, say
\begin{equation}
\label{eq:basicope}
V_k(z_1) V_{k'}(z_2) \rightarrow (z_1-z_2)^{\Delta_{k+k'}-\Delta_k-\Delta_{k'}}
e^{-{i\over 2}\theta^{ij} k_i k'_j} V_{k+k'} + \ldots
\end{equation}
with
$$
V_k(z) = :e^{ik\cdot x(z)}:
$$
or any operator obtained by multiplying this by conformal fields $\p x$,
fermions in the superstring, etc.

Since the action \rfn{wsa} is quadratic, the world-sheet physics is
entirely determined by the propagator $\vev{x^i(z)~ x^j(w)}$.
The boundary conditions which follow from varying the action
\rfn{wsa} are:
\begin{equation}
\label{eq:bndc}
g_{ij} {\p}_{n} x^j + 2{\pi} i B_{ij} {\ap} {\p}_t x^j
\vert_{{\p}{\Sigma}} = 0
\end{equation}
Now, taking $\Sigma$ to be
the upper half-plane with the coordinate $z= t + iy$, $y >
0$ we find the boundary propagator to be
\begin{equation}
\label{eq:bnprp}
\langle x^i (t) x^j (s)  \rangle
= - {\a}^{\prime} G^{ij} \log (t - s)^2 +{i\over 2}
{\t}^{ij} {\e}( t - s)
\end{equation}
where ${\e}(t) = -1, 0, +1$ for $t < 0, t = 0 , t > 0$
respectively, and
\begin{eqnarray}
\label{eq:opmb}
& G^{ij} = \left( {1\over{g + 2{\pi} {\ap}
B }}\right)^{ij}_{S} \\
\label{eq:optheta}
& {\t}^{ij} = 2{\pi} {\ap} \left(
{1\over{g + 2{\pi}{\ap} B}} \right)^{ij}_{A}
\end{eqnarray}
where $S$ and $A$ denote the symmetric and antisymmetric parts
respectively. From this expression we deduce :
\begin{equation}
\label{eq:cmtr}
[ x^i , x^j] : = T(x^i ( t- 0) x^j(t) - x^i ( t+0)
x^j(t)) = i {\t}^{ij}.
\end{equation}
Thus the end-points of the open
strings live on a noncommutative space with
$$ [ x^i, x^j ] =
i {\t}^{ij} $$
with ${\t}^{ij}$ a constant antisymmetric matrix.
Similarly, \rfn{basicope} becomes
\begin{equation}
\label{eq:ope}
{V_p} (t)
V_{q} (s) = ( t - s)^{2 {\ap} G^{ij} p_{i} q_{j}}
e^{-{i\over{2}}{\t}^{ij} p_{i } q_{j}} V_{p + q} (s)
\end{equation}
Indeed,
in this free world-sheet theory, it is no harder to compute the analogous
phase factor for an $n$-point function: it is precisely \refcyclicvertex.

In the formal limit $g_{ij}\rightarrow 0$, one finds from \rfn{opmb} and
\rfn{optheta} that $G^{ij}=0$ and $\theta^{ij}=2\pi\ap(B_{ij})^{-1}$.
Thus the dependence on the world-sheet
coordinates $s$ and $t$ drops out (the vertex operators have dimension
zero), and the OPE reduces to a conventional multiplication law.
For \rfn{ope}, this is the Moyal product \refmoyalproduct, and
by linearity this extends to the product of two general functions.

We have again found that background $B_{ij}$ leads to
noncommutativity.  A precise connection to the previous discussion of
toroidal compactification can be made by applying T-duality to turn
the D$2$-branes of the present discussion into D$0$-branes, and then
taking the zero volume limit of the torus.  This T-duality leads to
the relation $\theta^{ij}=2\pi\ap(B_{ij})^{-1}$.  However, the present
argument also works for $\IR^d_\theta$.

A rather similar argument shows that world-volumes of D-branes in the
Wess-Zumino-Witten model are described by field theory on ``fuzzy
spheres'' as described in \rfs{othernc} \cite{Alekseev:1999bs}.  An
interesting difference is that here $dB\ne 0$ and the corresponding
algebra is not associative, except in suitable limits.

\subsubsection{Deformation quantization from topological open string theory}

The idea that by considering only vertex operators of dimension zero,
a vertex operator algebra such as Eq. (\ref{eq:ope}) will reduce to an
associative algebra, is rather general and is best thought of in the
framework of topological string theory
\cite{Witten:1988ze,Dijkgraaf:1998a}.  For present purposes, these are
string theories for which correlation functions depend not on the
locations of operators but only on their topological arrangement on
the world-sheet, which is exactly the property we used of the 
$g_{ij}\to 0$ limit of \rfn{ope}.

One can use topological string theory to construct a deformation
quantization corresponding to a general Poisson structure,
generalizing the discussion we just gave.
We summarize this, following the work of \citeasnoun{Baulieu:2001}.
See also \cite{Cattaneo:2001bp}.

We start with the action \rfn{wsa}, not assuming that $g_{ij}, B_{ij}$
are constant and rewrite it in the first order form:
\bea
\label{eq:wsac} & S = {1\over{4\pi\ap}} \int_{\Sigma} ( g_{ij}
{\p}_{a} x^i {\p}^a x^j - 2{\pi}i{\ap} B_{ij} {\e}^{ab} {\p}_a x^i
{\p}_b x^j ) = \crt & \int p_{i} \wedge {\rm d} x^i - {\pi}{\ap}
\, G^{ij} p_{i} \wedge \star p_{j} - {\half}{\t}^{ij} p_{i}\wedge
p_{j}
\eea
where
$$
2{\pi}{\ap} G + {\t} = 2{\pi}{\ap} \left( g +
2{\pi}i{\ap} B \right)^{-1} $$
Now take the ${\ap} \to 0$
limit keeping ${\t}$ and $G$ fixed. The remaining part $\int p
\wedge {\rm d} x + {\half} {\t} p \wedge p$ of the action
\rfn{wsac} exhibits an enhanced gauge symmetry 
\cite{Cattaneo:1999fm},
\bea \label{eq:gsm} p_i \mapsto p_i - {\rm d}
{\l}_i - {\p}_i {\t}^{jk} p_j \, {\l}_k, \qquad x^i \mapsto x^i +
{\t}^{ij}{\l}_j
\eea
To quantize, this symmetry must be gauge fixed, which can be done
by standard BV procedures, leading to a topological string theory
with some similarities both with the type A and type B sigma
models.  Its field content is
conveniently described by promoting $p_i$ and $x^i$ to twisted
superfields, with an expansion in world-sheet differential
forms $d\sigma^i$ with components of all degrees $0,1,2$.
The original fields are the $0$-form part of $x^i$ and the $1$-form
part of $p_i$; the other components are the additional ghosts and
auxiliary fields of the BV framework.

A basic observable in this theory is
the three-point function on the disc,
\bea
\label{eq:crln}
\langle f(x(0))
g(x(1)) \left[ h(x)\chi\ldots \chi \right] (\infty) \rangle_{\t,
\Sigma = {\rm disc}} = \int_{X} f \star g \, h
\eea
for $f,g \in C(X)$, $h \in {\Omega}^{{\rm dim}X}(X)$.
As the notation indicates, this will be identified with the star
product of a deformation quantization associated to $\theta$.
This can be seen by developing the perturbation series in powers
of $\theta$; each term in the expansion can be expressed as
an explicit sum over Feynman diagrams, producing Kontsevich's
construction of deformation quantization.

An important advantage of this approach is that many properties of
the formalism have simple arguments from string theory.  In particular,
the associativity of the $\star$-product defined by \rfn{crln}
follows in a sense from associativity of the OPE.  This is best expressed
in terms of a more general set of Ward identities obeyed by
the string amplitudes in the theory, which allow making contact with
and generalizing the discussion in \rfs{defq}.

As an example of a new result derived from \rfn{wsac}, we give an
expression for the ${\t}$-deformed action of infinitesimal
diffeomorphisms $\delta x^i = v^i(x)$
on functions and covariant derivatives \cite{Baulieu:2001},
which can be used to make \rfn{genaction} generally covariant. 
One defines these in terms of the \con\ action on functions
$\delta f = v^i\p_i f$, with ${\t}$ transforming as a \con\ tensor,
$\delta {\t}^{ij} = v^k {\p}_k {\t}^{ij} - {\t}^{ki} {\p}_k v^j +
{\t}^{kj}{\p}_k v^i$.
In the TFT, this becomes
$$
\delta f = 
U_{v} f (x) 
= \langle f(x(0)) \oint \left( p_i v^i (x) \right)^{(1)} 
\left[ {\d} (x({\infty}) - x) \right] \rangle_{{\t}, {\Sigma}}
$$
where the integral is taken over an arc surrounding the point $0$ 
and ending on the boundary of the disc ${\Sigma}$.
This leads to
$$
U_{v} f (x) 
= v^i {\p}_i f(x) + {\t}^{kl} {\p}_k v^i {\p}^2_{li} f (x) + \ldots .
$$

\subsubsection{The decoupling limit}

Many of the deeper aspects of the connection to string theory require
a more careful treatment of the decoupling limit leading to \nc\ field
theory, as was first made by \citeasnoun{Seiberg:1999vs}.  

String theories have many more perturbative degrees of freedom than
field theories, so to get field theories one looks for controlled
limits in which almost all of these degrees of freedom go away,
usually by sending their masses to infinity.  In the context of
D-branes, the generic such limit takes $l_s\to 0$ and thus the string
tension $T=1/l_s^2$ to infinity, and simultaneously takes the
transverse distance $L$ between branes to zero holding $L T$ fixed.
This is the energy scale of the lightest open strings stretched
between branes, so in this limit we keep these degrees of freedom
while sending excited string energies to infinity as $1/l_s$.
Finally, one rescales the coupling constant to keep it fixed in the
limit.  This leads to a field theory of the lightest open strings,
which for flat D-branes will be MSYM.  Similar limits in other brane
theories can lead to more exotic results, as we will mention below.

However there are other massless states in the string theory as well,
the closed strings which lead to the gravitational sector, and we need
to argue that these decouple.  A naive but often correct argument for
this is that their couplings are gravitational and are suppressed by a
factor $G_N E$ which wlil also go to zero in this limit.  This
requires detailed consideration in examples, however.  It will turn
out to be true in the case at hand at least up to $3+1$ dimensions; we
will discuss a potential subtlety below.

The key new point in the present context is that the masses of open
strings in this limit are determined by the metric $G^{ij}$ defined in
\rfn{opmb}, which has nontrivial $B$ dependence.
This follows from the standard string theory relation between the mass
of a state and the world-sheet dimension of the corresponding vertex 
operator, which (as is visible in \rfn{ope}) is controlled by $G^{ij}$.
Thus the decoupling limit,
sometimes referred to as the Seiberg-Witten limit in this context,
takes $l_s\to 0$ while holding $G$ and ${\t}$ fixed.

The nontrivial relation between the original metric $g_{ij}$ and
$G^{ij}$ will show up at many points in the subsequent discussion.
One refers to $g_{ij}$ as the ``closed string metric'' and $G^{ij}$
as the ``open string metric'' as they each govern the kinetic terms
and the energies of the lightest states in their respective sectors.

Once this realization is made, the subsequent steps in the derivation
of the D-brane world-volume action go through without major changes
from the \con\ case, leading to \nc\ MSYM and even the \nc\ NBI action,
defined by taking \rfn{nbi}, replacing products with star products,
and setting the metric to $G^{ij}$ and $B=0$.

The remaining step is to determine the prefactor and thus the gauge coupling.
This follows directly if we accept that the Seiberg-Witten map of
\rfs{kin}.D.4 between \con\ and \nc\ gauge theories maps the \con\ action
\rfn{nbi} into the \nc\ action \rfn{nbi}, as we can just specialize to
$F=\hat F=0$.  This leads to the relation
\begin{equation}
\label{eq:couplingrel}
 {\sqrt{\det(G)}\over G_s l_s (2\pi l_s)^p}  
=
 {\sqrt{\det(g+2\pi l_s^2 B)} 
  \over g_s l_s (2\pi l_s)^p} .
\end{equation} 
which determines an ``open string coupling constant'' $G_s$ and
the corresponding \nc\ gauge coupling.  

The decoupling limit now will
take $l_s\to 0$ in Eqs. (\ref{eq:opmb}), (\ref{eq:optheta})
and (\ref{eq:couplingrel}), scaling the original string coupling as
$g_s \sim l_s^{(3-p+r)/4}$ to end up with \nc\ Yang-Mills
theory with finite parameters,
\be
\label{eq:gaugerel}
\theta = B^{-1}; G_{ij} = (2\pi l_s^2)^2 B_{ik} B_{jl} g^{kl} .
\ee

\subsubsection{Gauge invariance and the SW map}

An important point which can be seen by carrying out this discussion
more explicitly is the precise point at which \con\ gauge invariance
is replaced by \nc\ gauge invariance.  As we mentioned in 
\rfs{gt}.4,
there is a very general world-sheet argument for \con\ gauge invariance,
and indeed this argument is not incorrect; rather, one obtains \nc\ gauge
invariance by choosing different conventions, and there is a formal
equivalence between \con\ and \nc\ gauge theories.
\cite{Seiberg:1999vs,Andreev:1999pv,Seiberg:2000zk}.

The origins of gauge invariance can be seen in open bosonic string
theory.  In this theory, the vertex operator for a gauge boson is
\begin{equation}
\label{eq:gfvo}
e_{i}(p) :{\p}_t x^i e^{ip \cdot x} :
\, \longleftrightarrow \, A = A_{i} (x) dx^i
\end{equation}
as was already implicit in \rfn{boundact}.

The abelian gauge invariance,
\begin{equation}
\label{eq:nas}
{\d} A_{j} = \p_j {\ve} ,
\end{equation}
then follows by varying $A_j$ in \rfn{boundact}, by taking the
integral of the total derivative $\int dz \p_t( \ve)$ as zero.
Extra terms can appear in \rfn{nas}
if there are divergences when the operator
under consideration (say $V_i$)
coincides with its neighbors $V_{i-1}$ and $V_{i+1}$.
In Yang-Mills theory, this leads to contact
terms $V_{i-1} \ve - \ve V_{i+1}$ which become the nonlinear terms in
the gauge transformation law.

The previous discussion makes it very plausible that such an argument
will carry through in the \nc\ case with the star product appearing as
in \rfn{ope}.  However, making this argument precise requires choosing
a cutoff on the world-sheet, and different choices at this step can
lead to different results.  If one point-splits and uses the
propagator \rfn{bnprp}, one obtains the star product, but one can also
find prescriptions in which the term proportional to $\theta^{ij}$
goes away in the coincidence limit, leading to \con\ gauge invariance
instead.  In particular, this will be the result if one defines a
propagator and cutoff at $B=0$, treating the $B$ term in \rfn{wsa}
using world-sheet perturbation theory.

Physics cannot depend on this choice and in general a change of
renormalization prescription on the string world-sheet corresponds to
a field redefinition in space-time.  By the preceding arguments, this
field redefinition must be the Seiberg-Witten map of \rfs{gt}.4.  The
two descriptions are therefore equivalent, at least in some
formal sense, but they are each adapted to different regimes,
with \con\ gauge theory simpler for small $B$ and \nc\ gauge theory
simpler for large $B$.

Once we understand that the resulting gauge invariance depends on the
choice of world-sheet regularization, we can consider choices that
lead to different star products.  A simple choice to consider is one
which leads to the same star product \rfn{moyalphase}, but defined
using a parameter $\theta$ which does not satisfy \rfn{optheta}, which
would be obtained by treating part of the $B$ term in \rfn{wsa} as a
perturbation.  Denote this part as $\Phi$; it will enter \rfn{nbi}
as did $B$, so we would obtain a description in terms of a \nc\ action
depending on $\hat F+\Phi$.

This description can be obtained from string theory by a simple
generalization of the discussion above.  The result
is again Eqs. (\ref{eq:opmb}) and (\ref{eq:optheta}) but now with
an antisymmetric term in the metric on both sides (analogous to the
$B$ term in \rfn{wsa}).  Combining this with \rfn{couplingrel} to
determine the gauge coupling, we obtain
\bea 
& \left( {1\over{G + 2{\pi} l_s^2 \Phi }}\right)_{S} 
= \left( {1\over{g + 2{\pi} l_s^2 B }}\right)_{S} 
\\
& \left( {1\over{G + 2{\pi} l_s^2 \Phi }}\right)_{A} 
+ {1\over2\pi l_s^2} \theta
= \left( {1\over{g + 2{\pi} l_s^2 B }}\right)_{A} \\
& G_s = g_s 
\left({\det(G+2\pi l_s^2 \Phi)\over\det(g+2\pi l_s^2 B)}\right)^{1/2}
\eea
which determines the parameters in the \nc\ form of \rfn{nbi},
\begin{equation}
\label{eq:swnbi}
S_{NCNBI} = {1\over G_s l_s (2\pi l_s)^p} \int \Tr
 \sqrt{\det(G+2\pi l_s^2(\hat F+\Phi))} .
\end{equation}
Using \rfn{swmap}, one can show that for slowly varying background fields
$l_s^2 F << 1$,  the descriptions with different values of $\theta$
are equivalent.  

In the special case of $\Phi=-\theta^{-1}$, these relations simplify
even further to \rfn{gaugerel}, even at finite $l_s$.  As mentioned in
\rfs{ncflat}.1, this is what one gets if one defines the derivatives
as inner derivations, so this is what emerges naturally from matrix
model arguments \cite{Seiberg:2000zk}.

More generally, it has been shown that the techniques of Kontsevich's
deformation quantization can be used to write a Seiberg-Witten map for
arbitrary (non-constant) $\theta$ \cite{Jurco:2000fs}.


\subsection{Stringy explanations of the solitonic solutions}
\label{sec:strsol}

So far we have only discussed the simplest arrangement of Dirichlet
branes, $N$ parallel branes of the same dimension.  There are a
bewildering variety of more complicated possibilities, with branes of
lower dimension sitting inside of those of higher dimension,
intersecting branes, curved branes and so forth.

Not all of the possibilities are actually distinct, however.  For
example, a configuration of lower dimensional D-branes sitting inside
a higher dimensional D-brane, is topologically equivalent to and can
often be realized as a limit of a smooth configuration of gauge fields
on the higher dimensional brane.  Another example is that an
intersecting brane configuration, say of branes $A$ and $B$, can often
be described as a single object, a nontrivial embedding of brane $A$
in the higher dimensional space and thus by some configuration of
scalar fields on brane $A$.  Thus, the solitons and instantons in
D-brane world-volume gauge theory themselves have interpretations as
D-branes.  We refer to \cite{Polchinski:1996na,Polchinski:1998} for
an introduction and overview of this subject.

A great deal of this structure survives the limit taking string theory
to \nc\ field theory, leading to stringy pictures for the gauge theory
solutions of \rfs{sol}.  What is quite striking is that
the agreement is not just qualitative but even quantitative, with
brane tensions and other properties agreeing, apparently for reasons
other than supersymmetry
\cite{Dasgupta:2000ft,Gross:2000wc,Harvey:2000jt}.
Establishing this detailed agreement of
course requires a good deal of string theory input and we refer to
\citeasnoun{Nekrasov:2000ih} and \citeasnoun{Harvey:2001yn} 
for detailed reviews of this class
of results, but provide an introduction here.

The simplest examples are the D$(p-k)$-branes embedded in D$p$-branes.
There is a general result to the effect that a collection of $N$ D$p$-branes
carrying gauge fields with Chern character
$$
{\ch}(F) = \Tr e^F = N + \ch^{(1)}_{i_1i_2} + \ch^{(2)}_{i_1i_2i_3i_4} + \ldots
$$
carries the same charges as (and is topologically equivalent to)
a collection with no gauge fields, but with additional D-branes whose
number and orientation is given by the quantized values of the
Chern characters.  For example, $\ch^{(1)}_{ij}$ counts D$(p-2)$-branes
which are localized in the $i$ and $j$ dimension, and so forth.

Carrying this over to the \nc\ theory in the obvious way, the fluxon
of III.B is an embedded D$(p-2)$-brane, and the instanton of III.C is an
embedded D$(p-4)$-brane.  According to this identification and the
results of \rfs{sol}, the D$(p-2)$ should be unstable to
decay, while the D$(p-4)$ should be stable, and this is borne out by
computation in string theory; the destabilizing mode in 
\rfn{fluxoninstab} is a ``tachyonic open string.''

The agreement is much more detailed and striking than this.  Another
qualitative point is that $m$ coincident D$(p-2)$-branes must themselves
carry $U(m)$ gauge symmetry; this is manifestly true of the $m$-fluxon
solution \rfn{genvortex} with $U=S^m$.  
A similar discussion can be made for higher
codimension branes and even for processes such as the annihilation of
a D-brane with an anti-D-brane, which are quite difficult to study in
the original string theory.  A very general relation between the
topology of gauge field configurations in \nc\ gauge theory, and the
classification of D-branes in string theory, has been found in
\cite{Harvey:2000te}, based on the K-theory of operator algebras.

Even better, paying careful attention to conventions, one finds that
\rfn{stension} exactly reproduces the tension of the D$(p-2)$-brane in
string theory.  It is somewhat surprising that the \nc\ field theory
limit would preserve any quantitative properties of the solutions.
The agreement of the tension is related to a deep conjecture of 
\citeasnoun{Sen:1998a}, as explained by \citeasnoun{Harvey:2000jt}.
\citeasnoun{Witten:2000nz} has shown how the \nc\ field theory and
these arguments can be embedded in the framework of string field
theory, and much recent progress has been made in this direction
\cite{Gross:2001rk,Rastelli:2001rj,Shatashvili:2001ux}.

The monopole in \rfs{sol}\ provides another example, which is now
related to intersecting brane configurations.  One can start from
the $B=0$ description of a monopole in the $3+1$ $U(2)$ MSYM gauge
theory of $2$ D$3$-branes, which is a D$1$-brane ``suspended
between'' the branes, i.e. extending in a transverse dimension
perpendicular to the D$3$-branes, and with one end on each brane.
One can show that this configuration not only reproduces the
magnetic charge of the solution, but even obtain the full Nahm
formalism from D-brane world-volume considerations 
\cite{Diaconescu:1997rk}.

If one turns on $B$ along the D3-branes, the essential point is
now that the D1-brane is no longer perpendicular, it ``tilts.''
The endpoints of the D1-brane are magnetically charged with
respect to the U(1) gauge fields on the D3-branes. The background
B-field acts as a magnetic field, which pulls these charges apart.
Tilting of the D1-brane makes its tension work in the opposite
direction and stabilizes the configuration. This qualitative
picture is confirmed by the profile of the explicit analytic
solution \cite{Gross:2000wc}.

Finally, let us mention the work of \citeasnoun{Myers:1999ps}, which
shows that D$p$-branes in background Ramond-Ramond fields form bound
states which can be thought of as D$(p+2)$-branes whose world-volume is
a product space with a fuzzy $S^2$.  This noncommutativity plays a
number of interesting roles in M theory \cite{Myers:2001ks} and its
slightly different origin poses the challenge to find a broader
picture which would incorporate it into our previous discussion.

\subsection{Quantum effects and closed strings}
\label{sec:strq}

The string theory derivation of \nc\ gauge theory allows one to
compute its coupling to a background metric, at least for small
variations of the flat background, by adding closed
string graviton (and other) vertex operators in the world-sheet computations
discussed above.  
Results of this type are given by many authors; we mention
\cite{Das:2000ur,Liu:2001ps,Okawa:2000sh,Okawa:2001if}
as representative.  Analysis of these couplings to Ramond-Ramond closed
string fields \cite{Liu:2001pk,Okawa:2001mv,Mukhi:2001vx}
was an important input into the result \rfn{swmapthree}.

As a particularly simple example, the coupling to the graviton
defines a stress-energy tensor in \nc\ gauge theory,
which turns out to be precisely \rfn{gstress}.

The general type of UV/IR relation we discussed at length in
\rfs{quant} is very common in string theory.  Perhaps its simplest
form is visible in the computation of the annulus world-sheet with one
boundary on each of a pair of D-branes.  This amplitude admits two
pictures and two corresponding field theoretic interpretations: it
can be thought of as describing emission of a closed string by one
D-brane and its absorption by the other, a purely classical interpretation,
and it can equally well be thought of as the sum of one-loop diagrams over
all modes of the open string stretched between the pair of branes,
a purely quantum interpretation.  World-sheet duality implies that the
two descriptions must be equal.

As discussed by \citeasnoun{Douglas:1997yp}, this leads very generally
to the idea that for D-branes at ``substringy'' distances (i.e. with
separation $L$ as above satisfying $L << l_s$), conventional gravity
is replaced by quantum effects in the world-volume gauge theory.  In
general this leads to different predictions from Einstein gravity or
supergravity, but in special circumstances (e.g. with enough
supersymmetry or in the large $N$ limit) the substringy predictions
can agree with gravity.  Conversely, since $L/T$ is a world-volume
energy scale, taking $L>>l_s$ accesses the UV limit of gauge theory
amplitudes, and one sees that these are replaced by the IR limit of
the gravitational description.

Although the decoupling limit takes $L\rightarrow 0$ and is thus in
the substringy regime, one can ask whether nevertheless this potential
connection to gravity can shed light on the nature of UV/IR mixing.

A strong sense in which this could be true would be if the new IR
divergences could be described by adding additional light degrees of
freedom in the effective theory, which would be directly analogous to
the closed strings.  This picture was explored in
 \cite{Minwalla:1999px,VanRaamsdonk:2000rr}.  For example, if we
introduce a field $\chi$ describing the new mode, we can then
reproduce a singularity such as \rfn{phifouronePI} by adding

$$
\int \chi (\theta^{ik} \theta^{jl} G_{kl}\p_i\p_j)^{{d\over 2}-1} \chi
+ \chi \phi
$$
to the effective Lagrangian.  This particular example looks quite
natural in $d=4$; one can also produce different power laws by
postulating that $\chi$ propagates in a different number of dimensions
than the original gauge theory, as is true of the closed strings in
the analogy.

The main observation is then that, comparing with \rfn{gaugerel},
the kinetic term for $\chi$ contains precisely the closed string metric
$g^{ij}$, which is compatible with the idea.  Indeed, there are cases
in which this interpretation seems to be valid
\cite{Rajaraman:2000dw}, namely those in which the
divergence is produced by a finite number of closed string modes because
of supersymmetric cancellations, e.g. 
as in $\CN=2$ SYM \cite{Douglas:1996du}.

In general, however, the closed string picture is more complicated
than this and one cannot identify a simple set of massless modes which
reproduce the new IR effects
\cite{Gomis:2000bn,Andreev:2000rm,Kiem:2000wt,Bilal:2000bk}.
Furthermore the effective field theory required to reproduce higher
loop effects does not look natural \cite{VanRaamsdonk:2000rr}.

One can also argue that if this had worked in a more complicated
situation, it would signal the breakdown of the decoupling limit we
used to derive the theory from string theory \cite{Gomis:2000bn}.
This is because exchange of a finite number of closed string modes
would correspond to exchange of an infinite number of open
string modes, including the massive open strings we dropped in the limit.
Explicit consideration of the annulus diagram however shows that these
massive open strings do not contribute in the limit.

\subsection{AdS duals of \nc\ theories}
\label{sec:ads}

One of the beautiful outcomes of string theory is the description
of strongly coupled large $N$ gauge theories by the AdS/CFT correspondence
\cite{Maldacena:1998re,Aharony:1999ti}.
In particular, one expects
that the $D3$-brane realization of \nc\ gauge theory has a
supergravity dual in the large $N$, strong 't Hooft
coupling limit. This was found in
\cite{Hashimoto:1999ut,Maldacena:1999mh} and takes the form
\bea \label{eq:backnc} & {\rm d} s^2 = {\ap}
\sqrt{\l} \times \crt & \left[ U^2  ( - dt^2 + dx_1^2 ) +
{{U^2}\over{1 + {\l} {\Delta}^4 U^4}} ( dx_2^2 + dx_3^2 )
+{{dU^2}\over{U^2}} + d{\Omega}_5^2 \right] \crt & U^2 =
{1\over{{\l}{\ap}^2}} \left(  x_4^2 + \ldots + x_9^2  \right) \crt
 & e^{\phi} = {{\l}\over{4\pi N}}
{1\over{\sqrt{1 + {\l} {\Delta}^4 U^4}}} \crt & B = - {\ap} {{{\l}
{\Delta}^2 U^4}\over{1 + {\l} {\Delta}^4 U^4}}\  dx_2 \wedge dx_3
\eea 
It is dual to the ${\CN}=4$ $U(N)$ \nc\ gauge theory, with
the noncommutativity $[x_2, x_3] \sim {\Delta}^2$, and 't Hooft
coupling ${\l} = g_{YM}^2 N$. The transverse geometry is a
five-sphere of radius $R$, $R^2 = {\ap} \sqrt{\l}$. For
$U \ll {1/({\Delta {\l}^{1/ 4}})}$ the solution \rfn{backnc}
approaches the $AdS_5 \times S^5$ supergravity background, dual to
ordinary large $N$ ${\CN}=4$ super-Yang-Mills theory. However, for
large $U$, corresponding to the large energies in the gauge
theory, the solution differs considerably: the dilaton flows, the
$B$-field approaches a constant in $23$ directions, and the
$23$-directions collapse.

The large $U$ limit of anti de-Sitter space is a time-like boundary,
and in the usual AdS/CFT correspondence the boundary values of fields
are related to couplings of local operators in the gauge theory.  The
drastic modifications to \rfn{backnc} in this region have been argued
by many authors to be associated with the lack of conventional local
gauge invariant observables.  
\citeasnoun{Das:2000md} have argued
that the proportionality \rfn{dipolelength} between length and
momenta, characteristic of the open Wilson loop and exploited in
\rfn{wilsontwo}, emerges naturally from this picture.

See \cite{Danielsson:2000ze,Li:1999am,Berman:2000jw,Russo:2000mg} 
for further physics of this correspondence, and \cite{Elitzur:2000ps}
for a discussion of duality and Morita equivalence.

\subsection{Time-like $\t$ and exotic theories}
\label{sec:exotic}

So far we discussed the theories with spatial noncommutativity, which
arise from Dirichlet branes with a $B$ field along the spatial
directions.  There are also limits with timelike $B$ field, leading to
exotic \nc\ string and membrane theories, the \nc\ open string (NCOS)
theory \cite{Gopakumar:2000na,Seiberg:2000ms}, the open membrane (OM)
theory \cite{Bergshoeff:2000ai,Gopakumar:2000ep}
and the open D$p$-brane (ODp) theories 
\cite{Gopakumar:2000ep,Harmark:2000ff}.
These appear to evade the arguments against timelike noncommutativity
in field theory.

Let us start with the D$3$-brane in \IIb\ string theory in a large
spatial $B$ field.  This theory exhibits S-duality, which maps
electric field to magnetic field and vice versa. Since constant
spatial $B$-field is gauge equivalent to a constant magnetic field
on the brane, $S$-duality must map this background into one with
large electric field. This should, in turn, lead to space-time
noncommutativity, with ${\t}^{i0} \neq 0$.

Thus, combining accepted elements of M/string duality, one concludes
that our previous arguments against time-like noncommutativity must
have some loophole.  This is correct and one can in fact take the
large electric field limit for any of the D$p$-brane theories, not
just $p=3$.  However, the details are rather different from the
spacelike case.  An electric field in open string theory cannot be
taken to be larger than a critical value $E_{c}={1\over{{2\pi\ap}}}$
\cite{Burgess:1987dw,Bachas:1996kx}.  As one approaches this limit,
since the ends of the open string carry opposite electric charges, its
effective tension goes to zero, and any attempt to reach $E>E_c$ will
be screened by string formation in the vacuum.

So, one takes the limit $E \to E_{c}$ while keeping the effective open
string tension ${\ap}_{o} = {\ap} \sqrt{E^2 \over {E_c^2 - E^2}}$
finite. It turns out that while one manages to decouple closed string
modes, the open string excitations remain in the spectrum. The
resulting theory, NCOS theory, is apparently a true string theory with
an infinite number of particle-like degrees of freedom. This is
consistent both with the earlier arguments and with the general idea
that an action with timelike noncommutativity effectively contains an
infinite number of time derivatives, thereby enhancing the number of
degrees of freedom.

The NCOS theory contradicts standard arguments from world-sheet
duality that open string theory must contain closed strings.  How this
works can be seen explicitly in the annulus diagram; extra phases
present in the non-planar open string diagrams make the would-be
closed string poles vanish.
These effects apparently resolve the unitarity problems of 
\ref{sec:quant}.C as well.

Since this string theory decouples from gravity, it provides a system
in which the Hagedorn transition of string theory can be analyzed in a
clean situation, free from black hole thermal effects and other
complications of gravitating systems \cite{Gubser:2000mf,Barbon:2001tm}.


A similar limit can be taken starting with the M-theory fivebrane
(the ``M5-brane'').
Its world-volume theory is also a gauge theory, but now involving
a rank two antisymmetric tensor potential.  The membrane is
allowed to end on an M5-brane and thus parallel M5-branes come with
light degrees of freedom (open membranes) directly analogous to the
open strings which end on a Dirichlet brane.  In the limit that the
branes coincide, the resulting light degrees of freedom are governed
by a nontrivial fixed point theory in $5+1$
dimensions, usually called the ``$(2,0)$ theory'' after its supersymmetry
algebra.  A similar limit in \IIa\ theory leads to ``little string theory,''
these theories have recently been reviewed by \citeasnoun{Aharony:1999ks}.

To get a noncommutative version of the open membrane theory,
we start with $N$ coincident M5-branes
with the background 3-form strength and the metric: 
\bea
\label{eq:tfs} & H_{012} = M^3_{\rm p} tanh {\b}, \qquad H_{345} = -
\sqrt{8} M_{\rm p}^3
{sinh {\b} \over \left(e^{\b} cosh {\b} \right)^{3\over 2}} \crt
& g_{\m\n}
= {\eta}_{\m\n},
\quad \m, \n =
0,1,2 \crt
& g_{ij} = {1\over{e^{\b} cosh {\b}}} {\d}_{ij},
\quad  i, j =
3,4,5
\crt & g_{MN} = {1\over{e^{\b} cosh {\b}}} {\d}_{MN}\quad M,N=6,7,8,9,11
\eea
where $M_{\rm p}$ is the gravitational mass scale, and  ${\b}$ is
the  parameter to tune. The critical limit is achieved by taking
${\b} \to \infty$, while keeping $$ M_{eff} = M_{\rm p} \left(
{1\over{e^{\b} cosh {\b}}} \right)^{1\over 3} $$ -- the mass scale
for the membrane stretched spatially in the $1,2$ directions --
finite. In this limit $M_{\rm p}/M_{eff} \to \infty$ and so the
open membranes propagating along the M5-branes will decouple from
gravity.

Finally, one can start with the NS fivebrane in either of the type
\II\ string theories.  Dirichlet $p$-branes can end on the NS fivebrane,
with all even $p$ in \IIa\ and all odd $p$ in \IIb, leading to open
D$p$-brane degrees of freedom.  Each of these is charged under a specific
Ramond-Ramond gauge field, and by taking a near critical electric
background for one of these fields, one can again reach a decoupling limit
in which only the corresponding open brane degrees of freedom remain,
discussed in \cite{Gopakumar:2000ep,Harmark:2000ff}.

All of these theories are connected by a web of dualities analogous to
those connecting the \con\ decoupled brane theories and the bulk
theories which contain them.
For example, compactifying OM theory on a circle leads to the NCOS
$4+1$ theory; conversely the strong coupling limit of this NCOS theory
has a geometric description (OM theory) just as did the strong coupling
limit of the \IIa\ string (M theory).

\section{CONCLUSIONS}

Field theory can be generalized to space-time with noncommuting
coordinates.  Much of the formalism is very parallel to that of
conventional field theory and especially with the large $N$ limit of
\con\ field theory.  Although not proven, it appears that quantum \nc\
field theories under certain restrictions (say with spacelike
noncommutativity and some supersymmetry) are renormalizable and
sensible.

Their physics is similar enough to \con\ field theory to make
comparisons possible, and different enough to make them interesting.
To repeat some of the highlights, we found that \nc\ gauge symmetry
includes space-time symmetries, that nonsingular soliton solutions
exist in higher dimensional scalar field theory and in \nc\ Maxwell
theory, that UV divergences can be transmuted into new IR effects, and
that \nc\ gauge theories can have more dualities than
their \con\ counterparts.

Much of our knowledge of \con\ field theory still awaits a \nc\
counterpart.  Throughout the review many questions were left open,
such as the meaning of the IR divergences found in
Sec. \ref{sec:quant}.C, the potential nonperturbative role of the
solitons and instantons of Sec. \ref{sec:sol}, the meaning of the high
energy behavior discussed in Sec. \ref{sec:quant}.D, and the high
temperature behavior.

One central problem is to properly understand the renormalization
group.  Even if one can directly adapt existing RG technology, it
seems very likely that theories with such a different underlying
concept of space and time will admit other and perhaps more suitable
formulations of the RG.  This might lead to insights into nonlocality
of the sort hoped for in the introduction.  Questions about the
existence of quantized \nc\ theories could then be settled by using
the RG starting with a good regulated nonperturbative definition of
the theory, perhaps that of \cite{Ambjorn:1999ts} or perhaps along
other lines as discussed in \rfs{opalg}.

The techniques of exactly solvable field theory, which are so fruitful
in two dimensions, await possible \nc\ generalization.  These might be
particularly relevant for the quantum Hall application.

It is not impossible that \nc\ field theory has some direct relevance
for particle physics phenomenology, or possible relevance in the early
universe.  Possible signatures of noncommutativity in QED and the
standard model are discussed in
\cite{Mocioiu:2000ip,%
Arfaei:2000kh,Hewett:2000zp,Baek:2001ty,Mathews:2000we,%
Mazumdar:2000jc} and by \citeasnoun{Carroll:2001ws}, who work with a
general extension of the standard model allowing for Lorentz violation
(see \cite{Kostelecky:2001xz} and references there) and argue that
atomic clock-comparison experiments lead to a bound in the QED sector
of $|\theta| < (10 \TeV)^{-2}$.  A \nc\ ``brane world'' scenario is
developed in \cite{Pilo:2000nc}, and cosmological applications are
discussed in \cite{Chu:2000ww,Alexander:2001ck}.

This motivation as well as the motivation mentioned in the introduction
of modeling position-space uncertainty in quantum gravity
might be better served by
Lorentz invariant theories, and in pursuing the second
of these motivations it has been suggested in
\cite{Doplicher:1994zv,Doplicher:2001qt} that such 
theories could be defined by
treating the noncommutativity parameter $\theta$ as a dynamical variable.
The space-time stringy uncertainty principle of \citeasnoun{Yoneya:1987} 
leads to related considerations \cite{Yoneya:2000sf}.

While we hope that our discussion has demonstrated that \nc\ field
theory is a subject of intrinsic interest, at present its primary
physical application stems from the fact that it emerges from limits
of M theory and string theory, and it seems clear at this point that
the subject will have lasting importance in this context.  So far its
most fruitful applications have been to duality, and to the
understanding of solitons and branes in string theory.  It is
quite striking how much structure which had been considered
``essentially stringy'' is captured by these much simpler theories.

Noncommutativity enters into open string theory essentially because
open strings interact by joining at their ends, and the choice of one
or the other of the two ends corresponds formally to acting on the
corresponding field by multiplication on the left or on the right;
these are different.  This is such a fundamental level that it has
long been thought that noncommutativity should be central to the
subject.  So far, the developments we discussed look like a very
promising start towards realizing this idea.  Progress is also being
made on the direct approach, through string field theory based on
noncommutative geometry, and we believe that many of the ideas we
discussed will reappear in this context.

Whether noncommutativity is a central concept in the full string or M
theory is less clear.  Perhaps the best reason to think this is that
it appears so naturally from the M(atrix) theory definition, which can
define all of M theory in certain backgrounds.  On the other hand,
this also points to the weakness of our present understanding: these
are very special backgrounds.  We do not now have formulations of M
theory in general backgrounds; this includes the backgrounds of
primary physical interest with four observable dimensions.
A related point is that in string
theory, one thinks of the background as defined within the
gravitational or closed string sector, and the role of
noncommutativity in this sector is less clear.

An important question in \nc\ field theory is to what extent the
definitions can be generalized to spaces besides Minkowski space and
the torus, which are not flat.  D-brane constructions in other
backgrounds analogous to what we discussed for flat space seem to lead
to theories with finitely many degrees of freedom, as in
\cite{Alekseev:1999bs}.  It might be that \nc\ field theories can
arise as large $N$ limits of these models, but at present this is not
clear.

Even for group manifolds and homogeneous spaces, where mathematical
definitions exist, the physics of these theories is not clear, and deserves
more study.  As we discussed in section \ref{sec:math}, there are
many more interesting noncommutative algebras arising from geometric
constructions, which would be interesting test cases as well.

At the present state of knowledge, it is conceivable that, contrary
to our intuition both from the study of gravity and perturbative
string theory, special backgrounds such as flat space, anti-de Sitter
space, orbifolds, and perhaps others, which correspond in M theory to
simple gauge theories and \nc\ gauge theories, play a preferred role
in the theory, and that all others will be derived from these.  In
this picture, the gravitational or closed string degrees of freedom
would be derived from the gauge theory or open string theory, as has
been argued to happen at substringy distances, in M(atrix) theory, and
in the AdS/CFT correspondence.

If physically realistic backgrounds could be derived this way, then
this might be a satisfactory outcome.  It would radically
change our viewpoint on space-time and might predict that many
backgrounds which would be acceptable solutions of gravity are in fact
not allowed in M theory.  It is far too early to judge this point
however and it seems to us that at present such hopes are founded more
on our lack of understanding of M theory in general backgrounds than
on anything else.  Perhaps \nc\ field theories in more general
backgrounds, or in a more background-independent formulation,
will serve as useful analogs to M/string theory for this
question as well.

In any case, our general conclusion has to be that the study of \nc\
field theory, as well as the more mysterious theories which have
emerged from the study of superstring duality (a few of which we
mentioned in \rfs{exotic}), has shown that field theory is a
much broader concept than had been dreamed of even a few years ago.
It surely has many more surprises in store for us, and we hope
this review will stimulate the reader to pick up and continue the
story.

\section*{Acknowledgements}

We would like to thank Alain Connes, Jeff Harvey,
Albert Schwarz and Nathan Seiberg for
comments on the manuscript, and the ITP, Santa Barbara for hospitality.
This work was supported in part by DOE grant DE-FG05-90ER40559, and by
RFFI grants 01-01-00549 and 00-15-96557.

\bibliographystyle{apsrmp}
\bibliography{nc,cds,sw,connes}

\end{document}